\RequirePackage{fix-cm}
\documentclass[twocolumn]{svjour3}          % twocolumn
\smartqed  % flush right qed marks, e.g. at end of proof
\usepackage{graphicx}
\usepackage{lineno,hyperref}
\usepackage{amsmath, amsopn}
\usepackage{mathtools}
\usepackage{bm}
\usepackage{bbm}
\usepackage{relsize}
\usepackage[ruled,vlined,linesnumbered]{algorithm2e}
\usepackage{caption}
\usepackage{subcaption}
\usepackage{float}
\usepackage{xcolor}
\usepackage{afterpage}
\newcommand{\vect}[1]{\bm{#1}}

\DeclarePairedDelimiter\floor{\lfloor}{\rfloor}
\newcommand{\ind}{\mathbbm{1}}

\newcommand{\truster}[1]{\overset{\rightharpoonup}{#1}}

\newcommand{\trustee}[1]{\overset{\leftharpoonup}{#1}}

\begin{document}

\title{Personalized multi-faceted trust modeling to determine trust links in social media and its potential for misinformation management %\thanks{Grants or other notes
}
% \subtitle{Do you have a subtitle?\\ If so, write it here}

\titlerunning{Personalized multi-faceted trust modeling in social media and its potential for misinformation management}        % if too long for running head

\author{Alexandre Parmentier \and
        Robin Cohen \and
        Xueguang Ma \and
        Gaurav Sahu \and
        Queenie Chen
}


\institute{A. Parmentier, R. Cohen, X. Ma, G. Sahu*, Q. Chen \at
              \email{\{aparmentier, rcohen, x93ma, gsahu, h445chen\} @uwaterloo.ca} \\
              Cheriton School of Computer Science, University of Waterloo \\
            \and
            *Corresponding author
}

\date{}

\maketitle

\begin{abstract}
In this paper, we present an approach for predicting trust links between peers in social media, one that is grounded in the artificial intelligence area of multiagent trust modeling.
In particular, we propose a data-driven multi-faceted trust modeling which incorporates many
distinct features for a comprehensive analysis.
We focus on demonstrating how clustering of similar users enables a critical new functionality: supporting more personalized, and thus more accurate predictions for users.
Illustrated in a trust-aware item recommendation task, we evaluate the proposed framework in the context of a large Yelp dataset.
We then discuss how improving the detection of trusted relationships in social media can assist in supporting online users in their battle against the spread of misinformation and rumours, within a social
networking environment which has recently exploded in popularity.
We conclude with a reflection on a particularly vulnerable user base, older adults, in order to illustrate the value of reasoning about groups of users, looking to some future directions for integrating known preferences with insights gained through data analysis. % ~170 words
\end{abstract}

\section{Introduction}
\label{intro}
Online sources of information are increasingly relied upon by many.
According to yearly studies by the Pew research center, the percentage of American adults using the internet has jumped from 52\% in 2000 to 90\% in 2019 \cite{internet_use_pew_2019}.
In addition to the established institutions that have made the jump from paper and TV to the web, many new blogs, content aggregators, and social networks have become a vital source in the information diet: up to 62\% of American adults rely on information shared through social media for their news \cite{social_media_pew_2015}.
A study conducted in the wake of the 2016 American election found that, among American voters, Facebook ranked as the third most relied upon source for news about the election (after Fox News and CNN), far outranking local TV and newspapers and many national stations \cite{election_information_2017}.
It is clear that the power to influence and inform has shifted drastically away from traditional institutions and into the hands of individuals. 

While this democratization of information and influence may strike one as appealing, there are reasons to be concerned about this new paradigm.
According to Facebook, throughout the 2016 American election thousands of ads designed to incite panic over gun rights and LGBTQ~\footnote{The acronym stands for Lesbian, Gay, Bisexual, Transgender, and Queer.} rights were purchased by accounts believed to be funded by the Russian government, some of them specifically targeting voters in swing districts \cite{russia_facebook_2017}.
Also in 2016, a heavily armed man broke into a neighborhood pizza parlor during business hours and fired shots after having become convinced by an online conspiracy popular on Twitter that the basement of the restaurant was used by the Clintons and other Washington elite to murder and rape children \cite{pizzagate_fisher_2016}.
%In a 2019 report, the Southern Poverty Law Center has stated that the ability to propagate hateful rumors and rhetoric online is a key factor in the 30\% increase in the number of hate crimes reported per year in America since 2014 \cite{hate_Stack_2019}.

% As can be seen from the above examples, the radically open information space online allows equal opportunity to foreign government propaganda, paranoid conspiracies, unsubstantiated rumors, and hateful rhetoric\footnote{And these are only the types of content which are \textit{allowed} - online stalking, revenge porn, harassment and bullying all exist in a legal gray-area internationally, while other types of blatantly illegal and harmful activities are also facilitated by the internet.}.
% The ``top-down'' flow of information, where professionals working in reputable institutions provide advice to a few large media companies which then distribute it through well-known channels, has lost much of its relevance in the web era.
% Now, a ``bottom-up'' system has a large sway: individuals mixing opinion, fact and style jockey for relevance in a complex network of connections, where the most entertaining, provocative and attractive often win the most attention.
% Unfortunately, much of this information (especially that related to the examples we gave above) must be considered untrustworthy: that is, information which misleads, incites, and manipulates, which does not represent the truth, and is not helpful to the lives of the people who read it.

% from 1.1 first para
With an increasing number of individuals garnering attention
from provocative posts, it becomes critical to assess whether
the content that is shown to users can in fact be trusted.
One way to address the existence of untrustworthy information online is to deploy message recommender systems.
Rather than showing users a random sampling or chronologically ordered list of the content that has been added to the network since their last visit, artificial intelligence (AI) systems can be designed to reason about which messages should be shown to which users.
% There are a wealth of possible approaches to this problem.
% These systems can focus on the content of messages, using black-lists, natural language processing and image recognition to detect unwanted or disturbing content.
% They can model the reputations of content creators (the authors of messages and tweets), finding patterns and determining which authors have a history of pushing untrustworthy content into the network.
% Another approach is to focus the view on the network as a whole, noticing messages that are spreading quickly, or information that appears to be being spread in a manner which suggests coordination and subjecting it to further scrutiny.
% Alternatively, these systems could focus entirely on the preferences of content consumers, recognizing patterns in the reactions to content among the consumers and recommending messages which appear likely to fit into the model of a user's preferences. 
The subfield of multiagent trust modeling is especially relevant:
the future trustworthiness of an agent can be predicted based on reported
past experiences of peers with this agent.
While this research has traditionally been applied in contexts such
as selecting trusted sellers in e-marketplace environments
\cite{PTM_Zhang_2008}, a few efforts in recent times have focused on using the methods
for reasoning about reputable content in online social networks
\cite{social-trust_Sardana_2018}.
One promising new direction has been to recognize that multiple features
of the data may be relevant, and thus that a proper weighting of
these different contributing factors, when reasoning about trustworthiness,
is important. This is the basic premise of the very novel pursuit known as
multi-faceted trust modeling \cite{multi-faceted_mauro_2019,multi-faceted_fang_2015}.

In this paper, we first of all expand the horizons of multi-faceted
trust modeling in order to offer a more comprehensive treatment
of the different features under consideration.
We then introduce a very important new focus on supporting personalized
solutions of trust modeling for users.

% from 1.3
% We note that personalized predictions of trust are important
% but are difficult to achieve. As we argue in Section 3, the concept of trust must be fundamentally subjective.
% The giving of trust from one agent to another implies that the giver is willing to take on some amount of risk in their interactions with the receiver - this willingness is based on a myriad of personal factors (including the giver's sensitivity to risk).
% Therefore, when models of trust formulation are designed, it is important to include an aspect of personalization.
% A ``one size fits all'' approach to trust prediction, where a single model of trust formulation is applied to all agents in a network, is essentially ignoring the fact that trust is subjective.
% In Section \ref{chapter_PMFTM}, we experiment with incrementally personalizing a multi-faceted trust model.
We do this by adding an unsupervised clustering step before trust formulation models are fit, and learning a distinct model for each cluster of users.
This approach allows groups of similar users, who potentially express trust in similar ways, to have a model fit for their community, rather than receiving trust predictions that have been smoothed out to apply well to the entire population of the network.
As will be explained, we consider trust to be subjective and thus
believe it critical to move beyond the standard view of current multiagent
systems trust modeling which adopts, for all agents in the network, a
``one size fits all" approach to trust prediction; we reveal how recommendations
that support personalization can lead to improved predictions.

In order to demonstrate the value of our approach, we apply
our methods in the context of recommending items to users,
making use of a Yelp dataset of reviews which indicates user preferences.
We demonstrate the value of integrating multi-faceted trust modeling
which explicitly reasons about how to weight the different trust
indicators, and of supporting personalized predictions of trust links
when recommending content to users. Following our results, we reflect
further on how to extend our current algorithms and implementation, and
then explicitly discuss how the methods proposed in this paper can be used
towards helping to moderate online social network content for users.
We also illustrate the value of reasoning about
certain classes of users such as older adults, to consider solutions
which cater to the general needs and preferences of this user base.

\section{Background}
\label{section-background-multi-facet}
Trust modeling is a subfield of study within the artificial intelligence
research area of multiagent systems \cite{weiss2013multiagent}.
We consider trustworthiness as defined in \cite{survey-trust_Cho_2015},
namely the quality of being worthy of trust: in essence, the truster shows
willingness to take risk based on a belief that the trustee is expected to
exhibit reliable behaviour, drawing from an assessment of past experience.

\subsection{Multiagent trust modeling related work}
Multiagent trust modeling algorithms seek to predict the trustworthiness of another agent based upon first hand experience and reports provided by other agents in the environment, sometimes referred to as advisors~\cite{PTM_Zhang_2008}.

The beta reputation system (BRS) proposed by Jøsang and Ismail~\cite{BRS_Josang_2002} estimates reputation of selling agents using a probabilistic model.
The beta distributions are a family of statistical distribution functions that are characterized by two parameters $\alpha$ and $\beta$. The beta probability density function is defined as follows:

\begin{equation}
    beta(p|\alpha, \beta) = \frac{\Gamma (\alpha + \beta)}{\Gamma (\alpha) \Gamma (\beta)} p^{\alpha - 1}(1 - p)^{\beta - 1}
\end{equation}

where $\Gamma$ is the gamma function, $p \in [0, 1]$ is a probability variable, and $\alpha, \beta > 0$.
This function shows the relative likelihood of the values for the parameter $p$, given the fixed parameters $\alpha$ and $\beta$.

Ratings from peers (used to estimate the reputation of a seller) are binary in this model (1 or 0, to represent that the advisor considers the seller to be satisfactory or dissatisfactory in a transaction).
Individual ratings received are combined by simply accumulating the number of ratings supporting the conclusion that the seller has good reputation and the number of ratings supporting the conclusion that the seller has bad reputation.

The prior distribution of the parameter $p$ is assumed to be the uniform beta probability density function with $\alpha = 1$ and $\beta = 1$. The posteriori distribution of $p$ is the beta probability density function after observing $\alpha - 1$ ratings of 1 and $\beta - 1$ ratings of 0.

The reputation of the seller $s$ can then be represented by the probability expectation value of the beta distribution, which is the most likely frequency value, used to predict whether the seller will act honestly in the future.
The formalization of this is given as follows:

\begin{equation}
    Tr(s) = E(p) = \frac{\alpha}{\alpha + \beta}
\end{equation}

Zhang and Cohen~\cite{PTM_Zhang_2008} suggest  a  personalized trust model (PTM)  to  determine whom to  listen  to  among  a  network  of  buyers  and  sellers  in  thee-marketplace  domain.
In  particular,  they  address  whether  a buyer, \textit{b},  should  purchase  a  product  from  a  seller, \textit{s},  based on  a  combination  of global advice  from  other  buyers  (i.e., advisors, \textit{a}), and \textit{b}'s  own \textit{local} past experiences with \textit{s}.

The PTM global metric is further broken down to combine \textit{private} and \textit{public} trust estimates of advisors.
The intuition is that \textit{b} may have radically different expectations or preferences regarding \textit{s}'s product than \textit{a}, and so \textit{b} should have some notion of  how  much  to  trust \textit{a}.
To the extent that \textit{b} relies  on  past common  experiences  to  evaluate \textit{a}'s  trustworthiness, \textit{b} uses a \textit{private} trust metric to  incorporate \textit{a}'s recommendation.
To the extent that \textit{b} relies on \textit{a}'s similarity to the global rating of various sellers (i.e., how fair are \textit{a}'s ratings), \textit{b} uses a \textit{public} trust metric to incorporate \textit{a}'s recommendation.

In particular, \textit{b}’s private reputation according to \textit{a}, $R_{pri}(a,b)$, is modeled by the expectation of a beta distribution where $\alpha$ is the number of times \textit{a} and \textit{b} have agreed in the past about the reputation of other agents, and $\beta$ corresponds to how many times they have disagreed. 
The public reputation of \textit{b}, $R_{pub}(b)$, is again modeled by the expectation of a beta distribution, where $\alpha$ corresponds to the number of times \textit{b}’s advice has agreed with majority opinion, and $\beta$ the number of times it has not.
The final reputation of \textit{b} for\textit{a} is then a linear combination of the private and public reputation of \textit{b}, weighted by a factor \textit{w} which reflects how much comparable experience \textit{a} has had with \textit{b} (i.e. the number of agents commonly rated agents).

\begin{equation}
    T(a, b) = w R_{pri}(a, b) + (1 - w) R_{pub}(b)
\end{equation}

\subsection{Multi-faceted trust modeling}
Multi-faceted trust modeling (MFTM) is a flexible and data driven approach to trust modeling.
Inspired by work in the social sciences which have outlined the numerous variables which influence the formation of trust relationships \cite{integrative-trust_Mayer_1995}, MFTM incorporates arbitrarily many indicators of trustworthiness into a single (optionally context-dependent) trustworthiness score.
Operationalizing this core idea for trust and social tie prediction has been proposed by multiple researchers (e.g. \cite{multi-faceted_mauro_2019,multi-faceted_fang_2015,multidimensional_Kwon-2009,gilbert_predicting_2009,multi-faceted_Jia_2013}).
As is evident in these works, there is little agreement over whether this technique should be called multi-dimensional, multi-faceted or composite trust modeling, and this confusion has likely led to some difficulty in coordinating efforts in this research direction.
We use the term ``multi-faceted'', in keeping with the most recent works.

The defining feature of an MFTM is a customizable vector of trust indicators, where each indicator is a real number based on two agents:
\begin{equation}
\vect{\Psi}(a_1,a_2) = \langle \psi_1(a_1,a_2), \psi_2(a_1,a_2), ..., \psi_n(a_1, a_2)\rangle
\end{equation}
A ``trust indicator'' can be thought of as a piece of evidence for or against trusting an agent under a particular context.
For example, $\psi_1(a_1,a_2) = friendCount(a_2)$ may be relevant to assessing the reputation of $a_2$ in a domain where only popular and reputable agents can accrue large numbers of friends.
The indicators $\psi_i$ must be computable given $A$ and $E$ (the set of agents and their attributes and the history of events).
One important feature of MFTM is its flexibility to tune its parameters to different domains of use.
The customizability of MFTM is highly attractive for application to social networks, as it is rare to find explicit statements of trust encoded into the feature set of online environments\footnote{For example, social network designers could elicit explicit statements of trust from users, but such a feature is not currently popular online.}.
Instead, an arbitrary number of ``imperfect'' indicators of trustworthiness, such as popularity, friendship, reputation, interaction history, preference similarity and institutional credibility can be considered as each contributing to a final tally of trustworthiness.
Clearly, the underlying assumption of this model is that the existence of trustworthiness between two agents can be predicted based on a comparison of the attributes and behaviors of those agents.

The consideration of multiple indicators of trust can be viewed as an emulation of the way in which humans consider multiple sources of evidence when deciding to trust or not \cite{integrative-trust_Mayer_1995}.
For example, consider the problem of choosing an auto mechanic shortly after having moved to a new town.
In this case, one has no interaction history with any nearby mechanics and must weigh available evidence in order to choose which mechanic to trust.
In a simple case, one might only consider two pieces of evidence towards or against a mechanic: has any colleague recommended them ($\psi_1$), and have their prices been posted clearly online ($\psi_2$).
In this case both indicators are binary, and it seems likely that the mechanic $a_j$ who satisfies both indicators, $\Psi(\truster{a_i}, \trustee{a_j}) = \langle 1, 1 \rangle$, will be a good candidate to trust.
(Note that we use $\truster{}$ for the trustor and
$\trustee{}$ for the trustee at times in our discussion below,
for additional clarity).

In order to predict trustworthiness, the relevance of each indicator can be learned using an off-the-shelf machine learning technique given $A$ and $E$ to train with.
To do this, a trust link is chosen as a target of prediction : $y$ (e.g. explicit statements of trust/friendship, high degrees of preference alignment).
Then, given the set of existing implicit/explicit trust links, a machine learning model fits a classifier $\hat{f}$ to the function that determines how trust indicators are related to trust links $f: \Psi(a_1,a_2) \rightarrow y$.
For example, in the case where logistic regression % (see Section \ref{logistic_regression})
is used, $y$ will be binary and we have:
\begin{equation}
T_c(\truster{a_1}, \trustee{a_2}) = P_{A,E}(\truster{a_1}, \trustee{a_2}, c) =  \frac{1}{1+\exp^{-({\theta \cdot \Psi(a,b))}}}
\end{equation}
where $\theta$ is the vector of weights learned through the logistic regression process and $T_c$ is trustworthiness under context $c$.
(We believe that whether someone is trusted may truly vary according
to the context; for the remainder of the paper, we drop the
context variable $c$ in the equations).
We wish to emphasize that while logistic regression is an elegant and natural choice with some popularity in the literature (e.g. \cite{multi-faceted_fang_2015}), it is by no means the only choice.

The ability to define custom indicators appropriate to whichever application domain one is pursuing offers a tremendous amount of flexibility.
As we will show in Section \ref{chapter_PMFTM}, both highly generic as well as application-specific trust indicators can be defined.

Finally, we wish to emphasize how MFTM can be seen as a generalization of a number of existing trust modeling techniques. 
Primarily this is because many trust modeling techniques do in fact consider multiple sources of evidence, but they weigh or combine this evidence in a non-data-driven manner.
For example, the beta reputation system can be configured so that old advice is considered less important than new advice.
However, a method for specifying \textit{how much more} important newer advice should be treated compared to older advice is not specified.
A similar situation occurs in the Personalized Trust Model \cite{PTM_Zhang_2008}, where private and public reputation are weighed against each other.
The weighting function chosen has a good statistical justification\footnote{Based on the Chernoff bound theorem.}, but ultimately does not specify how error bounds should be chosen, and thus how exactly to weigh personal and private reputation.
MFTM can consider arbitrarily many sources of information, and learns the weights for them directly from data.
For example, PTM could be roughly replicated by treating private and public reputation as trust indicators, and learning an appropriate function for combining them.

Another example of how MFTM is data driven is that it does not specify which distributions should be used to model beliefs.
For example, both the foundational trust modeling Beta Reputation System (BRS) \cite{BRS_Josang_2002} and PTM rely heavily on the beta distribution.
% While this choice is statistically justified (assuming the behaviour of agents is governed by a random process\footnote{An assumption that we should not generally expect to hold in real world data sets.}), it also constitutes a form of bias, and is vulnerable to abuse.
% For example, if one knew that the BRS was being used to model seller reputation on an e-marketplace, they could increase profits and maintain a high reputation simply by only scamming every 10th customer, or by acting honestly for every small purchase and scamming the less frequent buyers of expensive items \cite{kerr2009smart}.
By allowing arbitrary machine learning methods to combine many forms of trust evidence into prediction, MFTM loses Bayesian rigor, but gains a large degree of flexibility and generalizability.

\subsection{Trust-aware recommendation systems}

Trust modeling in the context of recommender systems has been
examined by several researchers, dating back to the seminal
paper of O'Donovan and Smyth \cite{o2005trust}.
More recent work has examined such issues as addressing cold start
recommendation using trust modeling \cite{guo2015trustsvd}
or examining how to speed up trust-aware recommendation through
improvements from matrix factorization \cite{ijcai2019-191}.
In this paper, trust-aware recommendation arises
as a central element of the validation of our proposed framework.

To explain: one of the recurring challenges in the development of trust models is finding grounds for the validation of the accuracy of the models \cite{survey-trust_Cho_2015}.
Trust models aim to predict new trust links but independent agents may choose to follow or ignore these predictions.
It is therefore difficult to truly evaluate the effectiveness of models without deploying a system on an active service and measuring the real effects of trust link prediction.
As this is expensive and requires the cooperation of an active social network service, many models validate their effectiveness on data generated by an agent simulation instead (e.g. \cite{BRS_Josang_2002,PTM_Zhang_2008,blade_regan_2006,LOAR_Champaign_2011}).
While this is a useful approach for contrasting the effectiveness of various models and gives the researcher a large amount of control for simulating specific types of agent behavior, it clearly adds a layer of ambiguity between the reported effectiveness of the model and its potential for real world application.
In some cases, merely changing simulation parameters can defeat systems that had performed well on the simulations their creators had designed \cite{kerr2010treet}.

A rising trend in this field is to validate models by applying their predictions to a recommendation task (e.g. \cite{multi-faceted_fang_2015,multi-faceted_mauro_2019}): that is, using the trust model to predict novel trust links, $\hat{\Gamma}$, in a multiagent system (MAS), then feeding those predicted links into a trust-aware item recommendation system.
These trust-aware recommender systems incorporate both user-item rating behavior and user-user social/trust connections to better recommend items by leveraging the fact that social/trust connections exert influence on the preferences of agents (e.g. you are more likely to watch/enjoy a film a trusted friend recommends). %\footnote{See Section \ref{section_latent_factors}.}.
The logic of this two part process is that when a trust model is able to accurately predict trust links in the context of peer to peer item recommendation, then the resulting accuracy of the recommender system trained with those links will improve.

For the validation of the model we present in Section \ref{chapter_PMFTM}, we implement a task-aware item recommendation task. % on a Yelp dataset.
%We note that trust modeling and recommender systems have important
%similarities and differences.
%Consulting peers for recommendations is a process common to both enterprises.
%However, trust modeling usually considers an aspect of personal experience, distinct from mere similarity as used for collaborative filtering \cite{trust_in_rs_donovan_2005}.}
As will be explained in more detail in Section \ref{rec_eval}
we introduce two distinct trust-aware recommendation systems,
TrustMF \cite{trustmf_2013} and MTR \cite{multi-faceted_mauro_2019}.
MTR belongs to a class of recommenders based on k-nearest
neighbours~\cite{DBLP:books/mk/HanKP2011}.
This approach requires a good selection of the value of $k$ and an appropriate distance metric to determine closeness.
In section \ref{rec_eval} we provide more insights into how these were
chosen for our experimentation with MTR.
TrustMF belongs to a class of systems known as latent factor models.
To be clearer about how these systems operate, we provide below additional
explanation.
As will be seen, this method works well with data-driven trust
recommendation, in seeking to leverage the most relevant factors
of the users.

\subsubsection{Latent factor models for recommendation}
Latent factor models for recommendation are a popular approach to collaborative filtering based recommendation derived from matrix factorization technique called Singular Value Decomposition (SVD) \cite{svd_Sarwar-2000}.
Specifically, by applying an SVD technique,  a $m \times n$ matrix $R$ of rank $\ell$ can be decomposed into three matrices of rank $k \leq \ell$:
$R = Q \cdot S \cdot V$,
where $Q$ is $m \times k$, $S$ is $k \times k$ and $V$ is $k \times n$.
While $S$ has a number of interesting mathematical properties, in recommender system literature it is frequently ignored by substituting $U = Q \cdot S$.
%
% \begin{equation}
% \label{eqn_svd}
% R_k = U \cdot V
% \end{equation}

% This decomposition is guaranteed to exist and provide the best rank-k approximation of the
% matrix R with respect to the Frobenius norm \cite{svd_Sarwar-2000}.
% That is,  $R_k = U  \cdot V$ is the matrix that minimizes $||R - R_k||_F$ where $||A||_F = \sum^m_{i=1}\sum^n_{j=1} |a_{ij}|^2$.
% Phrased otherwise, this decomposition procedure can be used to ``compress'' a large matrix by splitting it into the two rank-k matrices which, when multiplied together, best reproduce the original matrix.

SVD can be applied to recommender systems when $R$ is the user-item matrix of review scores such that $r_{ij}$ is the rating user $i$ gave to item $j$. %(e.g. on a binary ``recommend or not'' scale, or on an ordinal 1-5 rating system). 
Naturally, this matrix is sparse - in practice, the vast majority of the entries in $R$ are unknown, as most users have only given feedback on a small number of items.
While SVD cannot be applied directly to a sparse matrix like $R$, we can imagine that the defined entries in $R$ comprise a subset of the entries in the (unknown) dense matrix $R'$ where every user has expressed an opinion on every item.
By SVD, $R'$, is guaranteed to have a minimal rank-k decomposition.
This line of reasoning serves as inspiration for the following loss function \cite{netflix_Koren-2009}:
\begin{equation}
\label{eqn_svd_loss}
\text{{\large $\min_{u_*, v_*}$}} \sum_{(i,j) \in \kappa} (r_{ij} - u_i^T v_j)^2 + \lambda(||u_i||^2 + ||v_j||^2)
\end{equation}
where $u_i$ is a length $k$ vector corresponding to user $i$ and $v_j$ is a length $k$ vector corresponding to item $j$, and $\kappa$ is the set of indices $(i, j)$ such that $r_{ij}$ is defined in $R$.
$\lambda$ simply controls the strength of the regularization penalty.
By optimizing Equation \ref{eqn_svd_loss}, one constructs matrices $\hat{U}$ and $\hat{V}$, where the $i$'th row of $\hat{U}$ is $u_i^T$ and the $j$'th column of $\hat{V}$ is $v_j$. Then, $ \hat{R} = \hat{U} \cdot \hat{V}$ is a matrix where the distance between defined members of $R$ and their corresponding entries in $\hat{R}$ has been minimized.
At the same time, estimates for every undefined entry in $R$ are present in $\hat{R}$. A user $i$ can then be recommended items where $r_{ij}$ is undefined (the user has not yet rated the item) but $\hat{r}_{ij}$ is high (the user is predicted to rate the item highly).

This approach is particularly amenable to the recommendation task, as it makes the optimization far more tractable. 
In particular, rather than grappling with the $O(mn)$ user-item ratings directly, the $O(k(m+n))$ values in $\hat{U}$ and $\hat{V}$ are all that need to be optimized.
This offers considerable performance improvements when $k << min(m, n)$  (in many applications there may be millions of users and items, but $10 \leq k \leq 100$ factors are sufficient for good modeling of the system \cite{netflix_Koren-2009}).
%Additionally, the running time of a single loop of the optimization equation is linear in the number of observed ratings ($\kappa$ above).

Koren et al. \cite{netflix_Koren-2009} describe the intuition behind this procedure in illuminating way.
For the task of recommending movies, we can imagine that each movie can be measured on $k$ dimensions. 
%For example: funniness, seriousness, amount of action, quirkiness, etc.
Each user will have some level of preference for various dimensions of a movie. 
%A user will enjoy a movie when the movie has high extension in the dimensions the user enjoys: this user prefers a mix of comedy and quirkiness, that user long dramas with character development.
Rather than explicitly defining these $k$ dimensions and laboriously categorizing each movie in this way, the SVD recommendation procedure infers factors directly from rating patterns.
These so-called ``latent factors" are essentially learned
via error minimization over available data regarding users and movies.

% While the actual distance between $R'$ and $\hat{R}$ is unknowable, this approach has been shown to perform well in numerous settings, including winning an international contest hosted by Netflix \cite{netflix_Koren-2009}.
% Further, minimizing Equation \ref{eqn_svd_loss} has been shown to be equivalent to maximizing the probability of latent factor matrices $U$ and $V$ given $R$ under a simple probabilistic model \cite{probabilistic-factorization_Salakhutdinov-2007}.

TrustMF is used to test the accuracy of our trust link
prediction algorithm in Section 3.
This system can be roughly characterized as combining the optimization described above with an optimization over a matrix of user-user trust links that shares a latent space with the user-item matrix.
%That is, in addition to learning matrices $\hat{U}$ and $\hat{V}$ describing user preferences and item factors respectively, a third matrix $\hat{W}$ representing user factors is learned.
%The matrix $\hat{U}$ now serves a dual purpose.
%Like before, the distance between $r_{ij}$ and $u_i^Tv_j$ is minimized, but in addition the distance between $l_{ij}$ and $u_i^Tw_j$ is minimized, where $l_{ij} > 0$ only when a trust link exists between users $i$ and $j$.
Conceptually, a user's preference for items shares a space with that user's preferences for trusting other users. 
Thus, the presence of trust links exerts an influence over the latent factors that are discovered, incorporating social trust into the recommendation process.
The ability to incorporate social trust data  makes this recommender system ``trust-aware''.

\section{Personalized Multi-Faceted Trust Modeling}
\label{chapter_PMFTM}

In this section we describe an experiment which explores the influence of personalization and context on a multi-faceted trust model.
We aim to demonstrate the benefit of learning trust formulation behaviours at the level of clusters of users, rather than on the entire population of agents.
We argue that this increase in resolution constitutes a form of personalization (albeit, performed at a group level rather than at an individual level).
In addition, we explore the impact of considering differing contexts of trust by testing the effect of predicting two types of trust links. 

At the heart of our solution is an effort to predict novel trust links in a social network by using machine learning methods to determine how to weight feature importance, and to approximate trust formulation procedures among groups of similar agents.
Our approach makes use of the flexibility of MFTM, which we demonstrate by combining features drawn from two existing proposals with our own novel features.
Evaluation is performed by measuring the error rates on a recommendation task that incorporates trust information.
This is performed on a data set collected from Yelp\footnote{\url{https://www.yelp.com/}}, a content rating site with social network features.

% \subsection{Use Case}

% As we have argued in Section \ref{section-background-multi-facet}, multi-faceted trust modeling provides a highly flexible and general framework for the application of trust models to various domains. 
% In this work, we argue specifically for the application of MFTM to social networks, as these domains are replete with possible trust-relevant indicators.

\begin{figure}
	\centering
	\includegraphics[width=0.5\textwidth]{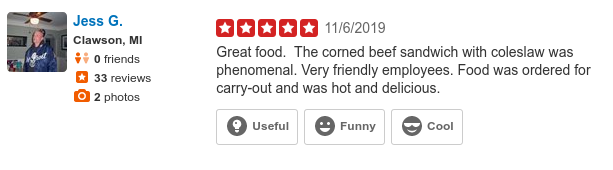}
	\caption{Example Yelp review with rating}
	\label{fig_example_review}
\end{figure}

%We used data from Yelp, a popular social-network and item-rating service to train and evaluate our models.
On Yelp, users can indicate binary social trust towards other users (friends) and submit ratings for products, businesses or websites (taken together, and following the trend in recommender systems literature, these entities are called ``items'') that they have experienced, indicating their satisfaction with that item.
These ratings are integers in the range $[1, 5]$, illustrated as stars, where higher numbers indicate a stronger recommendation.
An example 5-star review for Schwartz's Deli is presented in Figure \ref{fig_example_review}.
We used this data set particularly because it is amenable to validation of trust model effectiveness via a downstream item recommendation task.
Yelp was in fact used by \cite{multi-faceted_mauro_2019}, one of
the central multi-faceted trust modeling papers which motivated our work.

As will be seen in our experimentation below,
trust links for the Yelp environment will be predicted
both on the basis of friendship relations and
through the discovery of similar rating behaviour.

% \subsubsection{Choice of data set}

% TrustLink paragraph
% To be clearer on the concept of a trust link, we assume that each
% agent can model the trustworthiness of every other agent in the
% community, and those agents with sufficiently high trustworthiness
% can be trusted. When one agent believes another can be trusted, we
% view there being a trust link from the first agent to the second.
% In the context of item recommendation, this translates into the first
% agent being willing to accept the rating offered by the second.

% More information about the Yelp data set will be presented below in Section \ref{section_datsets}.

\subsection{Personalization}
The rationale for testing the effect of personalization is simple: we expect trust formulation procedures to vary from person to person, therefore learning approximations of trust formulation procedures may be more accurate on a more personalized scale.

% We have already explained that trust is a subjective phenomenon, and from personal experience we can verify that trust formulation procedures among humans vary.
% For example, some of us place a heavy importance on shared history and community involvement when choosing a car mechanic to trust, while others place importance on popularity and creative radio ads.
% As trust is inherently subjective, we do not impose the belief that one individual's trust formulation procedure is more correct than another's.

The inherent subjectivity of trust has important implication from a machine learning perspective.
In particular, it implies that trust predictors trained on large data sets representing the behavior of many individuals are not necessarily more correct than those trained on smaller groups.
While the predictor trained in the former case will likely have a higher accuracy across the broad population, it is essentially learning the ``average'' trust formulation procedure, potentially disadvantaging agent's whose preferences are not aligned with the population at large.
%As we expect trust formulation to vary significantly from individual to individual, a method of personalizing predictions is desired.

%The approach to personalization we take is to cluster users based on a) item and b) social preference similarity, then learn distinct predictors based on the data associated with each cluster of users.
We note that learning distinct predictors based on the data
associated with clusters of users was suggested by Fang et al. \cite{multi-faceted_fang_2015}, but was not pursued.
While our approach is useful for capturing the variance of trust formulation in smaller groups, it does not attempt to learn the preferences of individual agents.
We will discuss possible avenues for truly individual personalization of trust modeling and other approaches in Section \ref{future_chapter_4}.

\begin{figure}
	\centering
	\begin{subfigure}[b]{.49\textwidth}
		\includegraphics[width=\textwidth]{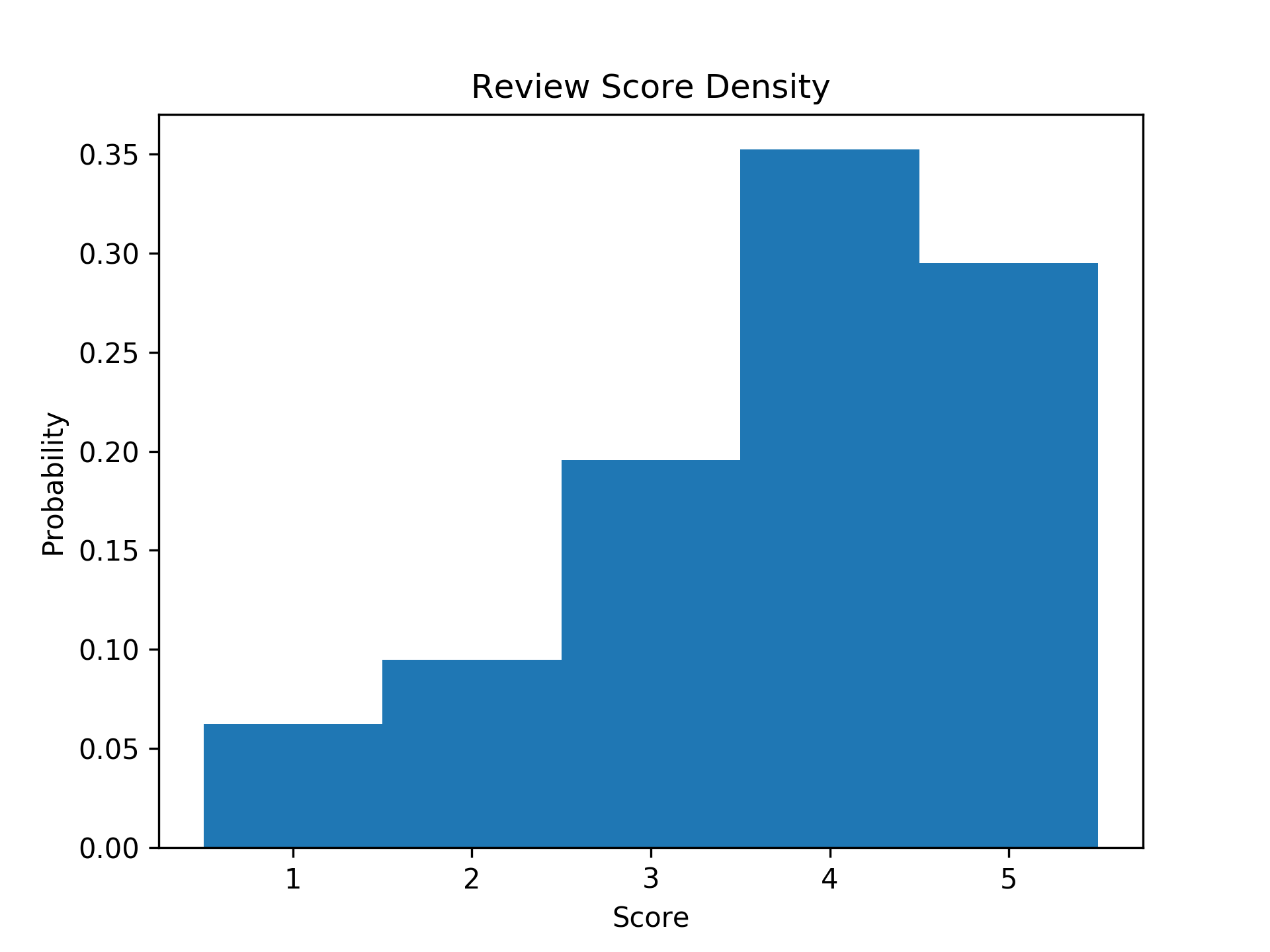}
		\caption{Review score distribution}
	\end{subfigure}
	\begin{subfigure}[b]{.49\textwidth}
		\includegraphics[width=\textwidth]{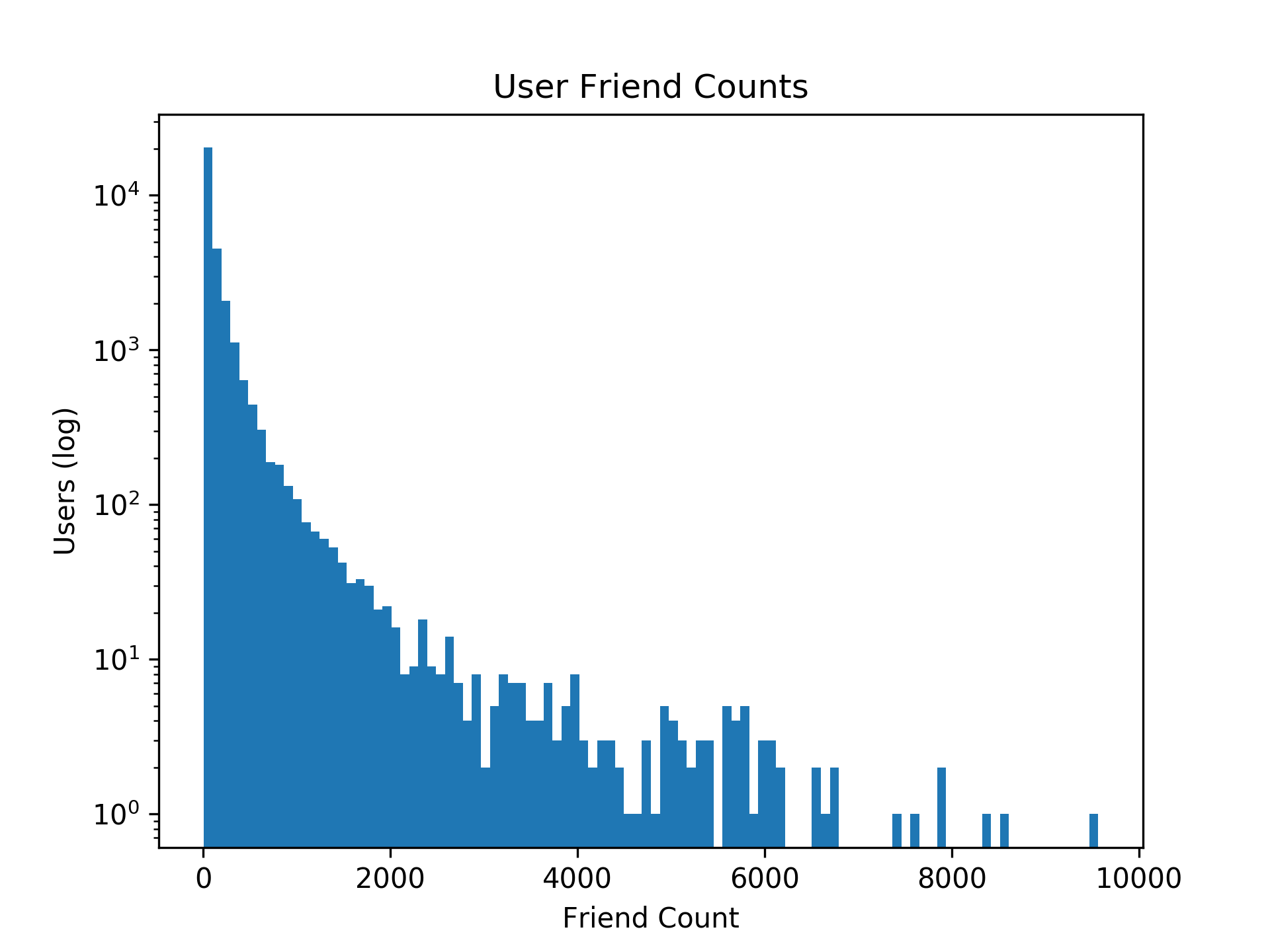}
		\caption{Friend counts}
	\end{subfigure}
	\caption{Review scores and friend counts}
	\label{fig_review_dist}
\end{figure}
	
\begin{figure}
	\centering
	\begin{subfigure}[b]{.49\textwidth}
		\includegraphics[width=\textwidth]{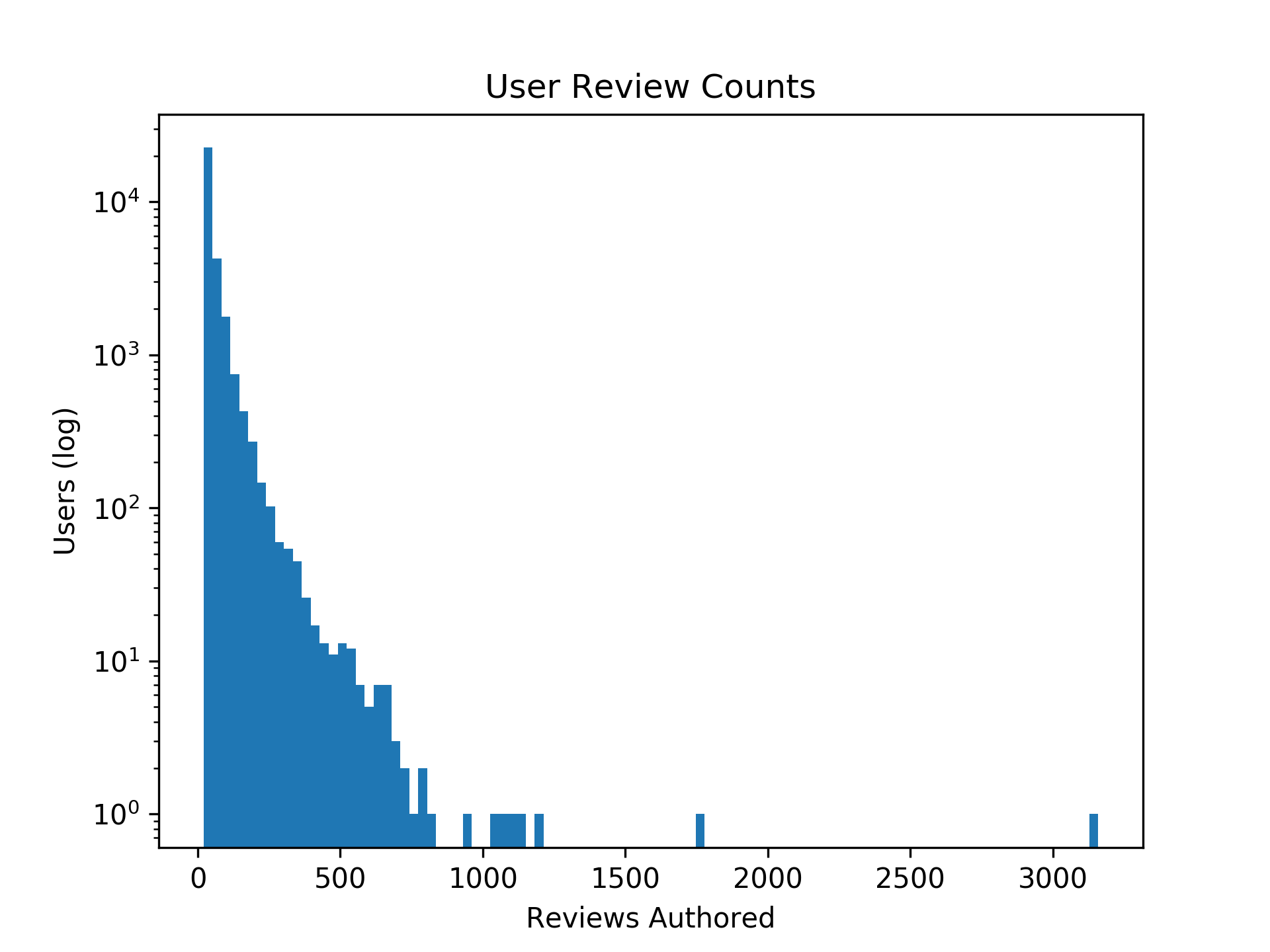}
		\caption{Number of reviews submitted per user.}
	\end{subfigure}
	\begin{subfigure}[b]{.49\textwidth}
		\includegraphics[width=\textwidth]{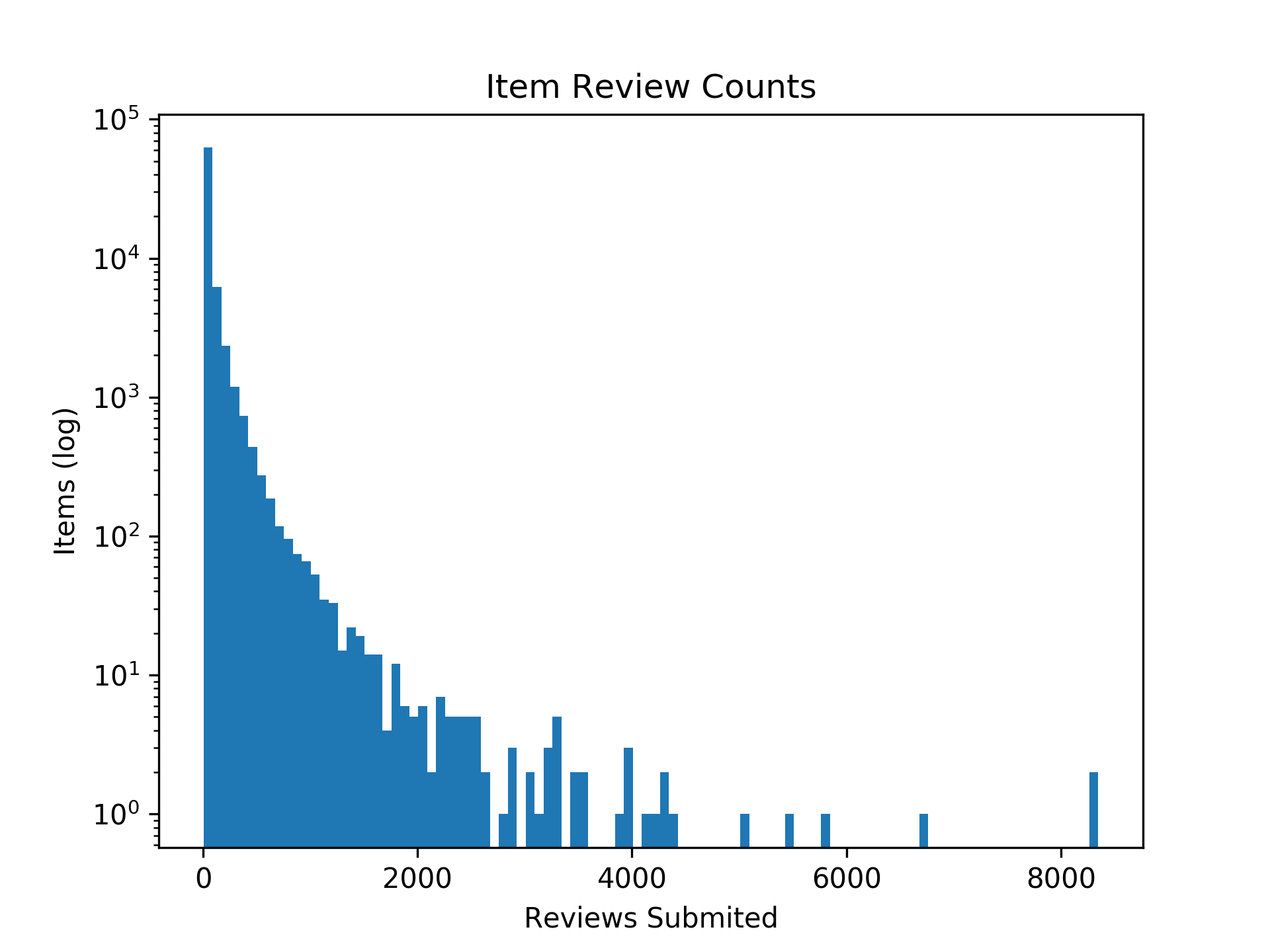}
		\caption{Number of reviews received per item.}
	\end{subfigure}
	\caption{Reviews submitted and received counts}
	\label{fig_user_review_hist}
\end{figure}

\subsection{Methodology}

In this work we test the effect of personalization or trust link prediction on an item recommendation task.
This is the case where agent $a_i$ encourages agent $a_j$ to invest resources into accessing or consuming item $k$ based on their own experience with it.
In particular, we test whether clustering agents and learning trust link predictors on the basis of clusters of similar agents, as opposed to  learning a single trust link predictor for the entire population of agents, can increase the accuracy of trust-aware item recommendation systems.
We also test the effect of altering the type of trust link prediction by either a) attempting to predict the presence of an explicit friend/trust link or b) predicting positive correlation in item review scores (i.e. two agents having scores that are the same).
\footnote{Predicting positive review score correlation between agents
directly may be helpful, since some research suggests
that friendship only correlates weakly with similarity
in reviewing behaviour \cite{librec_Guo_2015}.
This may be viewed as an implicit link between agents.}

% \newpage
\begin{table*}[h]
\centering
\begin{tabular}{p{2.5cm}|p{8.5cm}}
    \textbf{Experiment Name} & \textbf{Experiment Description} \\ \hline
	RealFriends & Perform no prediction of trust links whatsoever. Use the real, explicit trust/friend links in the data set.\\ \hline
	FriendPrediction & Predict trust/friendship links with no personalization step (i.e. learn one trust predictor for the entire population of agents).\\ \hline
	PrefPredict& Predict positive review score correlation with no personalization.\\ \hline
	PrefCluster-PrefPredict& Determine clusters of agents with similar preferences for items (i.e. positive item review score correlation) and predict positive review score correlation links for each cluster.\\ \hline
	PrefCluster-FriendPredict& Determine clusters of agents with similar preferences for items (i.e. positive item review score correlation) and predict trust/friendship links for each cluster.\\ \hline
	SocialCluster-PrefPredict& Determine clusters of agents with high overlaps in their social circles and predict positive review score correlation links for each cluster.\\ \hline
	SocialCluster-FriendPredict& Determine clusters of agents with high overlaps in their social circles and predict trust/friendship links for each cluster.
\end{tabular}
\caption{Experiment descriptions}
\label{table_experiments}
\end{table*}

Our final analysis will report the recommendation accuracy of 7 configurations, where each configuration uses an identical set of agents and recommendation procedures, but a different procedure for predicting the trust links between agents.
Each configuration is given a name reflecting which (if any) type of clustering was performed, and which type of trust link was predicted.
The complete list is presented in Table \ref{table_experiments}.

The entire procedure can be described sequentially, as follows.
Each step will be briefly explained, noting its inputs and outputs, then will be more carefully considered in subsections below.

When the first step is skipped, no personalization is performed. % (i.e. the first three items in the list above).
When the second step is skipped, no trust link prediction is performed. % (i.e. the first item in the list above).
The rationale of skipping certain steps is to compare and contrast the effect applying these steps has on the final accuracy of the recommendation task.

The overview presented below assists readers in clearly
understanding the interplay between the central processes of our
approach: clustering, trust link prediction, and recommendation evaluation,
before the details of our methodology are provided.

\begin{itemize}
	\item \textbf{Clustering}
	\begin{itemize}
		\item \textbf{Input:} All agents $A$ and an agent-agent similarity matrix, $S$.
		\item \textbf{Output:} An assignment of every agent to a cluster, $C$
		\item \textbf{Description:} Partition the agents into groups of highly similar agents. 
		We used social circle overlap (Jaccard Similarity) and review score correlation (Pearson Correlation Coefficient) as similarity measures.
		We developed two clustering methods for this step.
	\end{itemize}
	\item \textbf{Trust Link Prediction}
	\begin{itemize}
		\item \textbf{Input:} Clusters of agents, $C$, and trust indicator function $\Psi(a,b)$.
		\item \textbf{Output:} A matrix of trust link predictions, $\hat{\Gamma}$
		\item \textbf{Description:} For each cluster $c_l$ of agents a logistic regression learns a distinct MFTM trust prediction function for that cluster. 
		We experimented with predicting friendship links and positive review score correlation.
		Output a $|A| \times |A|$ matrix, $\hat{\Gamma}$, where, $\hat{\Gamma}_{ij} = 1$ if the classifier for the $i$'th agent's cluster predicts a trust link between agents $i$ and $j$ and $0$ otherwise.
	\end{itemize}
	
	\item \textbf{Recommendation Evaluation}
	\begin{itemize}
		\item \textbf{Input:} Agent-item rating matrix $R$, trust link prediction matrix, $\hat{\Gamma}$.
		\item \textbf{Output:} A agent-item matrix of predicted review scores, $\hat{R}$.
		\item \textbf{Description:} Given reviews present in the original data set and the predictions from the previous step train a trust-aware recommender system to predict review scores. 
		After training, we evaluate the correctness of the recommender on a reserved testing set using Mean Absolute Error (MAE) and Root Mean Squared Error (RMSE) metrics.
	\end{itemize}
\end{itemize}

\subsubsection{Clustering}
%Our main hypothesis involves exploring the effect that personalization can have on the effectiveness of trust link prediction.
%However, currently data sets make it difficult to test truly individual personalization.
We note that truly individual personalization is difficult to test.
This is because explicit elicitation of factors which influence trust on a personal level is rare on most services, and most agents have not participated in enough activity in order to accurately measure the patterns of their preferences implicit in their behavior.
Therefore, we focus on clusters of similar agents rather than considering each agent distinctly.
We posit that if personalization at this level of granularity is sufficient to increase the accuracy of our trust models, then we will have found evidence that some level of personalization is indeed useful for the trust modeling task, and will have motivated further research in the area.

Clustering procedures generally rely on the definition of a distance or closeness (alternatively, similarity) metric between elements to be clustered \cite{clustering_2005}. 
In this work, we tested two separate similarity functions.
Specifically, we tested clustering agents on the basis of preference similarity and social circle similarity, as defined in Equations \ref{eqn_preference_dist'} and \ref{eqn_social_dist'} respectively:
\begin{equation}
\label{eqn_preference_dist'}
\begin{gathered}
    prefSim'(a_i,a_j) = \\
    \frac{\sum_{k \in R_{i,j}} (r_{ik} - \bar{r}_{k}) (r_{jk} - \bar{r}_{k}) }
{\sqrt{\sum_{k \in R_{i,j}} (r_{ik} - \bar{r}_{k})^2}\sqrt{\sum_{k \in R_{i,j}} (r_{jk} - \bar{r}_{k})^2}}
\end{gathered}
\end{equation}
\begin{equation}
\label{eqn_social_dist'}
socialSim'(a_i, a_j) = \frac{|friends(a_i) \cap friends(a_j)|}{|friends(a_i) \cup friends(a_j)|}
\end{equation}
where $R_{i,j}$ is the set of items which both agents $a_i$ and $a_j$ have reviewed, $r_{ik}$ is the rating given by agent $a_i$ to item $k$, $\bar{r}_{k}$ is the average rating for item $k$, and $friends(a_i)$ is the set of agents $a_i$ has entered into mutual friendship with.
Put otherwise, we clustered agents on the basis of the Pearson Correlation Coefficient of scores they had given in reviews to items and on the basis of the Jaccard Similarity of their friend groups.

The choice of both metrics was motivated by a desire to extract metrics from our data which:
\begin{enumerate}
	\item Are relatively generic (i.e. could likely be applied to similar data sets).
	\item Could plausibly be argued to constitute a basis for determining which agents are similar enough that we might expect their trust formulation procedures to also be similar.
\end{enumerate}
Since our context of trust is based on recommending items, we argue that both criteria are met. 
For 1., we argue it is reasonable to assume that on any online service with an item review and recommendation component, it will be possible to calculate Equation \ref{eqn_preference_dist'}. 
Similarly, it is reasonable to assume that Equation \ref{eqn_social_dist'} will often be computable on these services, as it is widely believed that friend relationships are a useful tool for expressing preference alignment among agents in such domains \cite{trustsvd_2015}.
For 2., we argue that $socialSim'$ directly satisfies this criteria by its definition, as $socialSim'$ measures the observed similarity in the output of a trust-like relationship formation procedure (friendship).
For $prefSim'$, we argue that if two agents $a$ and $b$ have demonstrated a strong preference for similar items, then it is reasonable to conclude that their procedures for choosing who to trust for new recommendations should be similar. Thus, it is reasonable to cluster them under this context.
%\UPDATE{These are the two axes for clustering agents that we explore.}

While we have argued that these similarity metrics are relevant for our goals, they do present challenges as metrics for clustering algorithms.
Specifically:
\begin{itemize}
	\item Both metrics violate the triangle inequality.
	\item Both metrics can sometimes be undefined (when the denominator is 0).
\end{itemize}
As many clustering algorithms are defined over Euclidean spaces, these caveats represent significant restrictions of possible approaches. 
It is somewhat helpful that the second caveat can be addressed by simply substituting default values in the case where division by zero would occur.
Accordingly, we used the following metrics in our final procedure:
\begin{equation}
\label{eqn_preference_dist}
\begin{aligned}
    prefSim(a_i,a_j) = &
    \ 1 \ \textrm{if} \\ 
    & \ |R_{i,j}| < 4 \\
    & \ \textrm{or} \sum_{k \in R_{i,j}} (r_{i, k} - \bar{r}_{k})^2 = 0 \\
    & \ \textrm{or} \sum_{k \in R_{i,j}} (r_{j, k} - \bar{r}_{k})^2 = 0; \\
    & 1 + prefSim'(a_i,a_j) \ \textrm{otherwise} \\
\end{aligned}
\end{equation}

\begin{equation}
\label{eqn_social_dist}
    \begin{aligned}
    socialSim(a_i,a_j) = & \ 0 \ \textrm{if} \ |friends(a_i) \cup friends(a_j)| = 0; \\
    & socialSim'(a_i,a_j) \ \textrm{otherwise}
    \end{aligned}
\end{equation}

Note that $0 \leq prefSim(a_i, a_j) \leq 2$, where values below $1$ indicate a negative correlation. Therefore, the most appropriate default value is 1. Similarly, $0 \leq socialSim(a_i, a_j) \leq 1$, where values near $0$ indicate very few common friends between $a_i$ and $a_j$, thus, the most appropriate default value when neither agent has any friends is $0$.
In addition, in Equation \ref{eqn_preference_dist} we have also substituted a default value when $|R_{i,j}| < 4$.
This is because correlation tests produce noisy results with small data sets, making it prudent to choose a cutoff point under which no correlation metrics are considered.
Meanwhile, if this cutoff is too high, then potentially useful data is ignored to avoid error. 
We chose the cutoff at 4 arbitrarily.

While this at least leaves the similarity functions well defined, it also creates a situation where the vast number of pairs of agents have a default distance between them, as any two randomly picked agents in a large enough environment will be unlikely to have any interaction history.
This is a potential issue as it may causes clusters to appear significantly less cohesive than they actually are, e.g. in the case where agents $a$ and $b$ have ``default'' distance between them, but are both close to agent $c$.
In this case, $a$ and $b$ should likely be in the same cluster as $c$, even if they don't themselves appear to share any relationship.
% Note how this bears conceptual similarity to the concept of trust transitivity as introduced in Section \ref{section_background_tm_vs_rs}.

In addition to the challenges described above, our clustering task had the additional goal of finding relatively large clusters. This is because our ``downstream'' goal was to learn personalized classifiers for each cluster of agents. If clusters are too small, then the accuracy of classifiers will suffer.

Given these goals and constraints, our first attempt at a clustering was a simple, non-iterative greedy algorithm, shown in Algorithm \ref{alg_simple_cluster}. 
This algorithm takes as input the set of agents to be clustered, $A$, the similarity matrix between agents $S$ (where $S_{i,j} = sim(a_i, a_j)$ for some similarity function) and the desired size of clusters $\eta$.

\begin{algorithm}
	\KwData{agents: $A$, similarity matrix: $S$, cluster size: $\eta$}
	\KwResult{assignment of agents to clusters: $C$}
	$C \leftarrow \emptyset$\;
	\While{$|freeAgents(A, C)| > \eta$}{
		$a \leftarrow pickcentroid(A, C, S)$\;
		$c \leftarrow \{a\}$\;
		\While{$|c| < \eta$} {
			$next \leftarrow pickNext(c, A, C, S)$\;
			$c \leftarrow c \cup \{ next \}$\;
		}
		$C \leftarrow C \cup c$\;
	}
	$C \leftarrow C \cup freeAgents(A,C)$\;
	\Return{C};
	\caption{Greedy non-iterative clustering}
	\label{alg_simple_cluster}
\end{algorithm}

In the above, $freeAgents(A,C)$ returns the set of agents not yet assigned to a cluster in $C$ (unassigned agents), 
$pickCentroid(A, C, S)$ returns the unassigned agent with the greatest mean similarity to all other agents, and $pickNext(c, A, C, S)$ returns the unassigned agent with the greatest mean similarity to the agents in $c$. 

Roughly, Algorithm \ref{alg_simple_cluster} partitions the set of agents into at least $\floor{|A| / \eta}$ clusters of size $\eta$. 
It does this by picking the most central unassigned agent as the core of a new cluster $c_i$, then adding agents to that cluster in order of greatest mean similarity to agents already in cluster $c_i$ until $|c_i| = \eta$.
The process repeats for $c_{i+1}$, except only agents not already assigned to a cluster are considered.
This continues until less than $\eta$ agents remain unassigned, at which point all unassigned agents are added to a final cluster of unspecified size.

Clearly this algorithm is quite simple, but it is appropriate for the constraints outlined above. 
Firstly, all clusters of agents except for one will have a guaranteed minimum size $\eta$, allowing control over the minimum training data size for the downstream prediction task.
More importantly, it handles the non-Euclidean nature of the data by using the mean distance of all points in a cluster as a similarity metric, rather than a geometric center\footnote{This is inspired by the average linkage criterion used in hierarchical clustering algorithms \cite{clustering_2005}. We tested clustering this data hierarchically, but had little success producing clusters of reasonable size.}.

We improved this algorithm by transforming it into an iterative version listed below (Algorithm \ref{alg_cluster}).

\begin{algorithm}
	\KwData{agents: $A$, similarity matrix: $S$, cluster count: $k$, max iterations: $m$}
	\KwResult{assignment of agents to clusters: $C$}
	$C \leftarrow greedilyPartition(A, S, \floor{|A| / k})$\;
	\For{$i \in [0...m]$}{
		$S' \leftarrow computeClusterSims(A, C, S)$\;
		\For{$a_i \in A$}{
			$C \leftarrow assignToNearestCluster(a_i, S', C)$\;
		}
	}
	\Return{C};
	\caption{Modified $k$-means clustering}
	\label{alg_cluster}
\end{algorithm}

In the above, $greedilyPartition(A, C, S)$ assigns each agent to a cluster using the procedure outlined above in Algorithm \ref{alg_simple_cluster}. 
$computeClusterSims(A, C, S)$ computes a new similarity matrix, $S'$, between agents and clusters, where $S'_{i,j}$ is the average similarity between agent $i$ and all agents in the $j$'th cluster (other than themselves):
$$
S'_{i,j} =  \frac{1}{|c_j| - \mathbbm{1}(a_i \in c_j)}\sum_{a_k \in c_j} \mathbbm{1}(i \neq k)S_{k, i}
$$
where $\mathbbm{1}(cond)$ is the function which is equal to $1$ when $cond$ is true and $0$ otherwise.
$assignToNearestCluster(a_i, S', C)$ computes a modification of the current set of clusters $C$ by moving agent $a_i$ to the cluster $c_j$ that maximizes $S'_{i, j}$, that is, the cluster for which they have the highest average similarity with other cluster members, (with ties broken randomly).
This process repeats for a predetermined maximum number of iterations $m$.

This process is much closer to classic k-means clustering, again with the modification that distances between clusters and points must be calculated on the basis of mean distances rather than distances to the cluster's geometric center.
In addition, rather than picking random points to serve as initial cluster centers, the initial clusters are determined by a greedy partitioning method.
These modifications result in an algorithm that, in our experiments, tended to produce relatively large and cohesive clusters.
  
% Notably, a minimum cluster size is no longer guaranteed by this clustering method, which we have pointed out is a desirable feature for our later prediction task.
% We will describe our procedure for dealing with this in Section \ref{section_link_prediction}.

Both Algorithm \ref{alg_simple_cluster} and \ref{alg_cluster} require a parameter used to control the number of clusters ($\eta$ and $k$ respectively). 
When performing our experiments, we determined values for these parameters by running the clustering step multiple times over a range of parameters with a relatively low maximum iteration setting, then choosing the best performing parameter to proceed with.
\\

\begin{figure}
	\centering
	\begin{subfigure}[b]{0.49\textwidth}
		\includegraphics[width=\textwidth]{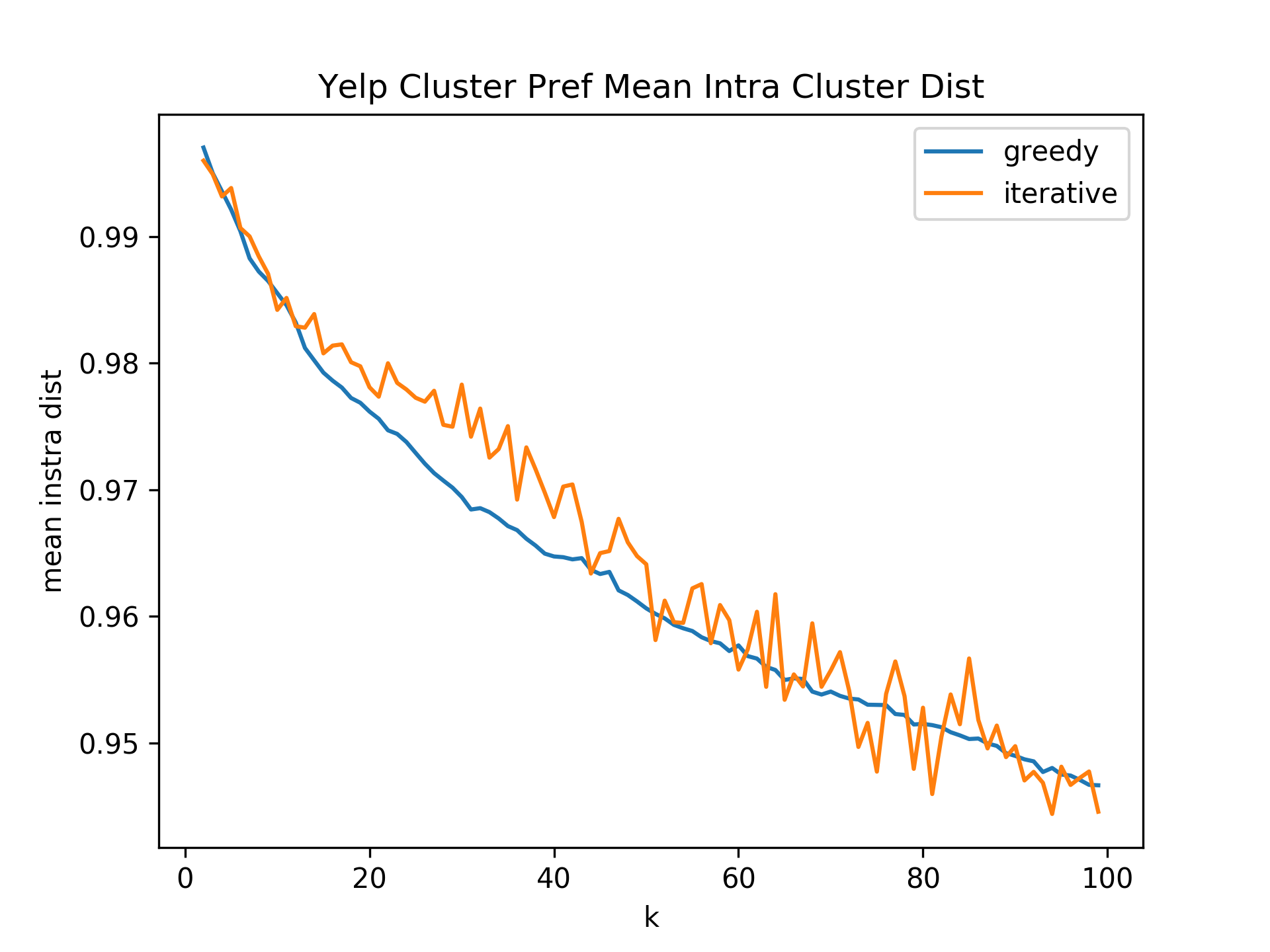}
		\caption{Intra-cluster distance (lower is better)}
	\end{subfigure}
	\begin{subfigure}[b]{0.49\textwidth}
		\includegraphics[width=\textwidth]{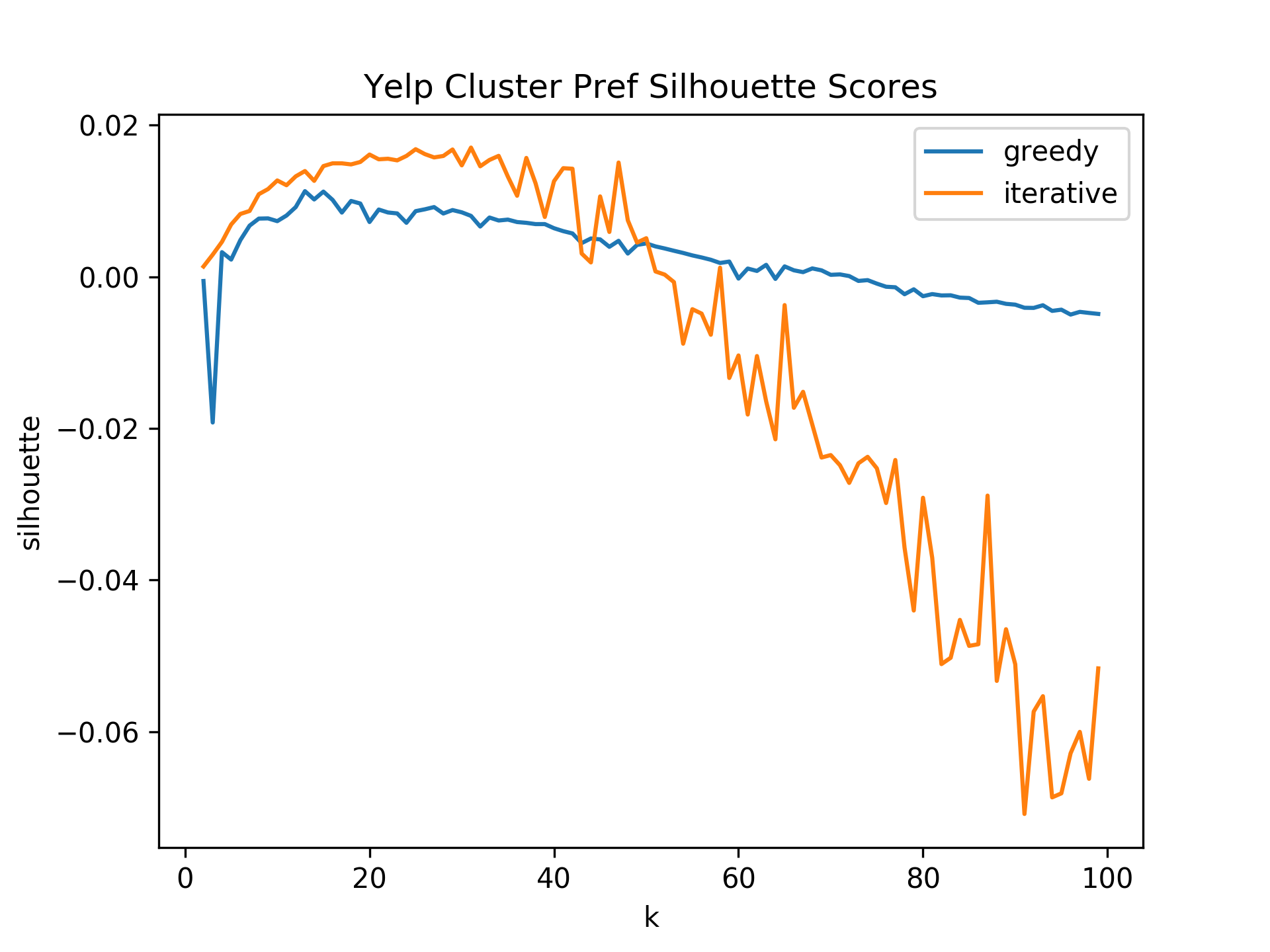}
		\caption{Cluster silhouette scores (higher is better)}
	\end{subfigure}
	\caption{Yelp data preference clustering results}
	\label{fig_clust_pref_k}
\end{figure}

\begin{figure}
	\centering
	\begin{subfigure}[b]{0.49\textwidth}
		\includegraphics[width=\textwidth]{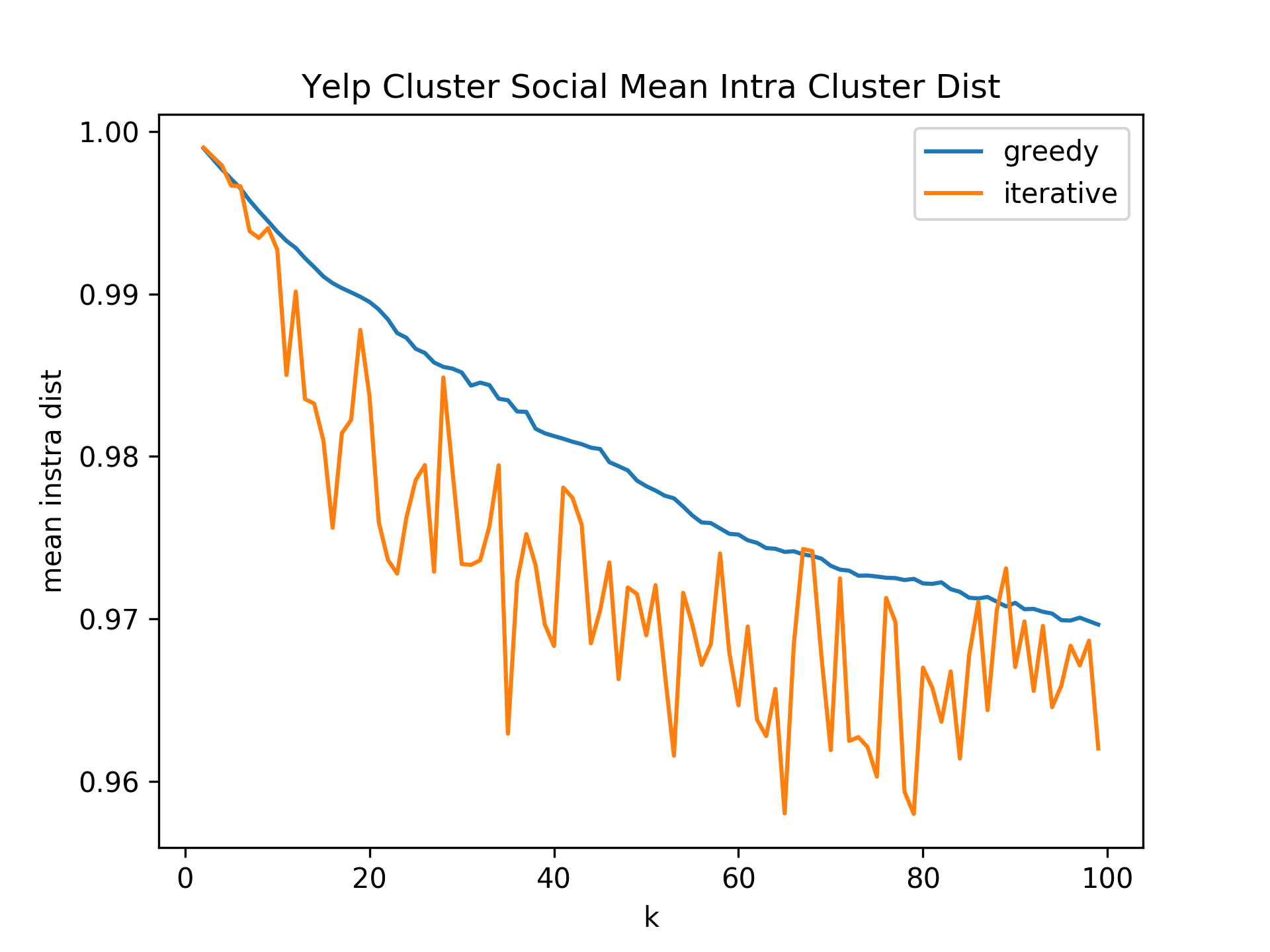}
		\caption{Intra-cluster distance (lower is better)}
	\end{subfigure}
	\begin{subfigure}[b]{0.49\textwidth}
		\includegraphics[width=\textwidth]{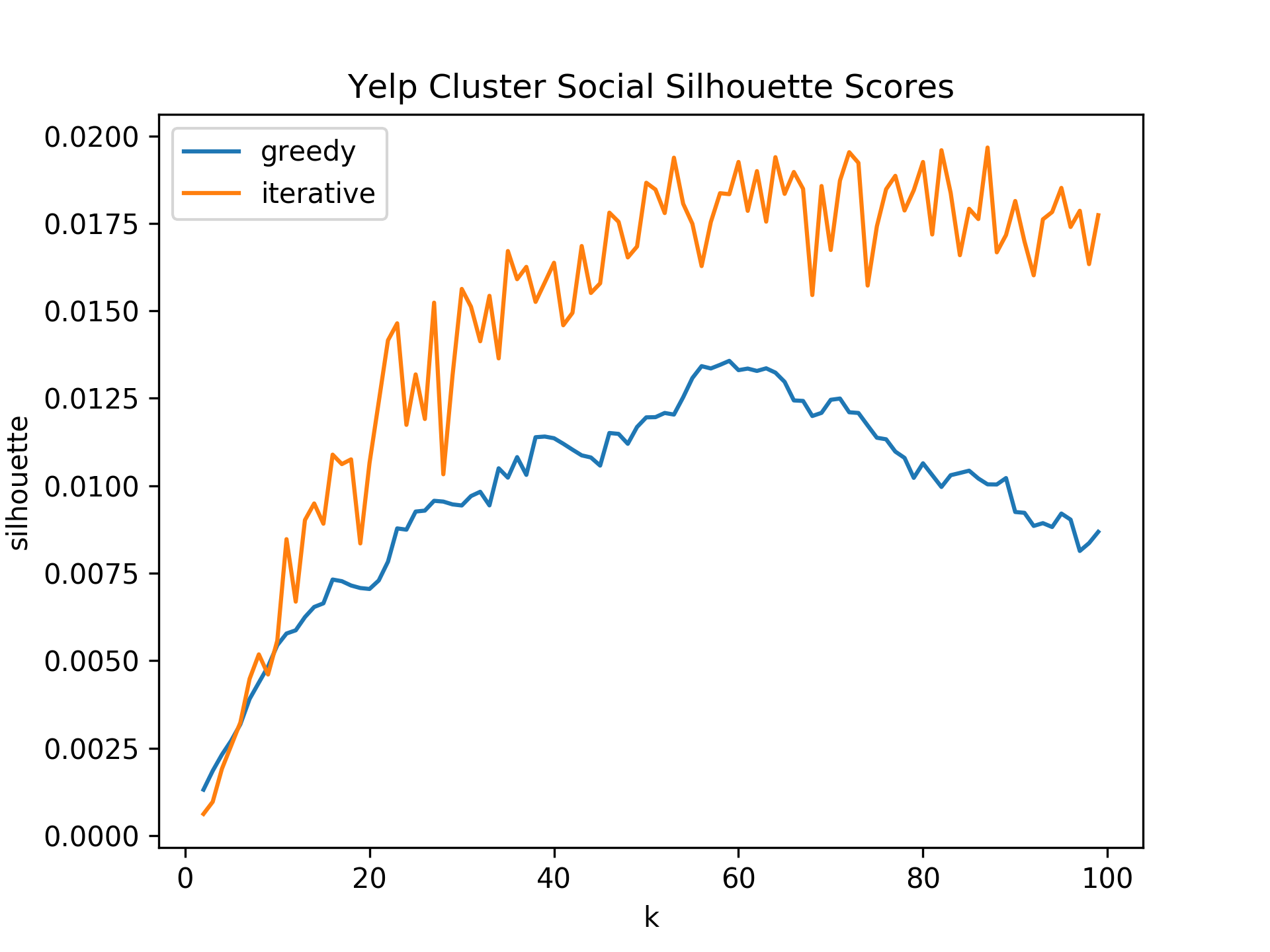}
		\caption{Cluster silhouette scores (higher is better)}
	\end{subfigure}
	\caption{Yelp data social clustering results}
	\label{fig_clust_social_k}
\end{figure}

\noindent {\textbf{Data set selection and filtering:}}
We pause to provide further details on the dataset used within
our experimentation, so that our continued description of the clustering process can be deepened by discussing it within this context.

Yelp is an ideal candidate to explore relationships between agents against a validation of recommended content.
It is a product review site and social network of crowd-sourced reviews targeting brick-and-mortar businesses.
In addition to writing reviews of services, users of the site can form mutual friendships and follow other users in order to receive the recommendations of these trusted users first.
Data describing users, reviews and businesses is made public by Yelp on a regular basis\footnote{\url{https://www.yelp.com/dataset}}.
The full data set from 2019 contained descriptions of 1,637,138 users, 192,609 businesses, and 6,685,900 reviews.

We filtered the data set both to reduce the massive amount of data and to narrow down the context of trust in focus.
Specifically, we only considered users who had reviewed at least 20 businesses that were tagged as restaurants.
This narrows the context of trust from ``recommending businesses or services'' to ``recommending restaurants'' and reduces the data set down to 30,721 users and 4,432,064 reviews concerning 74,560 businesses.
This filtering procedure was inspired by \cite{multi-faceted_mauro_2019}.

\begin{table*}[h]
	\centering
	\begin{tabular}{llllll}
	    \hline\noalign{\smallskip}
		& Mean  & Median & Mode & Min & Max  \\
		Friends Per User & 153.23 & 45 & 1 & 1 & 9564 \\
		Review Per User  & 49.66 & 33     & 20   & 20  & 3159 \\
		Average User Rating & 3.74 & 3.77 & 4 & 1.33 & 5 \\	
		Reviews Per Item & 59.33 & 17     & 3    & 3   & 8349 \\
		Global Review Scores    & 3.72  & 4      & 4    & 1   & 5   \\
		\noalign{\smallskip}\hline
	\end{tabular}
	\caption{Yelp Filtered Data Statistics}
	\label{table_yelp_data_stats}
\end{table*}

Statistics from the filtered data set are presented in the Table \ref{table_yelp_data_stats} and in the histograms in Figures \ref{fig_review_dist} and \ref{fig_user_review_hist}.
As can be seen, there is a relatively well spread out distribution of scores given to items\footnote{Compared to the popular Epinions data set, where nearly 80\% of reviews are 5 stars.}, centered around 4 stars.
This leads to a relatively difficult prediction task, as predicting the median review score is only correct in 35\% of cases.
Counts of friends and reviews are plotted on a logarithmic scale, and show a ``long tail'' distribution that is common in online phenomenon \cite{seely2008open}.

\noindent {\textbf{Clustering methods used with the data set:}}
Against the backdrop of the Yelp dataset, in Figures \ref{fig_clust_pref_k} and \ref{fig_clust_social_k} we illustrate the performance of clustering techniques as the number of clusters (k) is altered.
We measure cluster cohesiveness using two metrics, mean-intra cluster distance and silhouette score.
In the below, $dist(i,j)$ is the appropriate distance measure for the similarity function chosen, i.e. if $sim(i,j)$ is high when $i$ and $j$ are similar, then $dist(i,j)$ is low when $i$ and $j$ are similar.
Mean intra cluster distance is defined as follows:
\begin{equation}
meanintra(C) = \frac{1}{|C|}\sum_{c_i \in C}\sum_{j \in c_i}\sum_{k \in c_i} \frac{dist(i, j)}{|c_i|}
\end{equation}
That is, the average distance between all elements in a cluster and the other elements in that cluster, averaged over all clusters.

Silhouette score $s(j)$ for a single clustered point $j$ which has been assigned to cluster $c_i$ is defined as follows:
\begin{align}
a(j) &= \frac{1}{|c_i| - 1} \sum_{k \in c_i, j \neq k} dist(j, k) \\
b(j) &= \min_{\ell \neq i} \frac{1}{c_\ell} \sum_{k \in c_\ell} dist(j, k) \\
s(j) &= 
\begin{cases}
\frac{b(j) - a(j)}{max(a(j), b(j))} & if |c_i| > 1 \\
0 & otherwise
\end{cases}
\end{align}

That is, $a(j)$ is the average distance point $j$ has to other points in its cluster.
$b(j)$ is the minimum distance from $j$ to to any other point not in the same cluster as $j$.
$s(j)$ is the silhouette score for the point $j$. 
When a point exists which is not in the same cluster as $j$ but is closer to $j$ than the average point in $j$'s cluster, then the score is negative.
When the closest point to $j$ outside of its cluster is farther away than the average distance of point in $j$'s cluster, then the score is positive.
The silhouette score for a set of clusters $C$ is calculated by taking the average of a random sample of points from different clusters.
 
Both metrics capture a sense of the cohesiveness of a set of clusters, and can be used to judge the relative merits of different clustering schemes and parameter settings.
While both metrics are interesting, there are a few caveats to consider.
First, for this data set and clustering algorithm, it is expected that intra-cluster distance will decrease as cluster count increases.
This is shown to be true in the results reported.
Thus, this metric is better suited for showing the difference in performance between clustering methods than it is for choosing a value of $k$.
The silhouette metric does a better job of outlining the tradeoff for $k$ values, as it will punish a method for assigning two close points to separate clusters.
Therefore, a value of $k$ that maximizes the silhouette score over a range is a more appropriate guide to choosing a value of $k$.

Noticeably, performance on both scores is low in an absolute sense.
Silhouette score ranges from $[-1, 1]$, where positive scores are good and negative bad.
Our clustering algorithms achieve scores in the range of $[-0.06,$ $0.02]$ - a tiny portion of the possible range near 0.
Similarly, the scores for intra cluster dist should range from $[0, 2]$, where small scores are good, and our algorithms find scores in the $[0.95, 0.99]$ range.
Why is this the case?
The answer is related to the sparsity of defined links between agents when all $|A|^2$ possible pairs of agents are considered. 
As most agents do not know most other agents, and there is no basis for determining the similarity (distance) between them, in the vast majority of cases $sim(a_i, a_j)$ is equal to a default value for randomly picked $i$ and $j$.
Therefore, as both metrics take some kind of average of the distances between pairs of agents, the metrics will always be close to the default distance.

Should these low absolute scores deter us from this method? 
We argue they should not.
First, as we have briefly argued above, the nature of this data implies that average measures of cluster cohesiveness will always be close to a default.
Secondly, the trend lines show that appropriate choice of $k$ and cluster methodology can effect the sign and magnitude of silhouette scores in consistent ways - for example, when $k$ goes above $60$ in Figure \ref{fig_clust_pref_k} (b).
We take this as an indication that positive results are not simply a coincidence.

\subsubsection{Trust link prediction}
\label{section_link_prediction}

Our trust link prediction procedure was intended to combine what we perceived to be the best traits of the work of Fang et al. \cite{multi-faceted_fang_2015} and Mauro et al. \cite{multi-faceted_mauro_2019}.
Both works tested the effects of predicting trust links using Multi Faceted Trust Modeling (MFTM) on an item recommendation task.

In \cite{multi-faceted_mauro_2019}, a relatively large number of domain specific trust indicators are proposed for the Yelp data set; however, the importance of each trust indicator is not learned in a data driven way.
Instead, they selectively enable and disable indicators for each performance test, combining their values by simply taking the average of the enabled indicators. In \cite{multi-faceted_fang_2015}, a relatively small number of generic trust indicators are proposed for an Epinions data set.
The importance of each indicator is learned via logistic regression and performance is tested under a number of different sparsity conditions\footnote{We use logistic regression because of its simplicity and interpretability. Note that while this model can only learn a linear boundary between classes, using a non-linear model is also possible.
However simply using a non-linear model without clustering would not by itself lead to more personalized recommendations.}.

We will compare our work more closely to the works of Fang et al. and Mauro et al. in Section \ref{chapter_discussion}. 
In our work, we combined indicators proposed in both work and adopted a data-driven indicator importance weighting procedure.
%In addition, personalization for MFTM via clustering was proposed (but not tested) in \cite{multi-faceted_fang_2015}, an approach which we evaluate here.

%%%%%% TODO
\noindent \textbf{Trust indicator list:}
To quickly restate the goals of MFTM\footnote{A full treatment is given above in Section \ref{section-background-multi-facet}.}, we wish to define a vector of trust indicators over every pair of agents, $\Psi(\truster{a_i}, \trustee{a_j})$ then use machine learning to approximate the function $f: \Psi(\truster{a_i},\trustee{a_j}) \rightarrow y$, where $y$ is some type of trust link.

In Table \ref{table_yelp_indicators} we have listed all trust indicators that we calculated for Yelp data.
When an indicator was proposed in the works of either Fang et al. \cite{multi-faceted_fang_2015} or Mauro et al. \cite{multi-faceted_mauro_2019}, we have indicated this in the last column  (although some were also adjusted by us, as clarified below).
Some indicators are defined specifically for pairs of agents (e.g. the similarity of rating behavior for two agents), while others are defined on the basis of a single agent.
In Yelp data, the only explicit trust link present is mutual friendship.

%\begin{table}[ht]
%	\centering
%	\begin{tabular}{p{2.8cm}|p{10cm}|p{1.5cm}}
%		\textbf{Name} & \textbf{Description}          & \textbf{Source} \\ \hline
%		Benevolence   & Equation \ref{eqn_preference_dist}, the similarity in rating behavior between truster and trustee.                & Fang            \\ \hline
%		Integrity & How similar the trustee's ratings are to the global average. & Fang \\ \hline
%		Competence & How often the trustee's ratings are within an acceptable range of other agent's ratings & Fang \\ \hline
%		Predictability & How consistently the trustee's ratings are more/less positive than the truster's & Fang \\ \hline
%		Trust-in & The number of incoming trust links on the trustee & Fang \\ \hline
%		Trust-out & The number of outgoing trust links on the trustee & Fang \\ \hline
%		Distrust-in & The number of incoming distrust links on the trustee & Fang \\ \hline
%		Distrust-out & The number of outgoing distrust links on the trustee &
%		Fang \\ \hline
%		SocialJacc & $rel_{ab}$, Equation \ref{eqn_social_dist}, the Jaccard similarity in the truster and trustees friend sets  & Mauro \\ \hline
%		ItemJacc & Jaccard similarity relative to items reviewed. & \\ \hline
%		CategoryJacc & Jaccard similarity relative to categories of items reviewed  & \\ \hline 
%		AreFriends    & Are truster and trustee friends & \\ \hline     
%		AreFoF & Are truster and trustee friends of friends & 
%	\end{tabular}
%	\caption{Trust Indicators used for Epinions data.}
%	\label{table_epinions_indicators}
%\end{table}

Here, we describe some of these indicators in full detail, in order to illustrate some less obvious indicators.
Complete descriptions of the indicators not described here are available in \cite{multi-faceted_fang_2015} and \cite{multi-faceted_mauro_2019}.

\textit{Benevolence}: Already described in Equation \ref{eqn_preference_dist}.
In \cite{multi-faceted_fang_2015}, $\bar{r}_j$, the average rating given to item $j$ was replaced with $\bar{r_i}$, the average score agent $i$ gave to items. 
We made this replacement as a common rating behavior in the Yelp data set was for an agent to only submit 5 star reviews, causing frequent divisions by zero.
By comparing to the global average rating of an item, this behavior is no longer an issue. 
Intuitively, when $benevolence(a_i,a_j)$ is high, agents $a_i$ and $a_j$ may be inclined to trust each other's reviews, as they have reviewed items similarly in the past.

\textit{Competence}: A threshold value $\epsilon$ is used to determine how often the trustee's ratings where ``close enough'' to the ratings of other agents who had also rated those items to be considered ``correct''.
\begin{equation}
competence(a_i) = \frac{\sum_{j \in R_{i}}\sum_{k \in I_j}\ind(|r_{ij} - r_{kj}| < \epsilon)}{\sum_{j \in R_{i}} |I_j|}
\end{equation}
Where $R_i$ is all the items the $i$'th agent has rated and $I_j$ is the set of all agents who have rated item $j$ and $r_{ij}$ is the rating agent $i$ gave to item $j$. 
Competence is high when an agent's rating behavior is similar to the plurality of agents.
Since ratings on Yelp use a 5-star scale, we used the threshold value $0.5$.
Intuitively, when $competence(a_j)$ is high, $a_j$ may be trustworthy for agents who consider agreement with popular consensus to be important.

\textit{Predictability}: A threshold value $\theta$ is used to determine how often a trustee's preferences are consistently higher, lower, or similar to the truster's :
\begin{align}
n_u &= \sum_{k \in R_{i, j}} \ind(|r_{ik} - r_{jk}| \leq \theta) \\
n_n &= \sum_{k \in R_{i, j}} \ind(r_{ik} - r_{jk} < \theta) \\
n_p &= \sum_{k \in R_{i, j}} \ind(r_{ik} - r_{jk} < -\theta) \\
predictability(a_i, a_j) &= \frac{max(n_u, n_n, n_p)-min(n_u,n_p,n_n)}
{|R_{i,j}|}
\end{align}
where $n_u$, $n_n$, and $n_p$ count how many times the trustee rated an item about the same as the truster, lower than the truster, and higher than the truster respectively. 
Accordingly, predictability is lowest when $n_u = n_n = n_p$, meaning the trustee rates items better, worse, and equivalent to the truster in equal amounts.
This would mean there isn't a justification to expect that the trustee has a bias in any particular direction, relative to the truster.
Similar to Competence, we used a threshold value of $0.5$.
Intuitively, $predictability(a_i, a_j)$ may be important to $a_i$ deciding whether or not to trust $a_j$, as it is useful to know whether $a_j$'s ratings have a consistent bias compared to $a_i$.

\textit{Visibility}: The relative popularity of agent, taking into consideration how much content the agent has produced and the popularity of the most popular agent.

\begin{equation}
visibility(a_i) = \frac{appr(i)}{max_{a_j \in A}(appr(j) \times contr(i))}
\end{equation}

Where $appr(i)$ is the number of public ``appreciations'' an agent has received from other agents (e.g. likes) and $contr(i)$ is the number of contributions an agent has made (e.g. posts, reviews).
Intuitively, when $visibility(a_j)$ is high, $a_j$ may be trustworthy to agents who consider consistent popularity important.

Some of the indicators listed in Table \ref{table_yelp_indicators} were developed by us. 
For example, we normalized a number of the indicators proposed by \cite{multi-faceted_mauro_2019} by dividing by how many years the target user had been on the site.
This is useful for giving newer users a chance to compete with older users on certain attributes (e.g. how many ``fans'' they've accrued).
We also computed the Jaccard similarity between users with respect to the sets of items they had reviewed and the categories of items they had reviewed, reasoning that these indicators would help to outline the case when users has similar areas of interest.
Finally, we checked to see if pairs of users were friends of friends, a potentially useful feature for integrating trust transitivity into reasoning.

\begin{table*}[t]
    \centering
    \begin{tabular}{p{2.3cm}|p{7.8cm}|p{0.9cm}}
        \textbf{Name} & \textbf{Description}          & \textbf{Source} \\ \hline
        Benevolence   & Equation \ref{eqn_preference_dist}, the similarity in rating behavior between truster and trustee.                & Fang            \\ \hline
        Integrity & How similar the trustee's ratings are to the global average. & Fang \\ \hline
        Competence & How often the trustee's ratings are within an acceptable range of other agents' ratings & Fang \\ \hline
        Predictability & How consistently the trustee's ratings are more/less positive than the truster's & Fang \\ \hline
        SocialJacc & $rel_{ab}$, Equation \ref{eqn_social_dist}, the Jaccard similarity in the truster and trustees friend sets  & Mauro \\ \hline
        EliteYears & $elite_a$, the number of elite years the trustee has &  Mauro\\ \hline
        ProfileUp & $lup_a$, the number of compliments on the trustee's profile & Mauro \\ \hline
        Fans & $opLeader_a$, the number of fans the trustee has & Mauro \\ \hline
        Visibility & $vis_a$, the ratio of compliments received to amount of content produced by the trustee & Mauro\\ \hline
        GlobalFeedback & $fb_a$, the number of compliments the trustee's content has received & Mauro \\ \hline
        EliteNorm & EliteYears divided by trustee account age in years & \\ \hline
        ProfileNorm & ProfileUp divided by trustee account age in years & \\ \hline
        FansNorm & Fans divided by trustee account age in years & \\ \hline
        FeedbackNorm & GlobalFeesback divided by trustee account age in years & \\ \hline
        ItemJacc & Jaccard similarity relative to items reviewed. & \\ \hline
        CategoryJacc & Jaccard similarity relative to categories of items reviewed  & \\ \hline 
        AreFriends    & Are truster and trustee friends & \\ \hline     
        AreFoF & Are truster and trustee friends of friends & 
    \end{tabular}
    \caption{Trust indicators used for Yelp data.}
    \label{table_yelp_indicators}
\end{table*}

When computing these trust indicators, it appears necessary to consider all ordered pairs of agents in the environment, as trust is directed and can occur between any two agents.
This presents a significant computational bound on the number of agents that can be considered.
Discussion of this issue, and our approach to minimizing this impact is presented in Appendix \ref{section_computation}.
In brief, only pairs of agents where there is significant evidence that the pair have an overlap in interests / social circle are actually considered as candidates for novel trust link prediction.

%%%% TODO
\noindent \textbf{Classification process:}
The trust indicators listed in Table \ref{table_yelp_indicators} were used to predict two types of trust links 1) whether the truster had explicitly expressed trust in the trustee (friendship), and 2) whether the truster and trustee had a positive correlation in review scores. % (e.g. Equation \ref{eqn_preference_dist}). 
Note, in the case where expressed trust was the target of prediction, review score correlation \textit{was} considered as evidence (e.g. included in $\Psi(a_i, a_j)$) and vice versa, although the target of prediction was obviously not considered as evidence.

Following the example set by Fang et al \cite{multi-faceted_fang_2015}, we use logistic regression to learn functions that predict the presence of statistically likely trust links based on the vector of trust indicators computed between pairs of agents.
We refer to these functions as ``trust link classifiers'', as once learned, they classify each ordered pair of agents as either being linked by trust (the former should trust the latter) or not.
We used the SAGA solver logistic regression classifier included in the sklearn Python package \cite{scikit-learn} to learn these functions.

In the case where no clustering was performed, a single classifier was learned for all agents. 
When clustering was performed, a classifier was learned for each cluster of agents. 
Thus, each cluster specific classifier learns how the agents in the cluster form trust links in their role as trusters.
This makes obvious a substantial tradeoff to this approach to personalization: the more clusters are found (increasing cluster cohesiveness up to a point), the less data available to train machine-learning classifiers (decreasing prediction accuracy). 
We will discuss other potential approaches to personalization in Section \ref{chapter_discussion}.
For our purposes, we only learned cluster specific classifiers for clusters that had at least $100$ agents and at least $1000$ positive outgoing trust links.
When a cluster failed to meet these standards, it was assigned a generic classifier, trained on examples from a random sample of users across all clusters. 
We implemented this strategy in order to avoid training wildly inaccurate classifiers. 
When training the cluster-specific classifiers, all available data relevant to each cluster was used for training.
This is because we are not directly interested in how well each classifier is able to fit each cluster, only on whether this personalization process increases the accuracy of the downstream recommendation task.
Therefore, it is not necessary to reserve a test/validation set for any trust link classifier.
Note, at this point, test sets of ratings data \textit{are already} reserved for the recommendation task outlined in the next section.

A common problem in link prediction generally is the large class imbalance between positive and negative examples. 
Put simply, the number of negative examples of trust links in a community of agents %(agents that have nothing to do with each other, have never communicated, or who genuinely do not trust each other)
grows with $O(|A|^2)$, while positive examples %(friends, trust)
have much more conservative linear growth. 
This can be either because humans have an upper limit on how many others they will trust, or, like on Yelp, technical limitations are imposed on the number of allowed friends.
Compounding the problem is that there are two kinds of negative examples, which are often difficult to distinguish between. 
On the one hand, agents $a_i$ and $a_j$ may \textit{not} be friends simply because they have never met.
On the other hand, they may have interacted and prefer not to do so again in the future.

Therefore, it is necessary to devise a strategy for training classifiers to deal with this imbalance and ambiguity. 
One popular method, which we have used here, is to construct balanced training sets by including a random negative link for every positive link.
This method has the advantage of requiring no further tinkering to classifiers in order to accommodate a class imbalance.
The ambiguity in negative links is ignored as best as possible by simply sampling negative links randomly.

After training classifiers for each cluster, trust link prediction is performed by feeding the trust indicator vector $\Psi(a_i, a_j)$ to the appropriate classifier for the cluster of agent $a_i$. Ultimately a matrix of trust link predictions $\hat{\Gamma}$ is produced, where $\hat{\Gamma}_{ij} = 1$ if the classifier for the $i$'th agent's cluster predicts a trust link from $a_i$ to $a_j$ with probability greater than $0.5$. It is noted once again that predictions were only made for pairs of agents that were considered to be in the same neighborhood, as described in Appendix \ref{section_computation}.

\subsubsection{Recommender evaluation}
\label{rec_eval}

Two trust-aware recommender systems were used to measure the accuracy of the trust links predicted in the previous step: TrustMF and MTR.
Two systems were evaluated in order to reduce the risk (e.g. of using a a flawed implementation that skews results).
TrustMF leverages matrix factorization and gradient descent to optimize predictions of user-item ratings. %, as described above in Section \ref{section_latent_factors}.
We used the Librec implementation of TrustMF for our experiments \cite{librec_Guo_2015}.
MTR is a trust aware modification of a similarity based KNN recommendation model proposed by Mauro et al. \cite{multi-faceted_mauro_2019}.
Under this system, the predicted rating for an agent $i$ for an item $j$ is:
\begin{equation}
\hat{r}_{ij} = \bar{r}_i + \frac{\sum_{k \in N^\kappa_j(i)} inf_{ki}(r_{kj} - \bar{r}_k)}{\sum_{k \in N^\kappa_j(i)} |inf_{ki}|}
\end{equation}
where $\bar{r}_i$ is the mean score agent $i$ has given in ratings, $N^\kappa_j(i)$ is the set of the top $\kappa$ most influential agents on $i$ who have also rated item $j$, and
$inf_{ki}$ is the influence agent $k$'s recommendation exerts on agent $i$: a linear combination of the similarity between $k$ and $i$'s past rating behaviour and a trust metric.
In our case, $inf_{ki}$ is the probability that $k$ is trustworthy for $i$ according to the predictions of the trust model in the previous step
\begin{equation}
inf_{ki} = \beta \cdot \sigma(ik) + (1-\beta) \cdot \hat{\Gamma}_{ik}
\label{eqn_mtr_influence}
\end{equation}
where $\beta$ is simply a parameter for controlling the weight of trust modeling on the recommendation process.
When $\hat{\Gamma}_{ik}$ was undefined (e.g. in the case where $i$ and $k$ are not in the same neighborhood, see Appendix \ref{section_computation}), a value of $0$ was substituted.
We modified an implementation of a KNN based recommender system distributed in the Surprise library \cite{surprise} to test this method.

Accuracy of recommendation was measured by Mean Absolute Error (MAE) and Root Mean Squared Error (RMSE):
\begin{align}
MAE(\hat{R}) &= \frac{1}{|\hat{R}|} \sum_{r_{ij} \in \hat{R}} |\hat{r}_{ij} - r_{ij}| \\
RMSE(\hat{R}) &= \sqrt{\frac{1}{|\hat{R}|} \sum_{r_{ij} \in \hat{R}} (\hat{r}_{ij} - r_{ij})^2}
\end{align}
where $R$ and $\hat{R}$ are a set of real agent-item ratings and predicted agent-item ratings respectively, and $r_{ij}$ is the rating given by user $i$ to item $j$.
MAE simply captures the average unsigned error in predictions across all ratings, while RMSE is more heavily penalized for gratuitously erroneous ratings and less penalized for very nearly correct ratings.
These measures can be analogized to the mean and variance of a distribution over prediction error.
As measures of error, we prefer recommendations that minimize these measures.
Thus for all of the following graphs lower values are better.
The sensitivity of RMSE to outliers is a useful property for this application, as grossly inaccurate predictions can erode user trust in future recommendations.

\begin{figure}
	\centering
	\begin{subfigure}[b]{0.49\textwidth}
		\includegraphics[width=\textwidth]{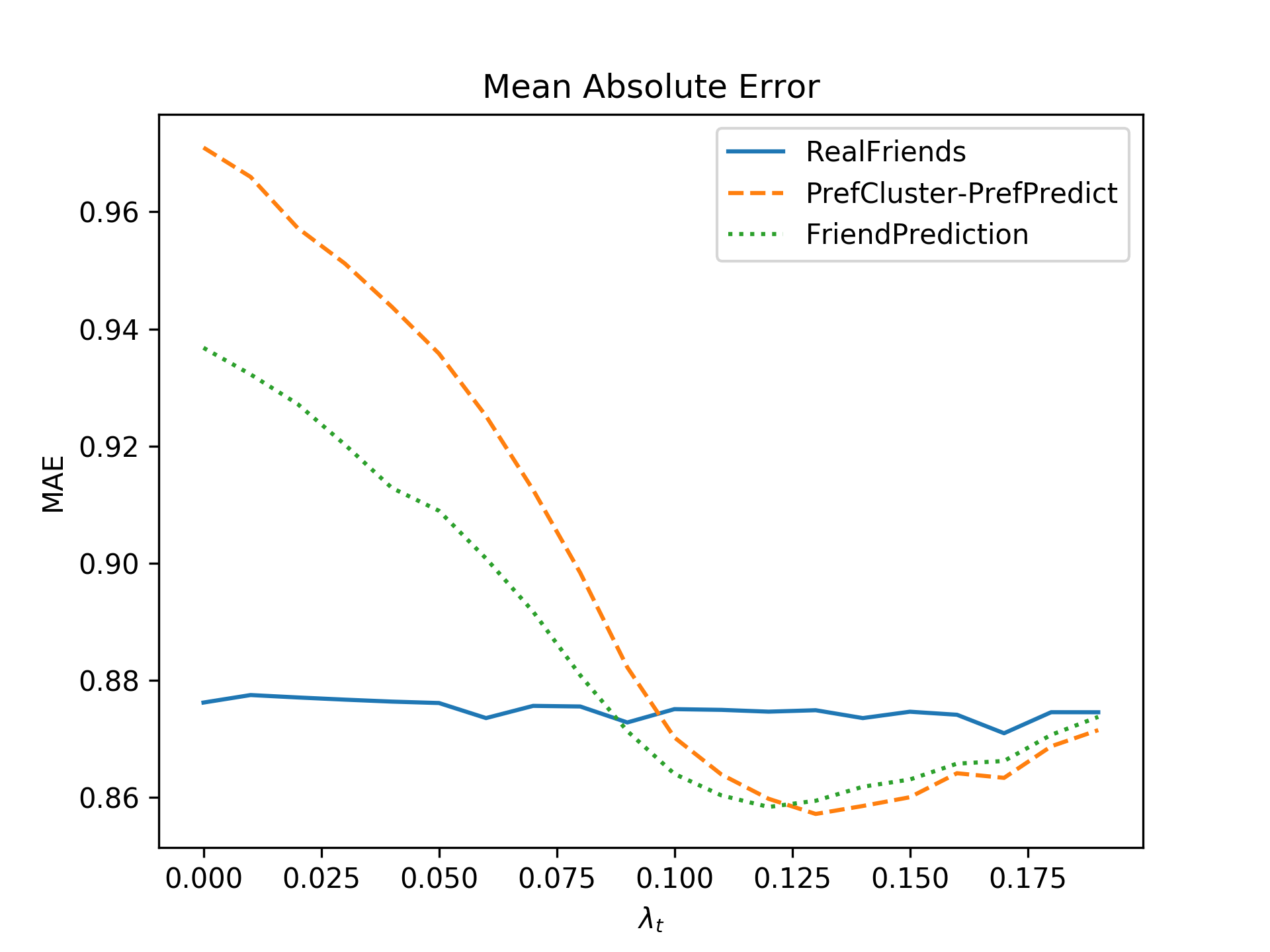}
		\caption{MAE as $\lambda_t$ increases.}
	\end{subfigure}
	\begin{subfigure}[b]{0.49\textwidth}
		\includegraphics[width=\textwidth]{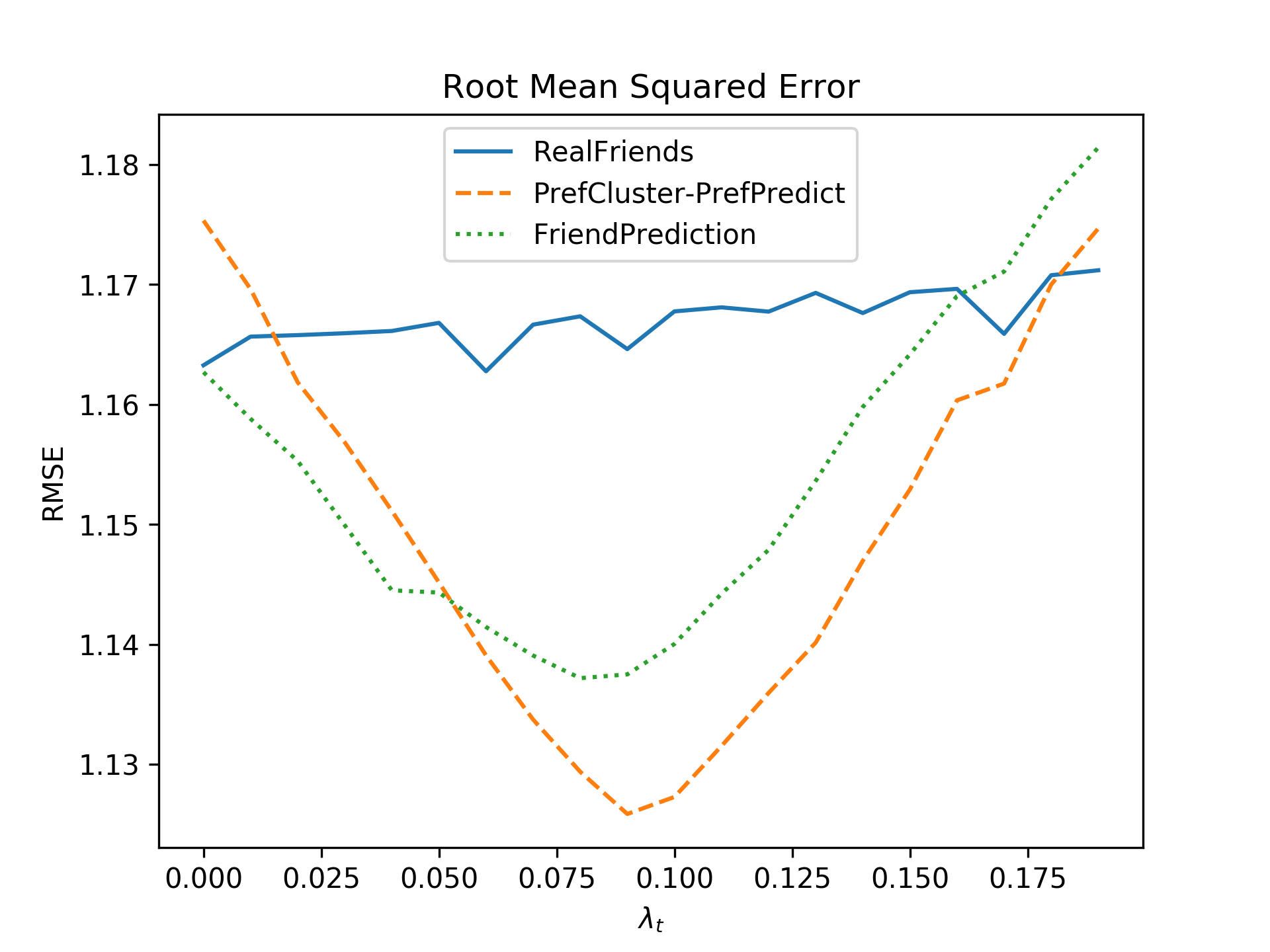}
		\caption{RMSE as $\lambda_t$ increases.}
	\end{subfigure}
	\caption{Tuning $\lambda_t$ for TrustMF}
	\label{trustmf-social-weight-tune}
\end{figure}

The latent factors model adopted by TrustMF was outlined already in Section 2.3.1.
TrustMF has the following significant hyperparameters: the regulation penalty $\lambda$, the weight given to fitting the user-user trust matrix (as opposed to the user-item rating matrix), $\lambda_t$, and the number of dimensions of the latent space $d$.
We kept the number of dimensions at the default of 10 and the regulation penalty at 0.01.
In order to determine an appropriate setting for $\lambda_t$, we sampled 10000 users from the filtered Yelp data set and plotted MAE and RMSE over the change in $\lambda_t$.
Results of this tuning are presented in Figure \ref{trustmf-social-weight-tune}.
For readability we have only plotted the best performing experiments from each of the main groups\footnote{See Table \ref{table_experiments} for the complete list of experiments.} (RealFriends as a baseline, FriendPrediction as a non-personalized (MFTM) example, and PrefCluster-PrefPredict as a personalized (PMFTM) example).
Each data point is the average of three runs with different random seeds, and an iteration limit of 200 epochs.
For these preliminary tests, we set the number of clusters at 10.

Recommendation accuracy changes little for the actual trust links in the data set (RealFriends) as the importance of trust for recommendation increases, but the MFTM and PMFTM lines form roughly convex curves, reaching a range of minimal values around $0.8 < \lambda_t < 0.125$.
For future experiments, we set $\lambda_t = 0.11$.

Figure \ref{trustmf-social-weight-tune} also serves as some encouraging early results, showing both that the impact of predicting trust links reduces recommendation error and a personalized approach can reduce this error further, given the correct weighting of trust importance.

\begin{figure}[t]
	\centering
	\begin{subfigure}[b]{0.49\textwidth}
		\includegraphics[width=\textwidth]{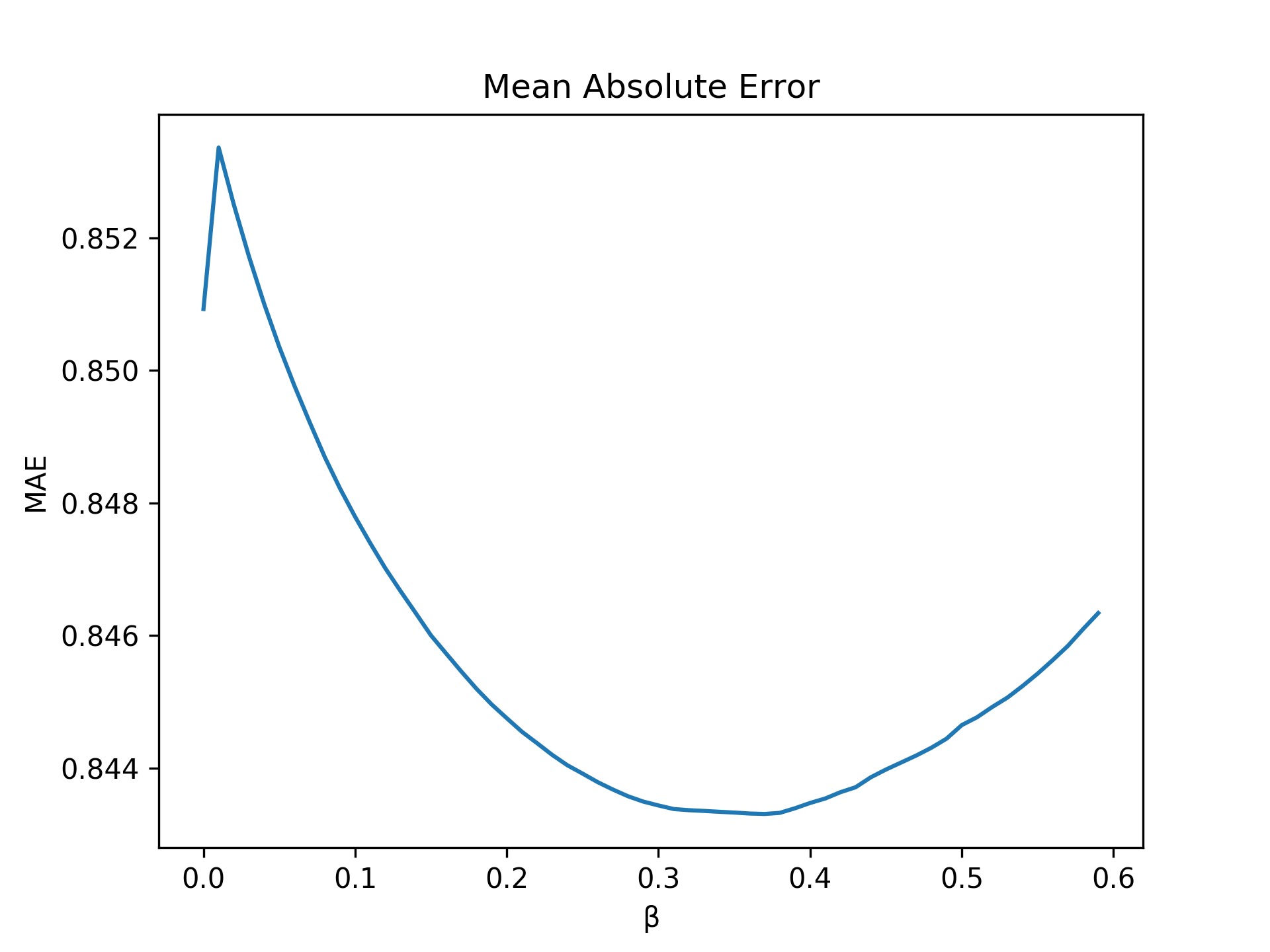}
		\caption{MAE as $\beta$ increases.}
	\end{subfigure}
	\begin{subfigure}[b]{0.49\textwidth}
		\includegraphics[width=\textwidth]{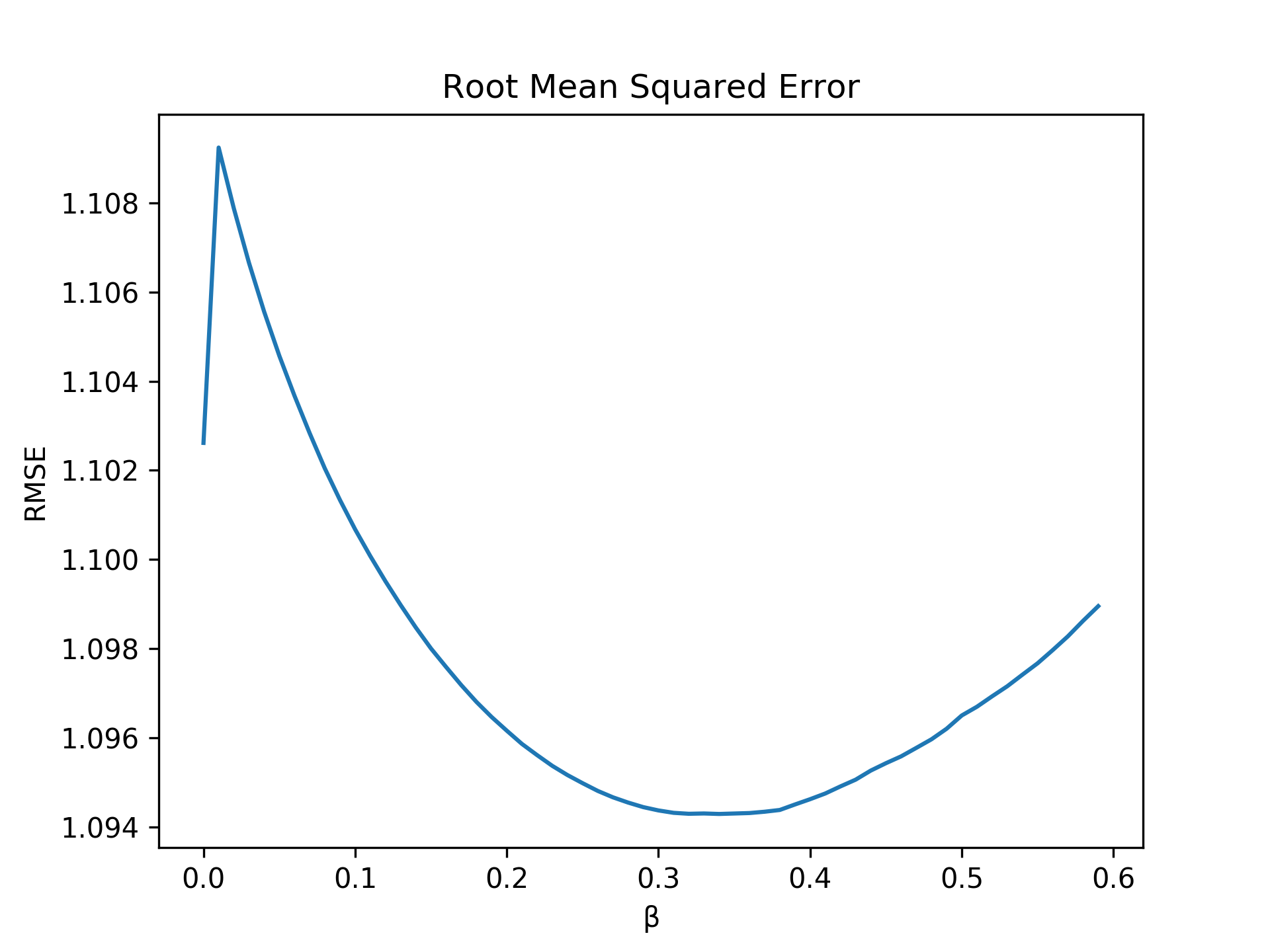}
		\caption{RMSE as $\beta$ increases.}
	\end{subfigure}
	\caption{Tuning $\beta$ for MTR}
	\label{mtr-social-weight-tune}
\end{figure}

MTR has two significant hyper parameters: the maximum neighborhood size for a user (e.g. the maximum number of peer recommendations that will be taken into consideration), $\kappa$, and a social weighting parameter, the value of $\beta$ in Equation \ref{eqn_mtr_influence}.
In their original work, Mauro et al. \cite{multi-faceted_mauro_2019} set $\beta$ at 0.1, but did not report on how modifying this variable effects recommendation accuracy\footnote{Note, $0 \leq \beta \leq 1$.}.
We used all Yelp users from the filtered data set and computed the recommendation accuracy as $\beta$ changed using a set of predictions based on a single social classifier (i.e. the FriendPrediction setup).
Results are illustrated in Figure \ref{mtr-social-weight-tune}, showing a clear tradeoff between only considering user-user similarity and incorporating trust.
Similar results were seen for the PrefPredict experiments.
Accordingly, future experiments were run with a value of $\beta = 0.3$.

We ran experiments with $\kappa$ set to 50, but also experimented later with modifying the value of $\kappa$ (to simulate sparsity).
As a reminder, this value is the maximum number of ratings that are considered when recommending to a user.

Test sets were created by reserving 20\% of each user's reviews.
For all figures in this section, except in the results shown in Figure \ref{trustmf-social-weight-tune}, these reviews were excluded from every step of the process\footnote{In earlier versions of this work \cite{alex2020pmft}, we only split reviews into test and train sets at the last step (recommender evaluation). This would allow, for example, the clustering step to use data to form clusters which was later being tested on.}, that is, the clustering and link prediction steps did not have access to these reviews.
Due to the computation time required to generate and evaluate many of these experiments, only the results reported in Table \ref{table_yelp_results} were cross validated.
In this case, 5-fold cross validation was used with respect to users, so each user had a distinct 20\% of their reviews reserved as a validation set for each of the folds.
This validation set was hidden from every step in the pipeline.
This validation approach is similar to the ones used in \cite{multi-faceted_mauro_2019} and \cite{multi-faceted_fang_2015}, where results were reported based on the average across folds of a 10-fold cross validation and the average across a complete leave-one-out cross validation, respectively.

%\TODO{Describe hyper parameter tuning}
%Both TrustMF and TrustSVD have a number of hyper parameters for controlling training. 
%In particular, they share dimension, max iterations, social importance, rating importance and regularization penalty.
%Dimensions are the number of latent factors each model learns, as described in Section \ref{section_latent_factors}.
%Max iterations and regularization penalty are standard optimization parameters: how many rounds of optimization to perform and how heavily to bias against high magnitude parameters.
%The most significant parameters are the social importance and rating importance parameters.
%Since both algorithms perform a dual optimization over the user-user trust matrix and the user-item rating matrix, it is necessary to choose values that encode the relative importance of each of the optimizations.
%
%In our experiments we kept dimensions at 10 and max iterations at 500 for TrustMF and 100 for TrustSVD (the latter is more computationally intensive). 
%Regularization and rating importance were held constant while social importance was varied and recommendation tasks with low iteration counts were run in order to find appropriate values for this parameter. Results of this hyper parameter search are illustrated in Figured \REFMIA.
%Once appropriate values for the social importance parameters were learned\TODO{...}

\subsection{Results}

\begin{figure}
    \centering
    \begin{subfigure}[b]{0.49\textwidth}
        \includegraphics[width=\textwidth]{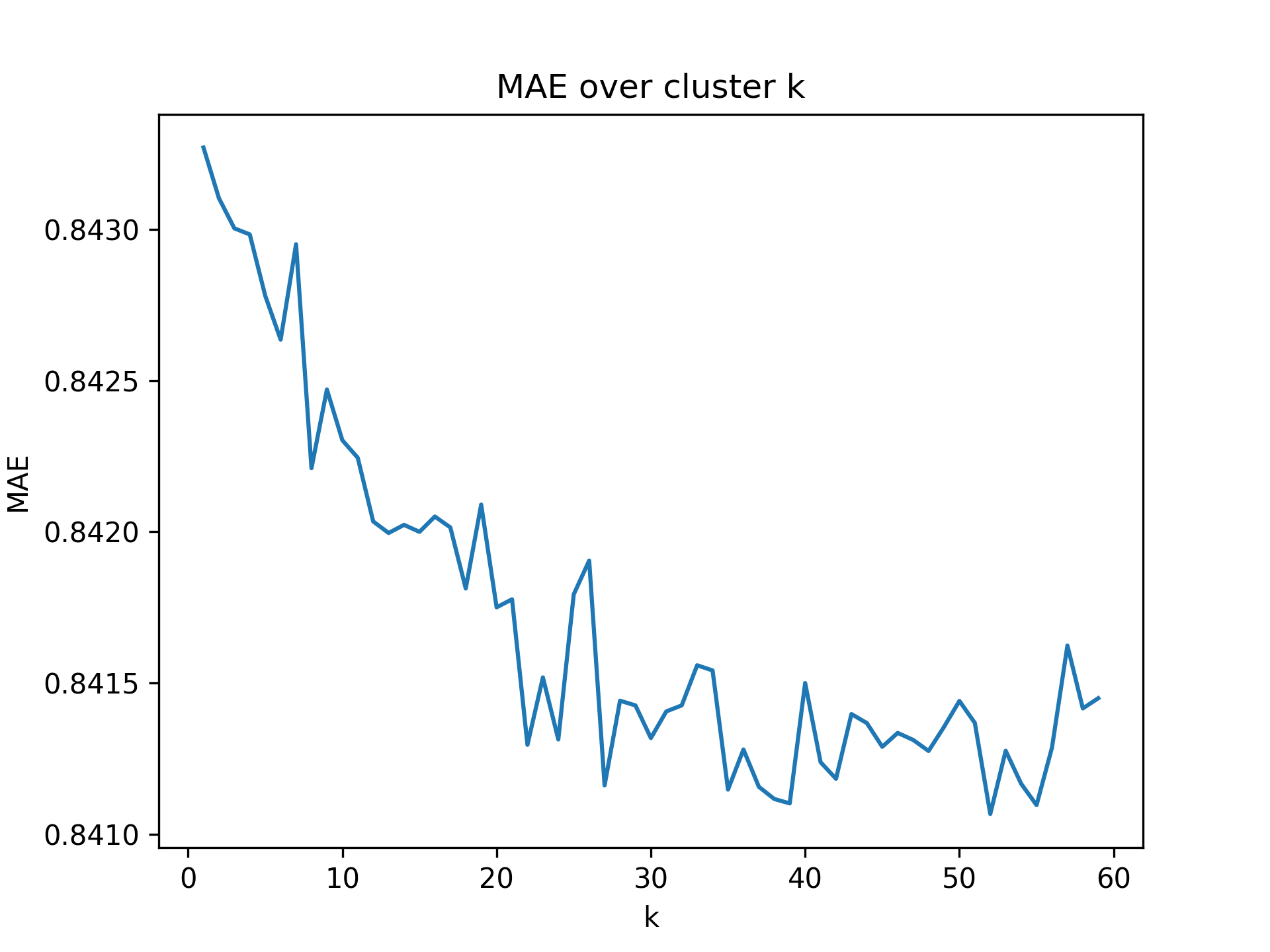}
        \caption{MAE as $k$ increases.}
    \end{subfigure}
    \begin{subfigure}[b]{0.49\textwidth}
        \includegraphics[width=\textwidth]{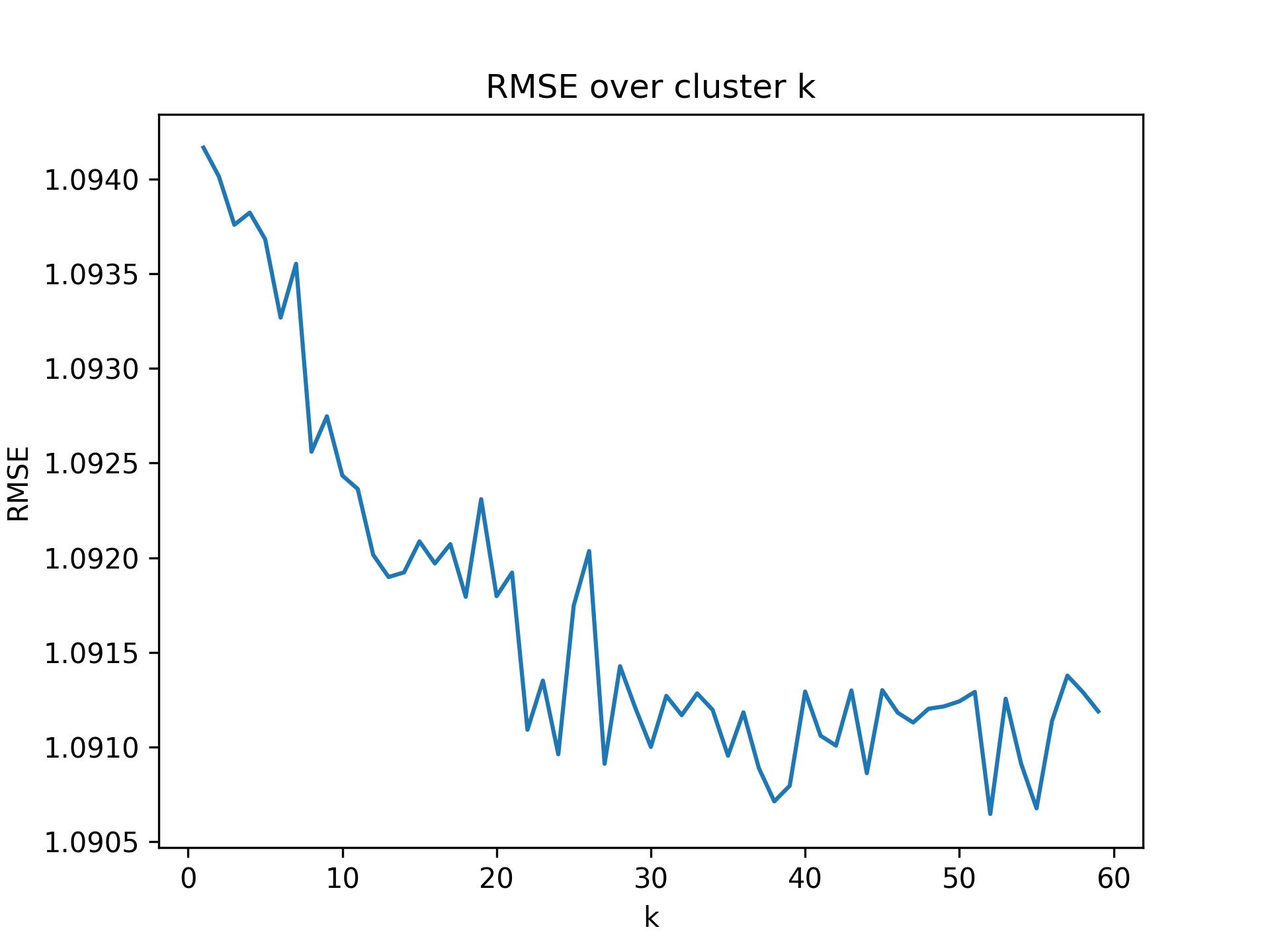}
        \caption{RMSE as $k$ increases.}
    \end{subfigure}
    \caption{Effect of $k$ on error for SocialCluster-FriendPredict with MTR}
    \label{social-social-yelp-over-k}
\end{figure}

\begin{figure}
    \centering
    \begin{subfigure}[b]{0.49\textwidth}
        \includegraphics[width=\textwidth]{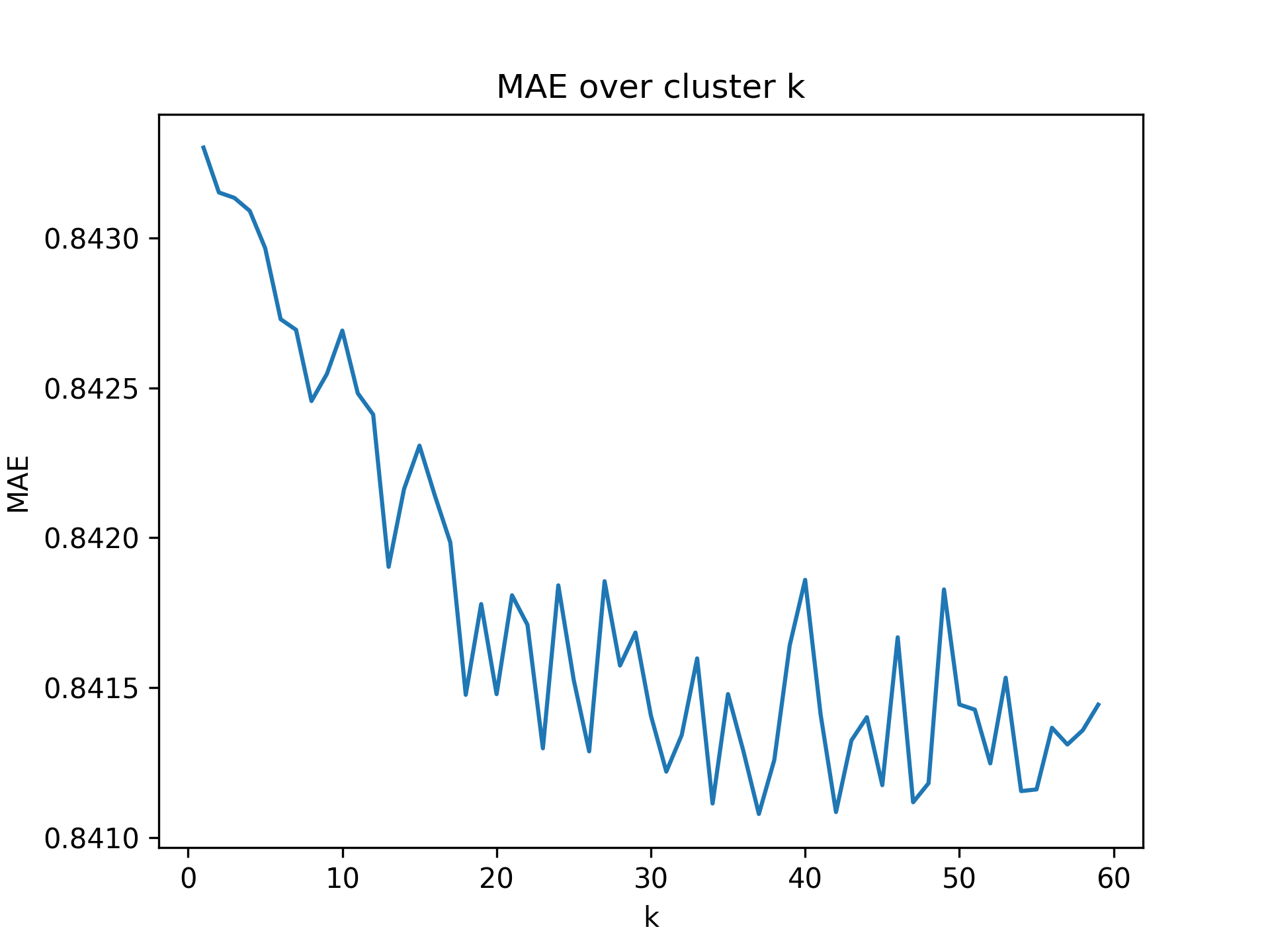}
        \caption{MAE as $k$ increases.}
    \end{subfigure}
    \begin{subfigure}[b]{0.49\textwidth}
        \includegraphics[width=\textwidth]{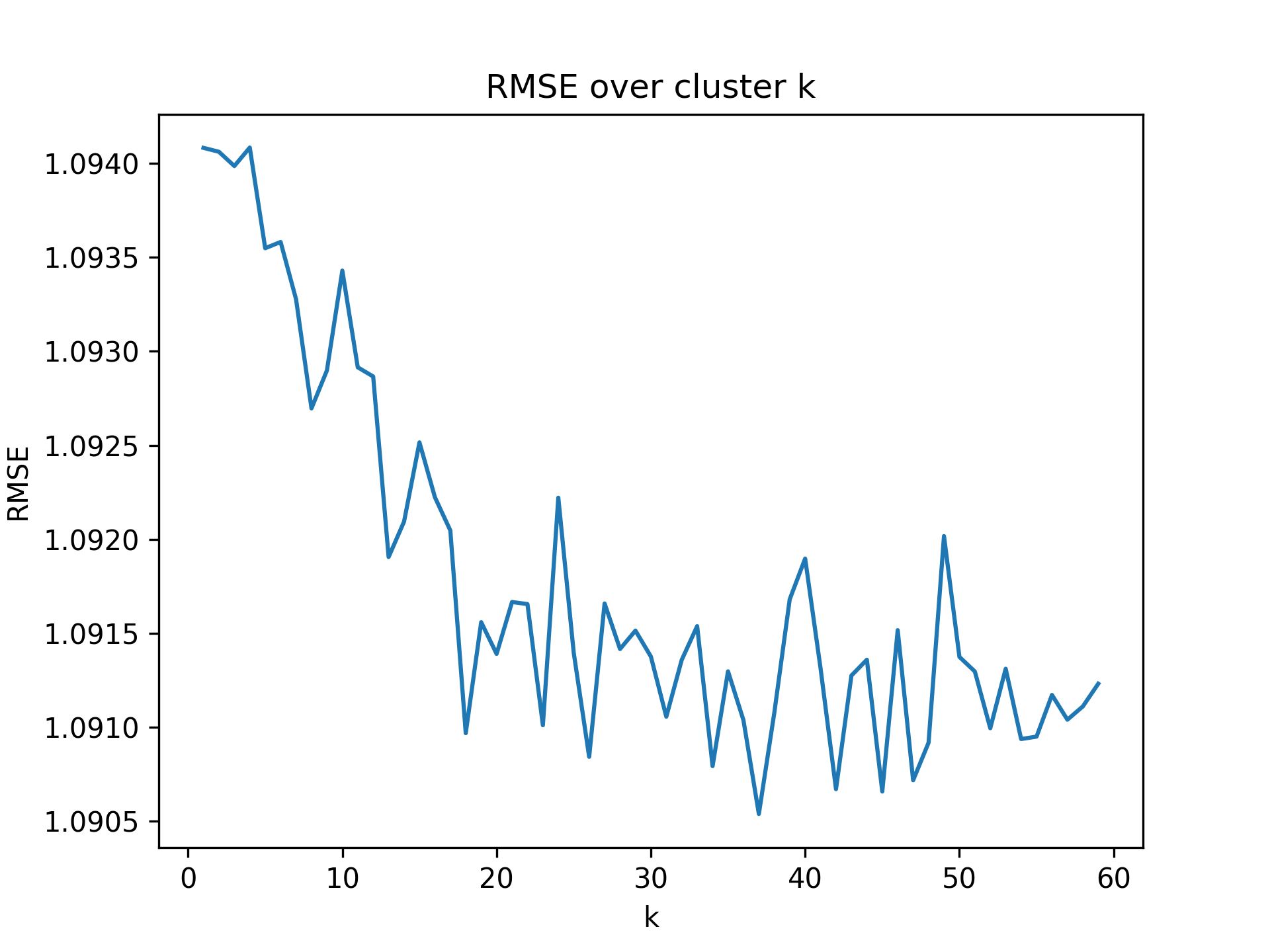}
        \caption{RMSE as $k$ increases.}
    \end{subfigure}
    \caption{Effect of $k$ on error for PrefCluster-FriendPredict with MTR}
    \label{pcc-social-yelp-over-k}
\end{figure}

\begin{figure}
    \centering
    \begin{subfigure}[b]{0.49\textwidth}
        \includegraphics[width=\textwidth]{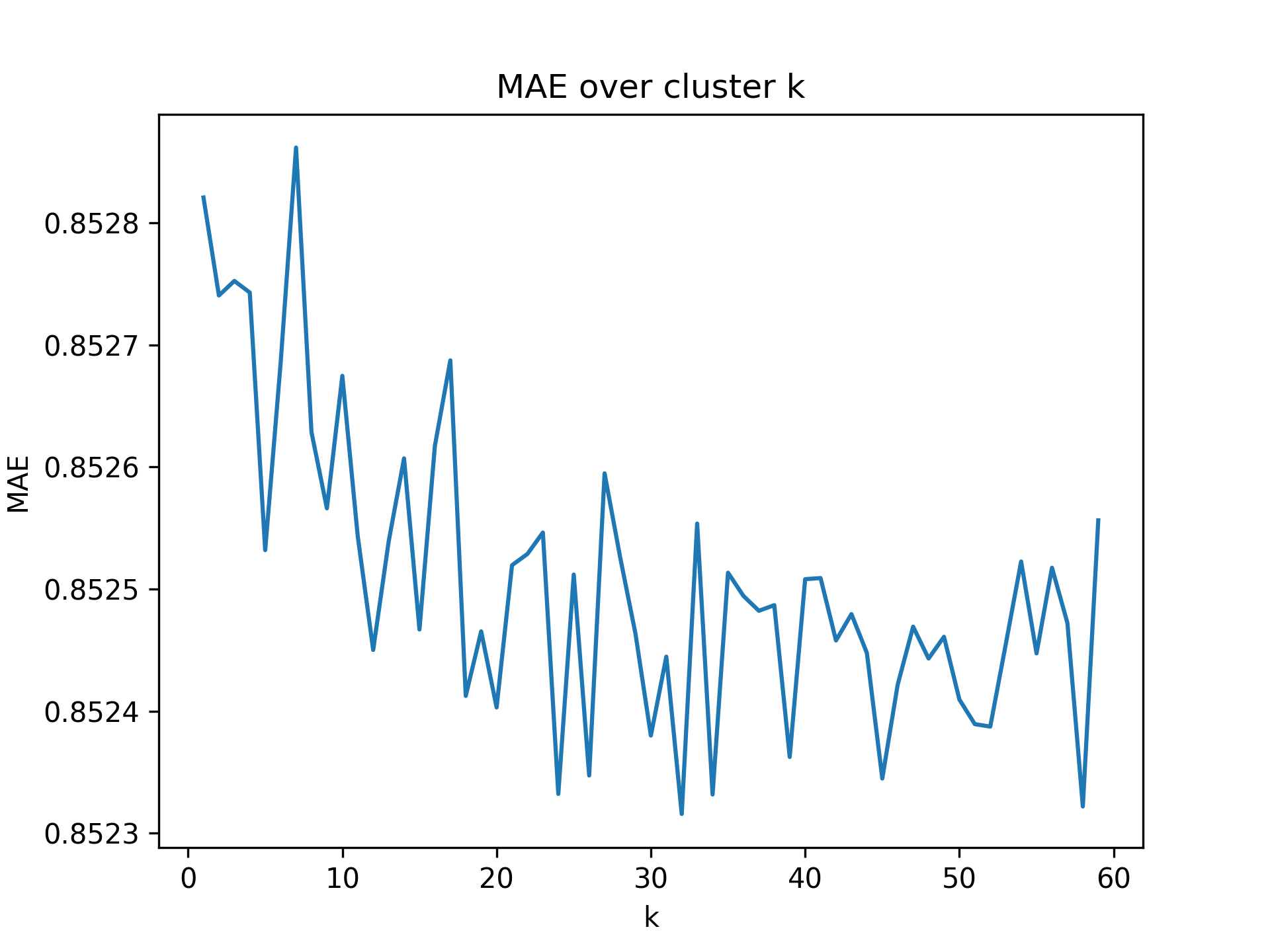}
        \caption{MAE as $k$ increases.}
    \end{subfigure}
    \begin{subfigure}[b]{0.49\textwidth}
        \includegraphics[width=\textwidth]{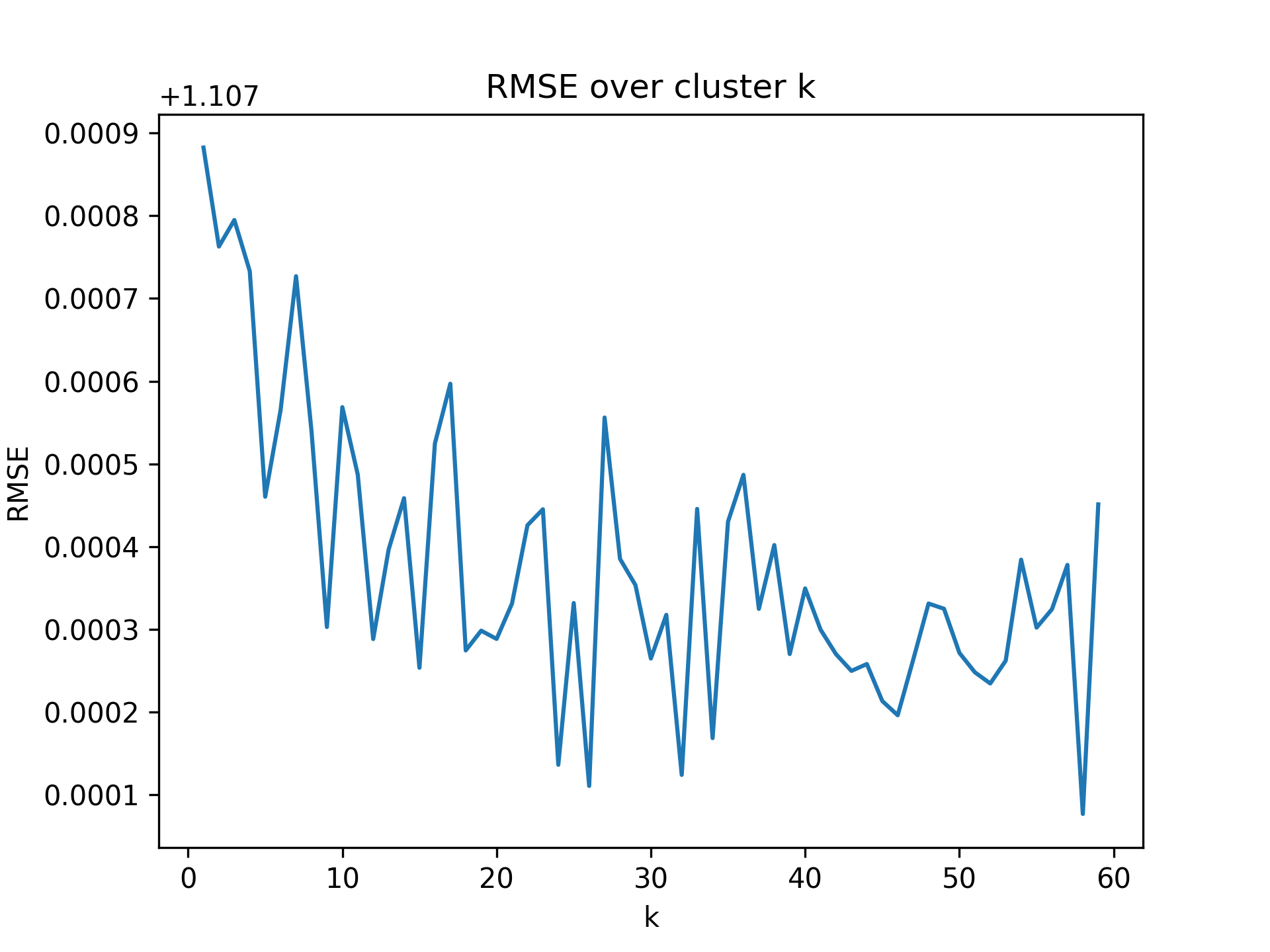}
        \caption{RMSE as $k$ increases.}
    \end{subfigure}
    \caption{Effect of $k$ on error for SocialCluster-PrefPredict with MTR}
    \label{social-pcc-yelp-over-k}
\end{figure}

\begin{figure}
    \centering
    \begin{subfigure}[b]{0.49\textwidth}
        \includegraphics[width=\textwidth]{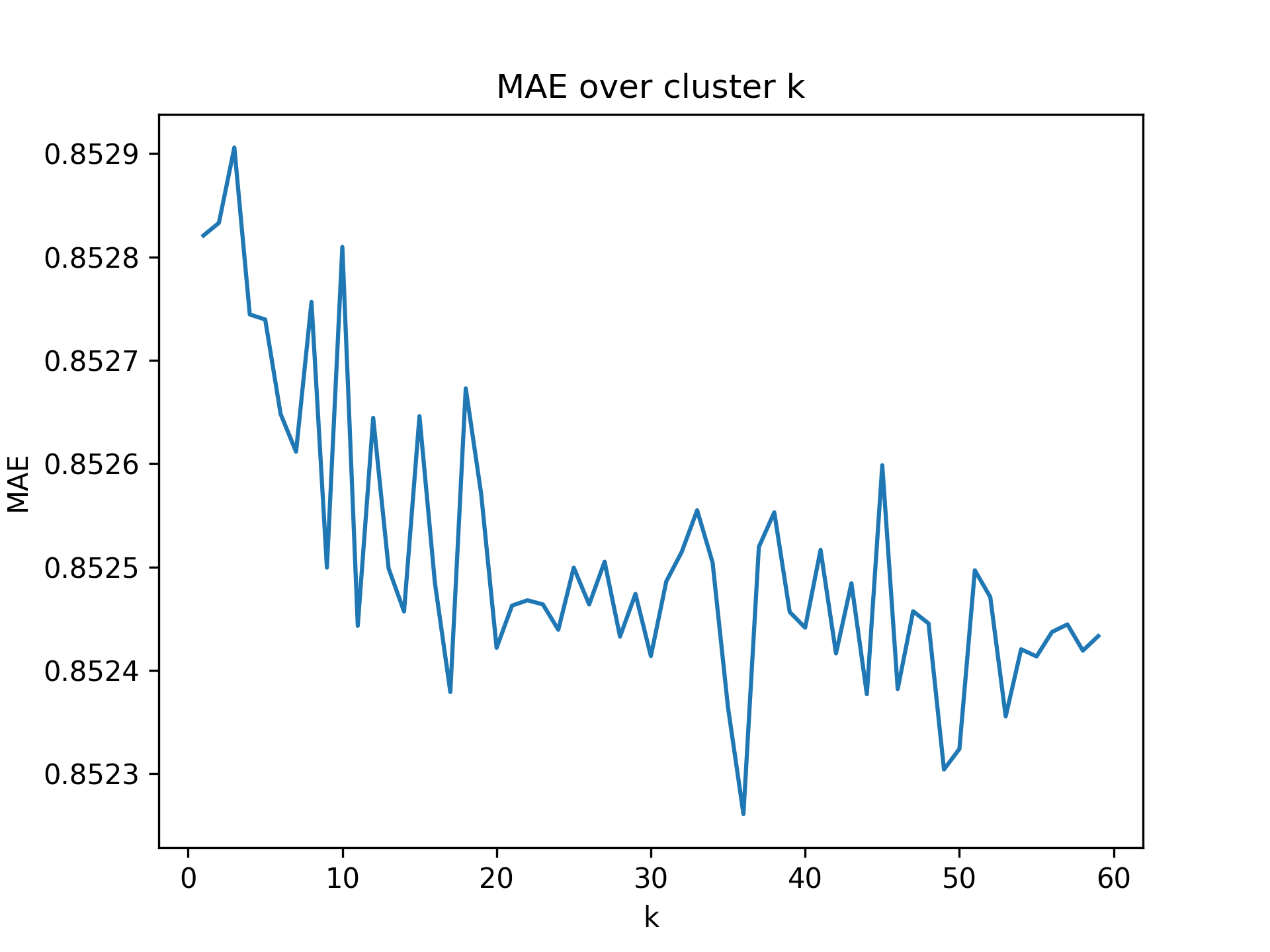}
        \caption{MAE as $k$ increases.}
    \end{subfigure}
    \begin{subfigure}[b]{0.49\textwidth}
        \includegraphics[width=\textwidth]{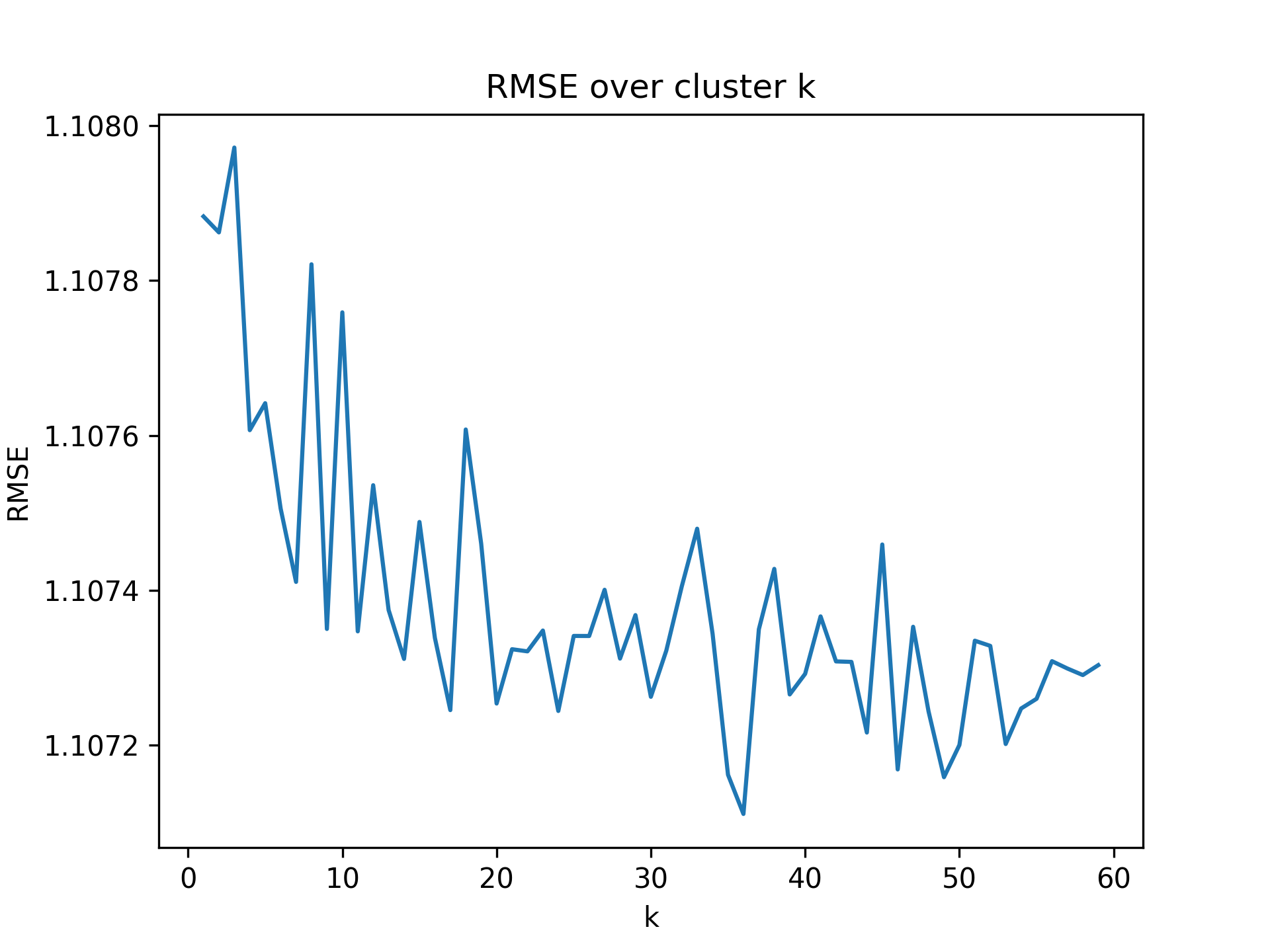}
        \caption{RMSE as $k$ increases.}
    \end{subfigure}
    \caption{Effect of $k$ on error for PrefCluster-PrefPredict with MTR}
    \label{pcc-pcc-yelp-over-k}
\end{figure}

In our original tests, we attempted to set the number of clusters, $k$, by evaluating the silhouette scores of clusters in a large range for each data set, then simply choosing $k$ based on whichever cluster count had the highest silhouette score.
Unfortunately this method was fickle, as the silhouette evaluation is based on a random sample of the clusters, and a single outlier could achieve a minimal score even if other nearby values of $k$ were not optimal.
Further, it is basically a heuristic to use cluster cohesion to choose the number of clusters, when ultimately we are interested in improving the personalized trust links.
Therefore, we iterated over a range of cluster values and repeated the entire experiment with each choice of cluster score, using the MTR recommender system.
Results are illustrated in Figures \ref{social-social-yelp-over-k} to \ref{pcc-pcc-yelp-over-k}.
In general, results show that as the number of clusters searched for ($k$) increases, the error in the task follows a consistent trend of reduction.
Note that when $k=1$, the situation is equivalent to a non-personalized approach (as only searching for a single cluster is equivalent to doing no clustering), and as $k$ increases the granularity of personalization increases.
When predicting whether two users should be friends or not, MAE can be lowered by 0.003 points by adding personalization, while when predicting aligned preferences it is only lowered by 0.0005 points.
These results are less impactful than the early results seen using the TrustMF classifier in Figure \ref{trustmf-social-weight-tune}.
That said, the results indicate a consistent trend of improvement as clustering based personalization is applied: a fairly consistent line of decrease in error can be observed in all lines.

\begin{figure}
    \centering
    \begin{subfigure}[b]{0.49\textwidth}
        \includegraphics[width=\textwidth]{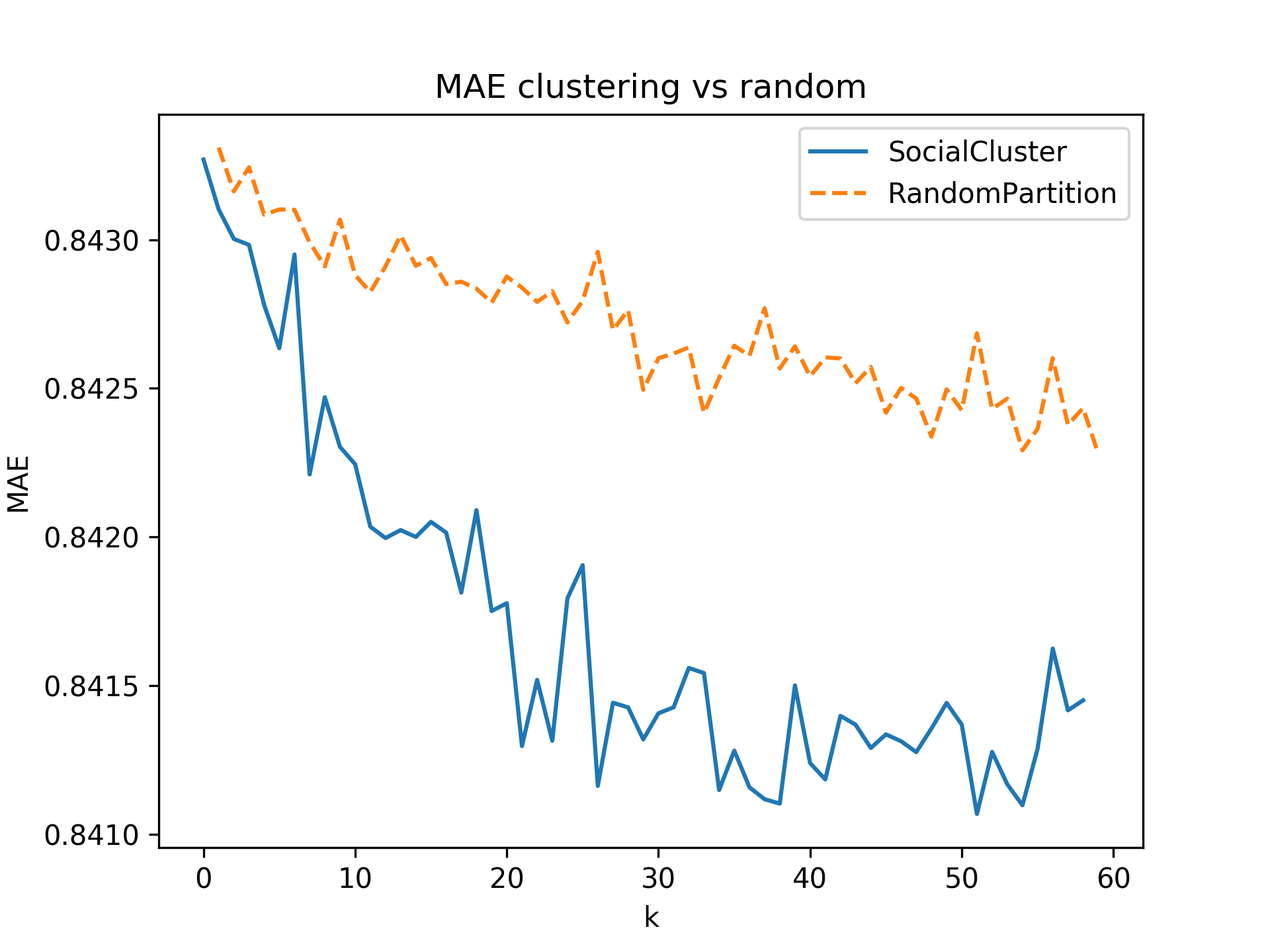}
        \caption{MAE as $k$ increases.}
    \end{subfigure}
    \begin{subfigure}[b]{0.49\textwidth}
        \includegraphics[width=\textwidth]{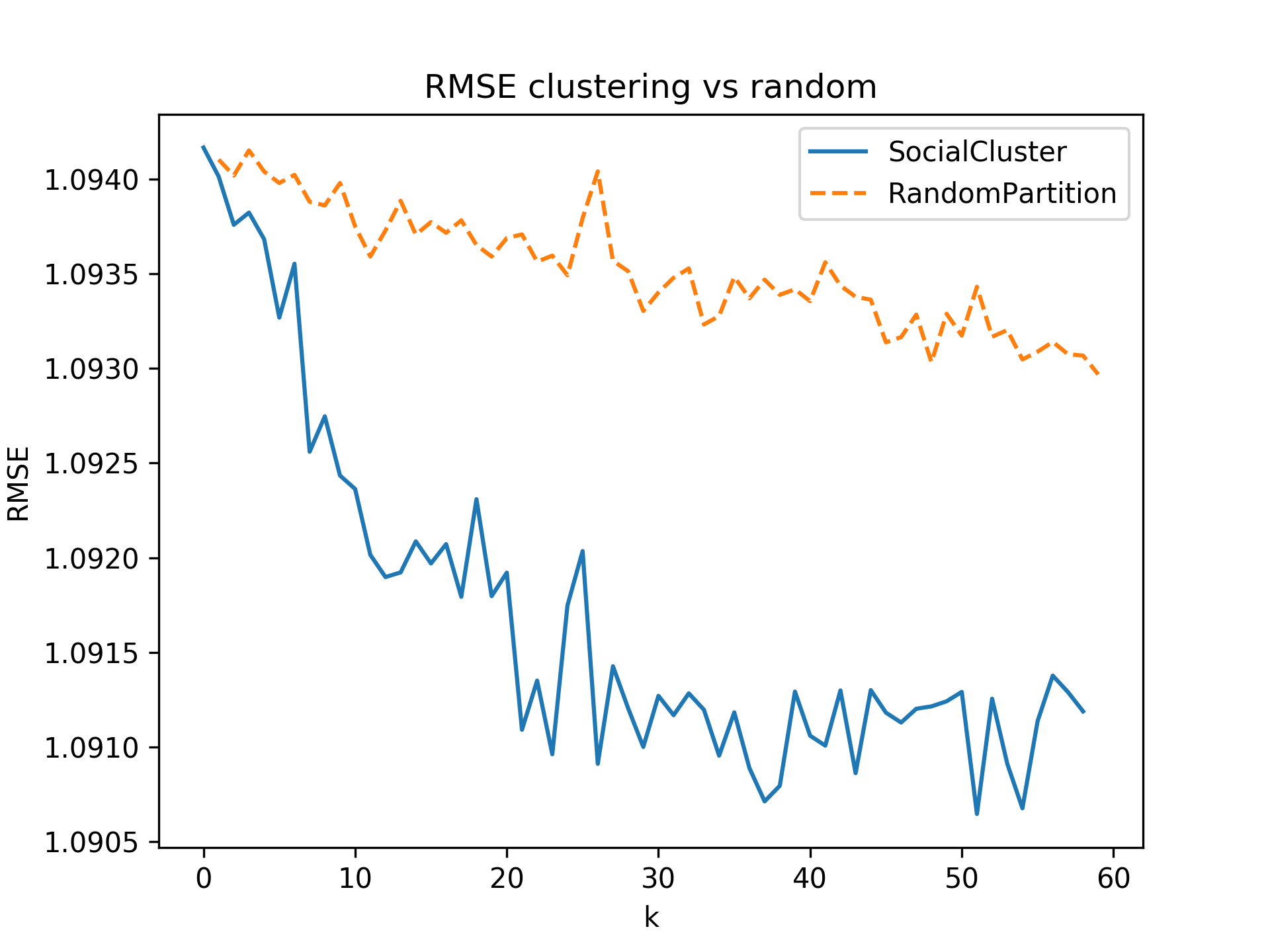}
        \caption{RMSE as $k$ increases.}
    \end{subfigure}
    \caption{SocialCluster-FriendPredict versus RandomCluster-FriendPredict.}
    \label{social_vs_random}
\end{figure}

We wished to verify that the results illustrated in these figures are indeed improving because our clustering technique was finding groups of similar users, which allowed the prediction techniques to learn more personalized classifiers for these groups.
For instance, it is conceivable that splitting users into groups and learning multiple classifiers is helpful regardless of the groups picked, as this procedure would be similar to bootstrap aggregating \cite{breiman1996bagging}, which allows simple classifiers to model multiple weak correlations in data.
To test this, we repeated the FriendPredict experiment, but clustered agents into $k$ clusters randomly\footnote{This random clustering essentially partitions the data set into $k$ random samples (a close emulation of bootstrap aggregation).}.
The results illustrated in Figure \ref{social_vs_random} compare this random clustering to clustering by social circle overlap.
This figure clearly shows that the reduction in error is largely due to the non-random clustering approach.
The solid and dashed lines start off identically at $k=1$ on the x-axis (no clustering) but as the numbers of clusters increases, the error decreases, for the case where social clusters are used.
We take this as evidence that the clustering technique is improving accuracy because clustering genuinely enables more personalized predictions, not simply because the number of models being learned has increased.

\begin{figure}
    \centering
    \begin{subfigure}[b]{0.49\textwidth}
        \includegraphics[width=\textwidth]{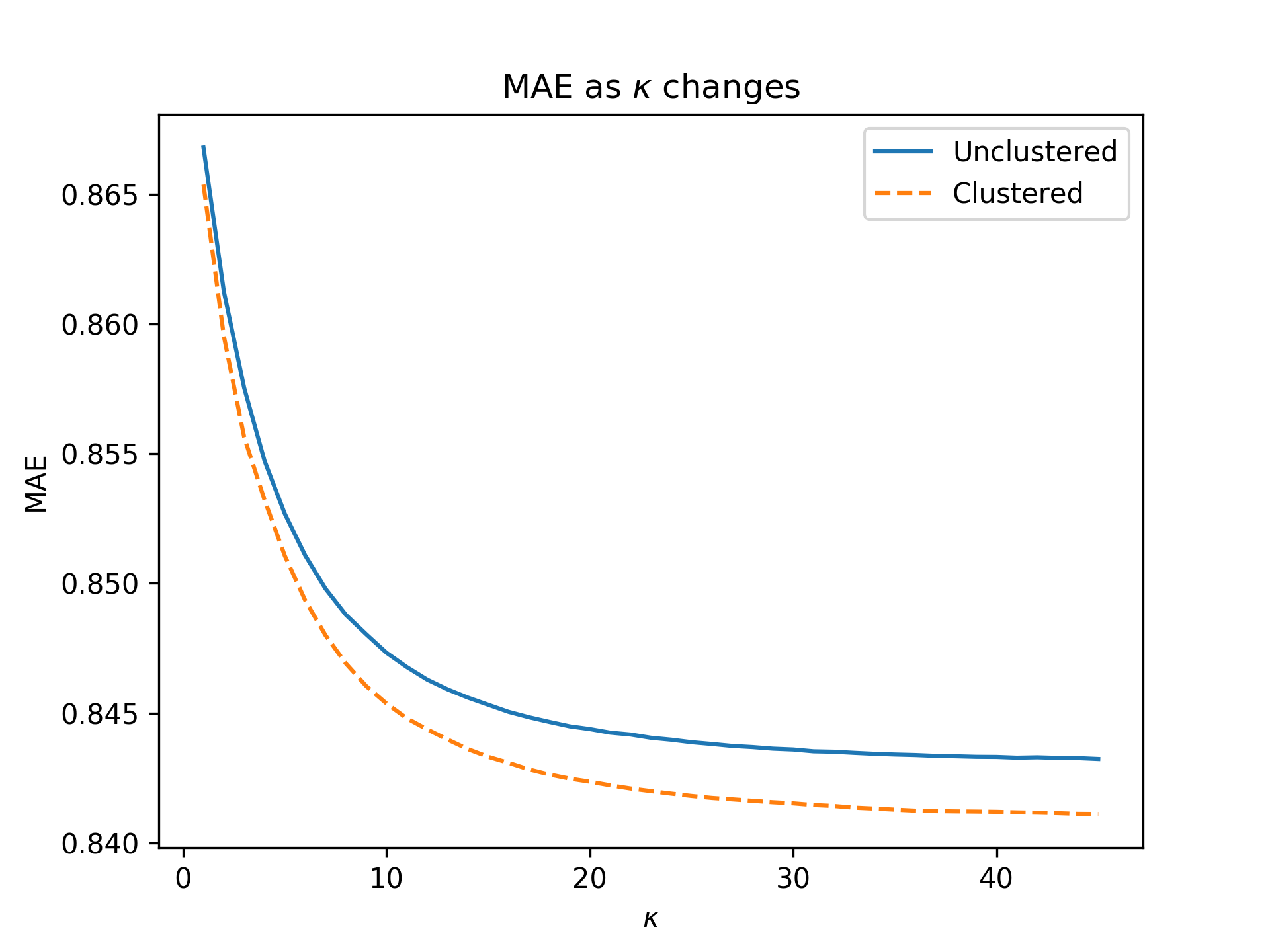}
        \caption{MAE as $\kappa$ increases.}
    \end{subfigure}
    \begin{subfigure}[b]{0.49\textwidth}
        \includegraphics[width=\textwidth]{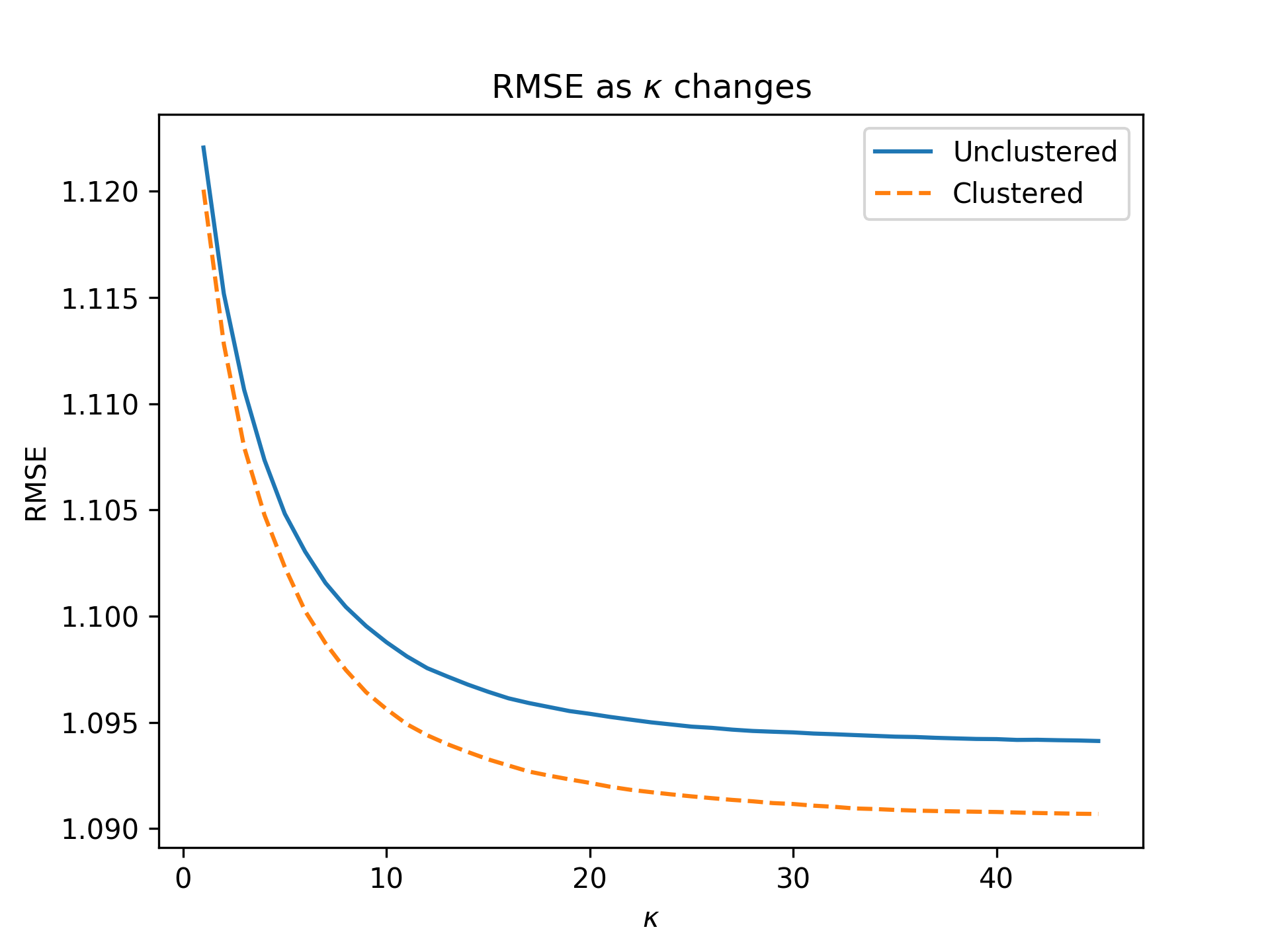}
        \caption{RMSE as $\kappa$ increases.}
    \end{subfigure}
    \caption{Effect of $\kappa$ on accuracy for MTR.}
    \label{kappa_change}
\end{figure}

We also experimented with modifying the $\kappa$ variable on MTR, effectively simulating sparsity, as this variable controls how many peer advisers can be considered for a recommendation.
Results illustrated in Figure \ref{kappa_change} compare a accuracy on the SocialCluster-FriendPredict task, comparing the error rates for a single cluster (unclustered) and for 55 clusters (clustered). 
Overall, the gap in error rate is most dramatic when a larger $\kappa$ value is used, but the advantage for a clustered approach applies over the range of values.

Finally, in Table \ref{table_yelp_results} we present the results of a 5-fold cross validation over user ratings for these tasks. 
Best results are bolded.
There are a number of interesting results.
First, the conceptually simple MTR system outperforms TrustMF across the board, despite the fact that the TrustMF system was allowed to run for a much greater period of time in order to reach convergence.
This gap in performance is often dramatic, for example, in the best cases for each system, MTR has a MAE that is 5\% lower than TrustMF, and a RMSE that is 1.6\% lower.

\begin{table*}[t]
    \centering
    \begin{tabular}{|l|l|l|l|l|}
        \hline
        \multicolumn{1}{|c}{} & \multicolumn{2}{|c|}{\textbf{MTR}}    & \multicolumn{2}{|c|}{\textbf{TrustMF}}     \\ \hline
        \textbf{} & \textbf{MAE} & \textbf{RMSE} & \textbf{MAE} & \textbf{RMSE} \\ \hline
        % NoSocial & .8674 & 1.1260 & .8935 & 1.1378 \\ \hline
        RealFriends & .8610 & 1.1196 & \textbf{.8879} & 1.1205 \\ \hline
        FriendPrediction & .8453 & 1.0960 & .9063 & 1.1179 \\ \hline
        SocialCluster-FriendPredict & \textbf{.8434} & \textbf{1.0932} &  .9120 & 1.1179 \\ \hline
        PrefCluster-FriendPredict & .8436 & 1.0936 & .9077 & 1.1189 \\ \hline
        
        PrefPredict & .8551 & 1.1105 & .8984 &  \textbf{1.1109} \\ \hline
        SocialCluster-PrefPredict  & .8551 & 1.1103 & .8987 & 1.1111 \\ \hline
        PrefCluster-PrefPredict  & .8551 & 1.1103 & .8987 &  1.1111 \\ \hline
    \end{tabular}
    \caption{Recommendation error results.}
    \label{table_yelp_results}
\end{table*}

On the better performing MTR recommender system, the best results are achieved when predicting friendship links rather than predicting preference correlation.
This makes, sense, as the MTR system already considers observable user preference (see Equation \ref{eqn_mtr_influence}), thus predicting new instances of aligned preferences is not likely to add much new information.
The best performing task, by a small margin, is the SocialCluster-FriendPredict task, which basically reconfirms the findings presented earlier in this section (with the added certainty of being averaged across folds).
Clustering did not have an appreciable effect (at least not at the scale of $10^{-4}$) for the preference alignment prediction task.

Note that, in the case where an improvement was seen, although the scale of the effect appears small it is clear from the previous graphs that this is not merely due to statistical variance. For example, Figure \ref{social-social-yelp-over-k}, clearly shows that the decrease in error between FriendPrediction and SocialCluster-FriendPredict is due to the increasing the number of clusters.

On TrustMF, results are more tightly grouped and there is quite little appreciable difference between experiments.
As this system is more conceptually complex than MTR, it is difficult to interpret exactly why this might be the case.
Interestingly, the best performing task for MTR is the \textit{worst} performing task on TrustMF.
Clustering does not have a positive effect in these experiments, and in the SocialCluster-FriendPredict task actually seems to harm the performance.
Our early experiments with this recommender system (presented in Figure \ref{trustmf-social-weight-tune}) suggested that there might be more interesting differences between approaches, but this was not the case in these final results.
We speculate that because these earlier results were computed using different techniques to select Yelp users (randomly selecting 10000 users versus our final strategy of selecting all 30000 users with more than 20 reviews submitted), the underlying distributions of ratings may have been different enough to cause this change.

\subsection{Conclusion}

In this work, we evaluated the effect that personalization via clustering had on the accuracy of a trust link prediction task.
We accomplished this by predicting novel trust links on a data set of Yelp users and measuring accuracy of these predicted trust links via an item recommendation task.
%In order to complete this, we proposed multiple approaches for clustering users of this network, and evaluated the accuracy of these clustering algorithms.
%Further, our link prediction approach combines techniques from the literature with novel features, resulting in a comprehensive MFTM solution.
%Finally, we used two trust-aware recommender systems from the literature in order to evaluate the utility of the predicted links.

Our results show that the option of predicting novel trust links results in better performance than using the explicitly stated trust links for the recommendation task.
Further, our results show a small but consistent improvement in recommendation accuracy when clustering is used to determine groups of similar agents and distinct trust prediction models are learned for each group of agents with the MTR recommender system (e.g. Figures \ref{social-social-yelp-over-k} to \ref{pcc-pcc-yelp-over-k}).
We showed that this improvement was not simply the result of the fact that more classifiers were trained, as randomly splitting users into groups does not improve accuracy nearly as much as the clustering technique does (Figure \ref{social_vs_random}).
While our early results with TrustMF inspired confidence, and we hoped to see more dramatic improvement in recommender accuracy from these experiments, the final results show that, while consistent, the improvements in accuracy from the procedures outlined here are small.
We will comment on ways these techniques could be improved, and potential avenues for future work, in Section \ref{future_chapter_4}.

%In addition to the experiments with personalization, our work combines techniques from the literature; we also make clear the applicability of MFTM to social networks.
In addition to the experiments with personalization (which explored multiple approaches for clustering users), we produced a comprehensive MFTM solution, combining techniques from the literature with novel features.
We also make clear the applicability of MFTM to social networks.
We experimented with predicting two types of trust links:
explicit friendship (the FriendPredict experiments) and implicitly stated preference alignment (the PrefPredict experiments) and evaluated the utility of the predicted links
derived by our methods, using two distinct trust-aware recommender systems.
We found that the preferred target of trust link prediction can vary with the desired use-case: it was not clearly preferable to predict friendship links or preference alignment links.
On the MTR system, which already strongly considers user preference alignment, our experiments performed better when predicting friend links, while on the TrustMF system predicting preference alignment between users produces (slightly) better recommendation accuracy.

\section{Discussion}
\label{chapter_discussion}
In this section, we first reflect on how our work compares to
those of other researchers, with respect to both multi-faceted
trust modeling and to personalizing solutions for trust modeling.
We then discuss how our methods for predicting trust links
using our particular personalized multi-faceted trust modeling
can serve as a useful starting point for handling misinformation
in social networking environments.

\subsection{Comparison with related work}

\subsubsection{Multi-faceted trust modeling}
Our work in Section \ref{chapter_PMFTM} was heavily inspired by the works of Mauro et al. \cite{multi-faceted_mauro_2019} and Fang et al. \cite{multi-faceted_fang_2015}.
All works have a similar structure: they propose a multi-faceted trust model and test it on a recommendation task on a data set harnessed from a site with item rating component.
We sought to extend these works by combining the best features from each of them while testing the effects of a personalization step to increase the accuracy of trust prediction.
In particular, Mauro's work developed a large set of trust indicators on the Yelp data set, while Fang's work proposed a smaller set of relatively generic indicators that could be used on the Epinion's data set. 
In our work, we combined these indicators when testing on the Yelp data set, with the goal of achieving a more comprehensive model of user to user trust formulation.
While Mauro's work proposes a large number of trust indicators, it does not seek to weight the importance of those indicators in a data driven manner: they instead experiment by taking a non-weighted average of a subset of the indicators.
Like in Fang's work, we have used a logistic regression to find weights for these indicators that fit the data set, believing this method to be a more principled approach to the problem.
% In Fang et al. \cite{multi-faceted_fang_2015}, the authors included the modeling of distrust, using the distrust links in the Epinions data set. 
In addition, we did some preliminary investigation of Epinions data in order to expand the environments examined under our approach. 
Conceptually, the approaches taken to personalize recommendations we undertook on the Yelp data would be easily transferable to this data set.
See Section \ref{future_chapter_4} for more details.

% The concept of clustering users before learning weights for trust prediction, as a method of personalization, was proposed parenthetically in Fang et al. \cite{multi-faceted_fang_2015}.
% Our work in Section \ref{chapter_PMFTM} was largely conceived as a test of whether or not this suggestion would indeed increase the accuracy of trust prediction. 

Another work which has relevance is that of Gilbert and Karahalios \cite{gilbert_predicting_2009}.
While not an artificial intelligence paper, the authors present a multi-faceted statistical analysis of the factors which affect tie strength between pairs of users in social media.
They found that a set of 74 variables collected from a the Facebook accounts of participants could be used to predict, with high accuracy, the answers these participants gave to survey questions designed to model social tie strength with their friends on Facebook (e.g. ``How comfortable would you feel asking this person for a loan?'').
This work presents strong evidence for the notion that trust (i.e. as an aspect of a strong social tie) can be predicted between agents based on relatively simple data extracted from interaction history on social media.

\subsubsection{Personalization Approaches}
\label{section_personalization}

Given the subjective nature of trust, it is clear that accurate trust models need to incorporate some level of agent-specific personalization.
It is feasible to model reputation or popularity without such personalization.
However, given the sparsity of data in most networks, the cold start problem\footnote{That is, the problem of giving personalized recommendations to a user who has just joined the network and has not expressed any beliefs, opinions or preferences.}, and computational limitations, it is not typically feasible to give each agent a completely distinct model.
Our approach to personalization in Section \ref{chapter_PMFTM} was to determine clusters of similar users and learn trust link classifiers on the basis of these clusters.

%This approach was suggested by Fang et al. \cite{multi-faceted_fang_2015}, but other approaches have been attempted in the trust modeling space.

The Personalized Trust Model developed by Zhang and Cohen \cite{PTM_Zhang_2008} (described in Section 2.1) can be seen as an extension of the Beta Reputation System \cite{BRS_Josang_2002} that computes both a private and public trust factors for integrating the advice of some other agent.
% An agent's private trust factor is based on the similarity in advising behaviour between the trustee and the truster, while the public trust factor is based on how similar the trustee's advising behaviour is to average advising behaviour (similar, but not identical, to the Competence trust indicator proposed in \cite{multi-faceted_fang_2015}).
% The final trust prediction for some truster-trustee pair is then a weighted combination of the private and public trust factors.
The weighting of these factors is based on the overlap in the number of common items the truster and trustee have advised on (rated), thus giving higher weight to private trust when there is more basis for comparing the two agents directly with respect to past behaviour.
Effectively, this system implements personalization by using generic predictions under uncertainty about individual preferences. %, and offering progressively more personalized predictions as more data becomes available.
However, their formula for assigning weight to personal and private trust factors is basically a heuristic, as the settings for appropriate error and confidence bounds are not derived in a data driven manner.
Our work attempts to implement personalization in a data-driven way, by identifying clusters of similar users and learning their trust formulation procedures at a cluster level\footnote{Fleming \cite{specific_fleming_2004} also proposes a progression in user modeling from assumptions about general users to ones about individuals but they also suggest an intermediate phase of learning more about groups, via stereotypes. Our consideration of clusters fits well within this vision.}.
%\TODO{Discuss Fleming increasingly specific user models?}

The usefulness of stereotypes towards improved trust modeling is an approach examined by other researchers who may also derive benefit from examining our data-driven methods.
The StereoTrust Model developed by Liu et al. \cite{liu_stereotrust_2009} implements personalization by allowing each agent to define its own grouping function for partitioning the set of other agents via stereotypes. 
For example, an agent may decide to stereotype based on stated interest, location, seniority, etc.
This is intended to model the subjective assumptions humans apply in every day life.
The agent then uses a trust estimation function (inspired by the Beta Reputation System \cite{BRS_Josang_2002}) to reason about their trust with respect to groups defined by stereotypes, rather than with respect to individuals. 
The trust an agent $\truster{a}$ has in another agent $\trustee{b}$ is then computed as a weighed average of the trust $\truster{a}$ has in all the groups that $\trustee{b}$ is a part of.

% Thus, by partitioning the set of agents in the environment into groups and reasoning about trust with respect to these groups, the data sparsity problem is reduced.
% However, it is unclear how the cold-start problem is alleviated by this system.
This system implements personalization by allowing each agent to specify its own stereotypes, although in practice it's not clear how this information would be elicited from real users. 
This approach relies on the notion that members of a group will act similarly, but by allowing individual agents to define groups arbitrarily, the usefulness of this notion is under a certain strain.
Without the ability to statistically analyze large amounts of data from the environment, it is unclear how individual agents could be expected to create stereotypes that define groups which actually have some cohesiveness of behaviour.
In their actual implementation, stereotypes were implemented based on rating similarity, so no agents actually had an opportunity to specify their stereotypes, and in practice this solution turns out to be a complex approach to reach the same end goal as, for instance, clustering users based on rating behaviour (as we have done).

\subsection{Towards improved handling of misinformation}
In order to improve the lives of users of social media, we have
presented an approach for predicting trust links between peers
within these networks. Our framework makes it possible to assess
whether the content created online is a good candidate to display
to a user or not (where options may include flagging messages
coming from sources that are not established to be well trusted).

Our work aims to improve online experiences by supporting distinct presentation of content to differing users, achieved by reasoning about relationships with peers and the concept of trust.
Our concern with trustworthiness of content relates well to companion efforts devoted to detect digital misinformation \cite{ciampaglia_2018,hui_2018,yang_2019}. 
There is a spectrum of possible outcomes when messages which are of questionable quality are shown to users, including special attention in contexts such as healthcare where the consequences may be more troubling \cite{ohashi_2017}.
Note that there will still be various options for actions to take, once trust modeling has provided some insights into messages of concern.

The methods we present here are designed to be self-contained algorithms which can be provided to any party which has the data at hand, to reason about trustworthiness. It would be possible, for instance, to have platform owners flag less trustworthy posts and individual users have agency to choose what kinds of information should be filtered for them.
Our algorithms would be able to indicate, for a particular user,
whether the other users in the environment are predicted to be
trusted (based on whether a trust link is predicted to exist between them).
There is then an interesting dilemma about whether top-down control of
the social network (e.g. dictated by government) or bottom-up management
of the content (e.g. under the control of individual users)
should be launched in order to take actions with respect to
the messages of agents with questionable trustworthiness.
While our model promotes a solution that is attuned to an individual's
preferences, in cases where these
may be in conflict with interests of the public (for instance, promoting hate)
a tension may exist in deciding where the control should lie.
If a decision is made to consult reputable outside sources
to determine acceptability, this could potentially be integrated into the trust
models to discourage inappropriate behaviour.
We do not propose an answer to this challenge of determining
appropriate control but merely acknowledge this as a concern
for anyone trying to address content recommendation in social media.

As our work has drawn out the value of personalized solutions, the models that we have presented should be flexible enough to support a variety of overall preferences with respect to final outcomes.
With respect to the array of concerns for this Special Issue, our work is best viewed as focused on computational approaches grounded in artificial intelligence methods which assist in the detection of misinformation and disinformation.
We introduce novel perspectives on this particular agenda for improving online social networks, through techniques for personalizing the analysis and with highlighting of the potential provided by performing trust modeling.\\

\section{Conclusions and Future Work}
\subsection{Summary}
% from 6.1
In this paper we considered the problem of improving the experience of users on social networks, particularly with respect to content overload and the propagation of untrustworthy information.
We argued that a trust modeling approach could be appropriate for social networks and could be used to enhance message recommendation systems.
We then outlined some of the issues involved in applying these models as they currently exist.
The types of trust models that can be applied need to be highly flexible, capable of capturing many different kinds of data, and personalizable.

We argued that a multi-faceted trust model was ideal for application to social networks.
This is because the multi-faceted model can incorporate arbitrarily many signals from the agents and their environment into a data driven model of how trust is apportioned by agents in an environment. 
We argued that this flexibility was a key feature, as it allows the model to adapt to many different kinds of social networks.
In Section \ref{chapter_PMFTM}, we designed a comprehensive MFTM and applied it to a large data set, including multiple new features and features proposed in previous works. 
We experimented with personalizing the predictions generated by a multi-faceted model, by clustering similar users and learning distinct models for each cluster of users.
We argued that although this approach is not ``truly individualized'' personalization, a data driven model like MFTM imposes a tradeoff between the number of users a model is learned for, as smaller numbers of users will have less data available to train classifiers with.
We showed that this approach can lower error rates in a downstream trust aware recommendation task.

In brief, our primary contributions with this work are to:

\begin{itemize}
    \item identify a critical challenge in applying trust models to social networks, namely to personalize trust prediction
    \item develop a clustering based approach to personalization on a large
   dataset, applying predictions to a downstream recommendation task
   and showing consistent improvements in error rates
   \item assemble a comprehensive array of multi-faceted trust indicators
   to incoporate into data-driven reasoning about trustworthiness
   as an advance to this method for multiagent trust modeling
   \item outline the potential for our approach to trust link prediction
  (reasoning either about rating behaviour of users or social circles)
    to assist in efforts to address misinformation in social networks,
    clarifying as well challenges which remain to be addressed
\end{itemize}

%\UPDATE{In the sections below,} we have evaluated a clustering based approach to personalization on a large data set, showing a consistent improvement in error rates when predictions were applied to a down-stream recommendation task, to demonstrate the value of our approach.
When misinformation abounds in social media, being able to judge which
sources are trusted is a critical step in assisting the users in
these networks to navigate the waters.
In the sections below we elaborate on ways to extend the models
we have developed, and how to assist users in handling misinformation
once our trust link prediction process has been run.
We follow this with some suggested steps forward to assist some of
the most vulnerable online users, older adults, illustrating how
reasoning with clusters of users, the centrepiece of our proposed
model, can be quite valuable in allowing unique experiences for this
user base.

\subsection{Future Work}
\subsubsection{Expanding upon Personalized Multi-Faceted Trust Modeling}
\label{future_chapter_4}

There are a number of ways the project of personalizing multi-faceted trust predictions can be extended.

\noindent \textbf{Clustering:}
For example, we spent considerable time in this paper explaining the difficulties involved in clustering points that represent agents in a social network.
While the approach we took was ultimately geometrically inspired, graph clustering algorithms could potentially offer a better fit to this type of data.
This is an especially attractive option, as the sparsity of defined similarities between agents when considered geometrically is a major issue for applying and accurately measuring performance of geometric clustering approaches.
There is also merit in examining hierarchical clustering methods, as certain types of data may fit this model well; however, parameters will need to be tuned to produce groupings with sufficient moderate sized clusters.
% We briefly experimented with the Markov Clustering (MCL) algorithm \cite{van2000graph}, however performance (as we measured it) was not significantly improved on the social clustering task, and somewhat worse when clustering by preference similarity.
% This can be because the Pearson Correlation Coefficient's notions of similarity, dissimilarity and neutrality can be expressed easily in the geometric setting (i.e. as 1, -1, and 0 respectively), but can not be expressed well in the MCL formulation, where only similarity and neutrality can be expressed. 
% We also experimented with hierarchical clustering methods, but found difficulty in tuning the parameters in order to produce groupings with many moderate sized clusters.
% While these particular approaches did not appear to be helpful in our experiments, we believe that methods more amenable to this type of data may exist, or at least be good subjects for future research.

Two other challenges related to the clustering aspect of this work are finding new methods of determining the optimal number of clusters ($k$), and considering other distance functions.
In this work, we ran the entire experiment from beginning to end (cluster, predict, recommend) many times in order to measure the effect of cluster count changes. 
% This approach would not be feasible on larger data sets.
% It may be that certain heuristics exist: for example, a minimum cluster size of 500 agents is enough to capture many distinct groups and will usually provide sufficient training data to train an accurate trust link predictor for that group of users.
Searching for new methods of determining $k$ which are more computationally tractable than an exhaustive search, or finding heuristics that can guide this search, would be a useful and interesting research project.
Second, we clustered agents in this work on the basis of social circle overlap (Jaccard similarity of trusted users) and preference similarity (Pearson Correlation Coefficient observed in train set ratings).
While we've argued that each of these are fairly natural metrics, it would be interesting to explore new metrics, including those based on implicit preferences (e.g. browsing behaviour), categories of interest (e.g. types of items enjoyed), and other biographical factors of the agents (e.g. geographic location, age).
Each of these can plausibly be argued to be indicative of some facet of agent similarity, which in turn may be correlated to similarities in trust formulation procedures.

Other clustering options that we could explore include
seeing whether meaningful clusters exist first using
a statistic such as the Hopkins Statistic
and delving further to determine the number of clusters
in the dataset using a metric such as the Bayesian
Information Criterion \cite{DBLP:books/mk/HanKP2011}. This then may help to
direct the choices for clustering that are used.

Our machine learning methods could also be broadened.
We settled on logistic regression as the central method used
but exploring further the use of other choices such as SVM~\cite{DBLP:books/mk/HanKP2011}
may help to determine whether additional robustness with performance
could be achieved.
Logistic regression is valuable as it admits a simple probabilistic
interpretation, is quick to optimize and the weight vector learned
is highly interpretable; however, there may be limitations of
dealing with linearly separable data,  making it more challenging to compute
interesting feature combinations.

\noindent \textbf{Experimental set up}: Our work does not consider dynamic changes in the network or agent preferences over time.
For example, our method did not consider agents who had no preference data associated with them, that is, new agents joining the network.
In practice, this could be handled by simply assigning generic predictions for agents who lacked sufficient preference data on which to cluster them.
A periodic re-training of the models would also allow the system to account for changing preferences over time.
This dynamic process of agents entering the network could be simulated for our experiments by leaving out a sample of users from the initial processing, then adding them after clusters have been created already.
% Our methodology for clustering should allow a moderate number of users to be added to existing clusters based on existing distance measures. 
% This would likely degrade the performance of the overall solution over time, as the cohesiveness of clusters would suffer by greedily assigning new users to the best existing clusters.
% The periodic retraining mentioned above would then be applied.

Another area for possible expansion is in our use of the personalized cluster classifiers.
We did not learn a classifier for a cluster when that cluster had less than 1000 positive examples of outgoing trust links and 100 agents in it.
This step was taken to avoid learning very inaccurate classifiers, but some of the classifiers learned still fit the data related to the cluster significantly worse than a classifier trained on larger sample of random agents.
Therefore, it is worth exploring better ways of combining the ``local'' (cluster specific) predictions with the ``global'' predictions, similar to the procedure taken in the Personalized Trust Model \cite{PTM_Zhang_2008}.
Perhaps the weight given to a local trust model could be based on the difference in accuracy between the fit of that model to the agents it represents and the accuracy a generic classifier would achieve for those agents.
This way, local irregularities could still be learned, but in cases where data is sparse, a little help from a generic classifier can nudge predictions towards a more accurate final outcome.
This approach could also be taken to enable more ``truly individual'' personalization.
For users with a large amount of activity (thousands of friends and other users to compare preferences with), a single-user classifier could be trained, and the results of this classifier combined linearly with a cluster or global classifier, allowing truly individualized personalization, and a gradual ramp up from generic to individual solutions as more data becomes available.

In our work, we excluded users from experimentation who had fewer than 20 reviews.
This filtering procedure was inspired by Mauro et. al \cite{multi-faceted_mauro_2019}, but it imposes certain biases on the following evaluations.
Under this procedure, only the opinions and activities of the most active users are taken into account (only about 2\% of Yelp users have submitted at least 20 reviews).
In our earlier experiments, we sampled users randomly, and the results from this time tended to show a more dramatic difference between personalized and non personalized approaches using TrustMF (e.g. in Figure \ref{trustmf-social-weight-tune}).
It would be valuable to experiment with different procedures for sampling users from this data set.

There is also merit in examining how our model operates in other social networking contexts.
Epinions is a reasonable second case for us to explore, as it was also examined by \cite{multi-faceted_fang_2015}.
We conducted a preliminary study of Epinions data sets and noticed that the chance of a randomly picked review score being 5 (the highest) is over 70\%, while on Yelp the distribution is much more spread out, with the highest probability being only 35\% on a score of 4.
With this kind of bias in the data, we would expect even better score accuracy on the score prediction task when applying our methods.
It is also interesting to note that Epinions users typically have fewer friends (trusted users) and that with Epinions users submitting ratings to written text (rating others' reviews), there is vastly more feedback to examine.
All of these differences may provide greater insights into the conditions under which our model has the most value.
Expanding our study to other more elaborate datasets will also help to shed
light on the scalability of our particular approach.

There may be additional challenges when examining other social networks.
While many recent projects in trust modeling focus on data from social networks with a significant item rating component (as it is convenient to measure trust-aware recommendation accuracy on a set of reserved ratings as a proxy measure for the quality of novel predicted trust links), we acknowledge that many popular networks such as Twitter and Facebook lack a significant item rating component. 
In cases like these, it would likely be necessary to engage in a user study (like the one in \cite{gilbert_predicting_2009}) and survey actual users whether the predicted trust links appeal to them or not.
This work would be useful, especially if a data set can be publicly released, as more data where preferences are explicitly indicated by users (rather than inferred) will be a boon to future trust modeling research.

\subsubsection{Integrating with efforts to address digital misinformation}
Integrating our proposed approaches directly into the larger effort aimed at combating digital misinformation would be a rich area for future work.
Some subtopics which would be especially valuable to explore
include connecting to efforts on detecting content which has
been generated by bots \cite{ferrara2014bots}.
Our methods may be able to provide more insights into bot detection algorithms
or our algorithms may be able to adjust their predictions based on
information revealed to us about suspected bot nodes within the network.
It would also be useful to adjust our predictions of trust links
and the use of these outcomes towards addressing misinformation,
in view of the networking behaviour in the social media environment.
Work such as that of Tong et al. \cite{tong_2018}
conducts an analysis of how rumours spread amongst the network's peers.
They suggest where to seed factual information in order to increase
the odds of halting false information. Shao et al. \cite{shao2018anatomy}
also proposes a way to limit attention to those nodes which
are most critical for the flow of information.
Cho et al. \cite{cho2019uncertainty} reflect both on stemming false informers and
promoting true informers by examining more closely how beliefs
of users are updated over time, considering various types of network centrality.
What we are able to
learn about trusted links, together with a study of the accompanying
network relationships, may provide important insights for where to
focus effort aimed not just at identifying misinformation but also
at stemming its tide.

\subsubsection{Exploring new avenues with the use of data sets}
A persistent difficulty in applying trust models to social networks has been in finding appropriate data set and evaluation procedures.
Most previous attempts to apply trust models to real social networks have relied on networks that included a significant content rating component, such as Yelp, Epinions, and FilmTrust.
These networks are attractive primarily because of the ease of harvesting objective test sets from the data extracted from them.
How to tell if two agents should \textit{really} trust each other?
Simply check the correlation between the ratings they have given to content - if it's positive, they should trust each other.

One concern is the fact that some of these datasets, such as that
of Epinions, represent dated information (with the site now defunct). 
The most popular networks (Facebook, Reddit and Twitter) may be overlooked because they lack a significant content rating component.  
The issue of securing publically available datasets for
some of these platforms arises at times as well.

There may be value to making more of an effort in the future to
interrogate actual users of systems.
We note the work of Gilbert and Karahalios \cite{gilbert_predicting_2009}, where 35 participants were recruited for the experiment.
The authors had access to the data that the participants agreed to share with them - it was not necessary to convince Facebook to produce a data set for the researchers.
After the statistical analysis, a qualitative analysis was performed to help contextualize the errors in the system by interviewing the participants.
Cooperating more closely with the users of online social networks in this way will likely be impactful for future trust modeling research.

\subsubsection{Considering vulnerable users: the case of older adults}
Another direction for future research that would be especially
valuable to explore is making use of information about the needs and
preferences of a cluster of users that are known independently
before performing the data analysis proposed here to determine
trust relationships (of value in assisting users in coping with
misinformation).

Below we sketch our current thoughts for how to integrate
this prior information about the user base into the overall solution.
In this approach, we could in fact have some preconceived notions of the user
at hand due to what the user modeling community refers to as stereotypes \cite{bootstraping_burnette_2019}
(what that entire class of users is likely to generally prefer).

We have thought of this direction forward as part of our particular
interest in offering support for certain groups of users who are
especially vulnerable online. One such community is that of older adults.
It would be valuable to be able to carefully advise this
demographic about misleading content, and personalized trust
link prediction via clustering may thus be of use.
We have begun to examine the special considerations of this user base
when it comes to misinformation; we note that other research has already identified notable differences for older adults in social media \cite{coelho2016survey,lehtinen2009,wylie2014misinfo}. 
The framework presented in this paper could expand to integrate prior knowledge of its users in the following way.

Suppose we had a group of older adult users.
Per the algorithm of Section 3, the unsupervised learning method
could assign them to the same cluster, suggesting the same weight
of their trust indicators (if they have similar trust profiles).

For Trust Link prediction, besides defining trust indicators as we discuss
in Section 3.3.3, we could also take the general preference of the older adults
into consideration, allowing some finer granularity to the reasoning.
In other words, we could consider factors which generally
influence the preferences of this particular user base.
Ideally, predicting a trust link from this set of users to a certain
peer should be predicated both on what the algorithm of Section 3
suggests from its data-driven analysis and also by the known prior preferences
of the user base.

%After that, we would make recommendations based on the trust link prediction.
%At a high level, the process would be: Given information with positive feedback from older adult user1, we will recommend this information to user2 who has a high probability of
%an existing trust link with user1.

A more detailed view of how to expand the overall process (our preliminary ideas for doing so) is as follows:

\begin{itemize}
    \item We have some priors about the needs of users who fit certain stereotypes (e.g. older adults). We'd like to help those particular users.
    \item We do the clustering based on the methods in Section \ref{chapter_PMFTM}.
    \item We examine the clusters and observe that a large portion of users in some cluster embody one of our known stereotypes, e.g. in some cluster over half the users are older adults.
    \item We combine data driven with stereotype based predictions for those clusters where large proportions of the users embody a particular stereotype, therefore applying our prior about the needs of a stereotype group to a cluster of users that seem to largely embody that stereotype.
\end{itemize}

One way this process could be operationalized once
clustering is performed as in Section \ref{chapter_PMFTM}, with clusters of older adults located:

\begin{itemize}
	%\item \textbf{Clustering}
	%\begin{itemize}
	%	\item \textbf{Input:} Older adults $A={a_1,...,a_n}$ and distance measure function: $D: A\times A \rightarrow R$
	%	\item \textbf{Output:} An assignment of every older adult to a cluster, $C$
	%\end{itemize}
	%
	\item \textbf{Trust Link Prediction}
	\begin{itemize}
		\item \textbf{Input:} Trust indicators of older adults $I_1(a_i,a_j),...,$ $I_m(a_i,a_j)$ and preference effective function: $f: I -> \{0,1\}$.
		\item \textbf{Output:} $T(a_i, a_j)$, the score of $a_i$ trusting $a_j$'s recommendation.
		\item \textbf{Process:} $Predictor1$ takes all indicators and makes a prediction. $Predictor2$ makes an independent prediction on indicators with $f(I)=1$. Get $T(a_i, a_j)$ by combining the two prediction results.
	\end{itemize}
	
	\item \textbf{Information Recommendation}
	\begin{itemize}
		\item \textbf{Input:} Trust-link Prediction $T(a_i, a_j)$ and information scores from $a_j$'s feedback, $s_j$.
		\item \textbf{Output:} Recommendation scores for $a_i$, $R(a_i, s_j)$
	\end{itemize}
\end{itemize}

It would be interesting to delve further into these options for
combining prior knowledge and trust link prediction, in order to
provide richer recommendations to users.
While an approach such as this could be used for any subgroup with known preferences, we feel it especially worthwhile to continue to learn more about the specific needs of older adults, and would explore our new direction with respect to this user base, as a first step.

% \begin{figure}
% % Use the relevant command to insert your figure file.
% % For example, with the graphicx package use
%   \includegraphics{example.eps}
% % figure caption is below the figure
% \caption{Please write your figure caption here}
% \label{fig:1}       % Give a unique label
% \end{figure}
% %
% % For two-column wide figures use
% \begin{figure*}
% % Use the relevant command to insert your figure file.
% % For example, with the graphicx package use
%   \includegraphics[width=0.75\textwidth]{example.eps}
% % figure caption is below the figure
% \caption{Please write your figure caption here}
% \label{fig:2}       % Give a unique label
% \end{figure*}
% \begin{table}
% % table caption is above the table
% \caption{Please write your table caption here}
% \label{tab:1}       % Give a unique label
% % For LaTeX tables use
% \begin{tabular}{lll}
% \hline\noalign{\smallskip}
% first & second & third  \\
% \noalign{\smallskip}\hline\noalign{\smallskip}
% number & number & number \\
% number & number & number \\
% \noalign{\smallskip}\hline
% \end{tabular}
% \end{table}

\section*{Declarations}

\section*{Funding}
Not applicable.

\section*{Conflicts of interest/Competing interests}
Not applicable.

\section*{Availability of data and material}
The dataset used in this work can be downloaded from \url{https://www.yelp.com/dataset} (the 2019 full dataset).

\section*{Code availability}
Algorithms as explained in the paper.

\begin{acknowledgements}
We are grateful to the reviewers for their valuable
feedback on an earlier version of this paper.
\end{acknowledgements}


% \newpage
\bibliographystyle{spmpsci}      % mathematics and physical sciences
\bibliography{mybibfile}   % name your BibTeX data base

% \newpage
\appendix
\section{Computation of Trust Indicators}
\label{section_computation}

In this appendix we discuss the challenge of the computation of
trust indicators between pairs of agents and adjustments that we made.
The trust indicator function $\Psi(a_i, a_j)$ is expected to be computed for all ordered pairs of agents.
Of course, as there are $O(n^2)$ possible pairs of agents, this rapidly becomes a computational issue as the number of agents considered grows.
In our experiments we worked with groups of agents where $|A| \approx 30000$, implying approximately 900,000,000 pairs - a large but tractable computation on modern consumer hardware. 
However, the unfiltered Yelp data set contains descriptions of 1,637,138 agents, and we can be assured that other large online environments contains many millions of users.
At this scale, the $O(n^2)$ computation time becomes a serious barrier, and storing the trillions of resulting vectors for further processing would likely be extremely costly.

However, it is not necessary to consider \textit{every} possible pair of agents. 
For example, if $a_i$ and $a_j$ have never interacted in any meaningful way and share no known interests -- in sum, we have no evidence of any way they might know or be interested in each other -- then it is likely safe to conclude, without any complex trust modeling, that they need not trust each other.
Further, we can conclude that the lack of a trust link between them is most likely the result of ignorance rather than opinion. 
To analogize, the potential trust relationship between a university professor in China and a wheat farmer in Canada need not be explicitly modeled and computed if no evidence can be found that the two may in fact share a communication channel or desire to interact in the future.

Thus a solution to the computation barrier presents itself: simply defining a neighborhood function, $N(a)$, on individual agents and only computing trust indicators and trust predictions between pairs of agents in the same neighborhood.
So long as computing $N(a)$ is efficient, the execution time of computing all \textit{relevant} trust indicator pairs then becomes linear with a constant bounded by the maximum neighborhood size.

The definition of $N(a)$ can be very liberal and still result in a substantial speed up.
For example, when computing trust indicators for the Yelp and Epinions data set, we used:
\begin{equation}
\begin{gathered}
a_j \in N(a_i) \; \iff \\
|R_{ij}| > 0 \vee friends(a_i, a_j)\; \vee friendOfFriend(a_i, a_j)
\end{gathered}
\end{equation}
That is, $a_j$ is in the neighborhood of $a_i$ if they have both reviewed at least one item in common, if they are friends, if they are friends of friends, or if they are friends of friends.

Applying this neighborhood function drastically reduces the number of pairs of agents that need to be considered in the following stages.

\end{document}